\documentclass[12pt, a4paper]{article}
\pdfoutput=1

\usepackage{amsmath}
\usepackage{amsfonts}
\usepackage{amssymb}
\usepackage{graphicx, rotating}
\usepackage{epsfig}
\usepackage{array}
\usepackage{latexsym}
\usepackage{graphicx}
\usepackage{xcolor}
\usepackage{amsmath,bm,amssymb}
\usepackage{cite}
\usepackage{slashed}
\usepackage{hyperref}
\usepackage[utf8]{inputenc}
\usepackage[export]{adjustbox}
\usepackage{multirow, verbatim}

\hypersetup{colorlinks, citecolor=bluscuro, linkcolor=black, urlcolor=bluscuro}
\definecolor{rossos}{cmyk}{0,1,1,0.55}
\definecolor{bluscuro}{rgb}{0.15, 0.2, .85}
\definecolor{bluchiaro}{cmyk}{1,.3,0.,0.1}

\begin{document}

\pagenumbering{gobble}
\begin{titlepage}

\begin{figure}
\href{https://lisa.pages.in2p3.fr/consortium-userguide/wg_cosmo.html}{                                                     
\includegraphics[width = 0.2 \textwidth, right]{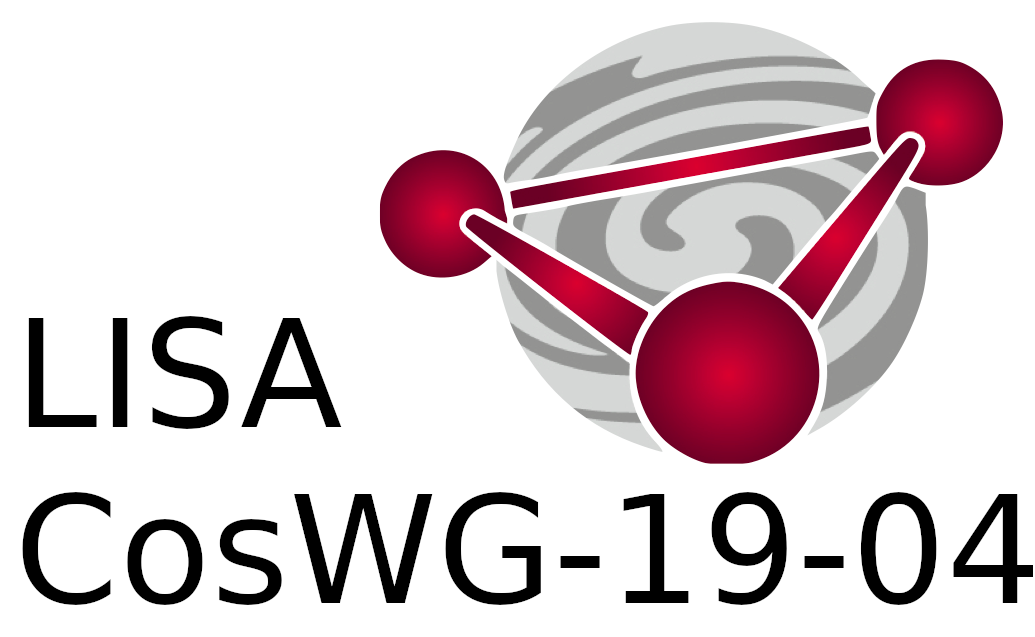}
}
\end{figure}

\begin{flushright}
\small
DESY 19-159\\
IPPP/19/27 \\
HIP-2019-14/TH \\
MITP/19-066\\
IFT-UAM/CSIC-19-139
\end{flushright}
\vspace{.3in}

\renewcommand*{\thefootnote}{\fnsymbol{footnote}}
\setcounter{footnote}{1}

\vspace{-0.75cm}
\begin{center}
{\Large\bf
Detecting gravitational waves from cosmological  phase \\
\vspace{.2cm} transitions with LISA: an update
}\\
\bigskip\color{black}
\vspace{1cm}{
  {\large
Chiara Caprini$^a$,
Mikael Chala$^{b,c,}$\footnote{Corresponding author: mikael.chala@ugr.es},
Glauber C.~Dorsch$^d$,
Mark Hindmarsh$^{e,f}$,
Stephan J.~Huber$^f$,
Thomas Konstandin$^{g,}$\footnote{Project coordinator: thomas.konstandin@desy.de},
Jonathan Kozaczuk$^{h,i,j,}$\footnote{Corresponding author: jkozaczuk@ucsd.edu},
Germano Nardini$^k$,
Jose Miguel No$^{l,m}$,
Kari Rummukainen$^e$,
Pedro Schwaller$^n$,
Geraldine Servant$^{g,o}$,
Anders Tranberg$^k$,
David J. Weir$^{e,p,}$\footnote{Corresponding author: david.weir@helsinki.fi}
  }\\
  \texttt{For the LISA Cosmology Working Group}
\vspace{0.3cm}
} \\[7mm]

\bigskip

\begin{abstract}
{We investigate the potential for observing gravitational waves from cosmological phase transitions with LISA in light of recent theoretical and experimental developments. Our analysis is based on current state-of-the-art simulations of sound waves in the cosmic fluid after the phase transition completes. We discuss the various sources of gravitational radiation, the underlying parameters describing the phase transition and a variety of viable particle physics models in this context, clarifying common misconceptions that appear in the literature and identifying open questions requiring future study. We also present a web-based tool, \texttt{PTPlot}, that allows users to obtain up-to-date detection prospects for a given set of phase transition parameters at LISA.}
\end{abstract}

\newpage

{\it $^a$ {Laboratoire Astroparticule et Cosmologie, CNRS UMR 7164, \\
Universit\'e Paris-Diderot, 75013 Paris, France}}\\
{\it $^b$ {Institute of Particle Physics Phenomenology, Physics Department, \\
    Durham University, Durham DH1 3LE, UK}}\\
{\it $^c$ {CAFPE and Departamento de F\'isica Te\'orica y del Cosmos, \\
Universidad de Granada, E-18071 Granada, Spain}}\\
{\it $^d$ {Center for Astrophysics and Cosmology \& PPGCosmo, \\
 Universidade Federal do Esp\'irito Santo, 29075-910, Vit\'oria, ES, Brazil}} \\
{\it $^e$ {Department of Physics and Helsinki Institute of Physics, PL~64, \\
    FI-00014 University of Helsinki, Finland}} \\
{\it $^f$ {Department of Physics and Astronomy, University of Sussex, Brighton BN1 9QH, UK}} \\
{\it $^g$ {DESY, Notkestr.  85, 22607 Hamburg, Germany}} \\
{\it $^h$ {Department of Physics, University of California, San Diego, La Jolla, CA, 92093, USA}} \\
{\it $^i$ {Department of Physics, University of Illinois, Urbana, IL, 61801, USA}} \\
{\it $^j$ {Amherst Center for Fundamental Interactions, Department of Physics, \\ 
		University of Massachusetts, Amherst, MA, 01003, USA}} \\
{\it $^k$ {Faculty of Science and Technology, University of Stavanger, 4036 Stavanger, Norway}} \\
{\it $^l$ Departamento de Fisica Teorica, Universidad Autonoma de Madrid, \\
Cantoblanco, 28049, Madrid, Spain} \\
{\it $^m$ Instituto de Fisica Teorica, IFT-UAM/CSIC,Cantoblanco, 28049, Madrid, Spain} \\
{\it $^n$ { PRISMA$^+$ Cluster of Excellence and Mainz Institute for Theoretical Physics, \\
Johannes Gutenberg-Universit\"at Mainz, 55099 Mainz, Germany}} \\
{\it $^o$ II. Institute of Theoretical Physics, University of Hamburg, D-22761 Hamburg, Germany} \\
{\it $^p$ {School of Physics and Astronomy, University of Nottingham, Nottingham NG7 2RD, UK}}
\end{center}
\bigskip

\vspace{.4cm}

\renewcommand*{\thefootnote}{\arabic{footnote}}
\end{titlepage}
\pagenumbering{arabic}

\tableofcontents

\section{Introduction}

\newcommand{\bit}{\begin{itemize}}
\newcommand{\eit}{\end{itemize}}

\newcommand{\mpl}{m_{\text{P}}} 
\newcommand{\Mpl}{M_{\text{P}}} 

\newcommand{\Hc}{H_*} 
\newcommand{\HN}{H_\text{n}} 
\newcommand{\Nb}{N_\text{b}} 
\newcommand{\nb}{n_\text{b}} 
\newcommand{\Rc}{R_\text{c}} 
\newcommand{\Rstar }{R_*} 
\newcommand{\Tc}{T_\text{c}} 
\newcommand{\vc}{v_\text{c}} 
\newcommand{\Ec}{E_\text{c}} 
\newcommand{\Sc}{S_\text{c}} 
\newcommand{\TN}{T_\text{n}} 
\newcommand{\vw}{v_\text{w}} 
\newcommand{\tN}{t_\text{n}} 

\newcommand{\NucRatVol}{\mathcal{P}} 
\newcommand{\LatHea}{\mathcal{L}} 
\newcommand{\EneDen}{e} 
\newcommand{\TraAno}{\theta} 
\newcommand{\Pre}{p} 

\newcommand{\OmVac}{\Omega_\text{vac}} 

\newcommand{\OmGW}{\Omega_\text{gw}} 
\newcommand{\OmGWscaled}{\tilde\Omega_\text{gw}}
\newcommand{\OmGWnow}{\Omega_{\text{gw},0}}
\newcommand{\fitfun}{C}
\newcommand{\fp}{f_\text{p}} 
\newcommand{\fpnow}{f_{\text{p},0}} 
\newcommand{\zp}{z_\text{p}} 

\newcommand{\tShock}{\tau_\text{sh}}
\newcommand{\tTurb}{\tau_\text{tu}}
\newcommand{\tOn}{\tau_\text{v}}
\newcommand{\tCorr}{\tau_\text{c}}

\newcommand{\fluidL}{L_\text{f}}
\newcommand{\fluidV}{\bar{U}}
\newcommand{\cs}{c_\text{s}}

\newcommand{\vCJ}{v_\text{CJ}} 
\newcommand{\AdiInd}{\Gamma} 

\newcommand{\blu}{\color{blue}}
\newcommand{\roig}{\color{red}}
\newcommand{\eq}[1]{Eq.~(\ref{#1})}
\newcommand{\lag}{\mathcal{L}}
\newcommand{\lp}{\left(}
\newcommand{\rp}{\right)}
\newcommand{\nn}{\nonumber}
\newcommand{\pslash}{\!\not\! p}
\newcommand{\kslash}{\!\not\! k}
\newcommand{\qslash}{\!\not\! q}
\newcommand{\deslash}{\!\not\! \partial}
\newcommand{\Dslash}{\!\not\!\!  D}
\newcommand{\vev}[1]{\langle {#1} \rangle}
\newcommand{\be}{\begin{equation}}
\newcommand{\ee}{\end{equation}}
\newcommand{\bea}{\begin{eqnarray}}
\newcommand{\eea}{\end{eqnarray}}

\newcommand{\TeV}{\,\mathrm{TeV}}
\newcommand{\GeV}{\,\mathrm{GeV}}
\newcommand{\MeV}{\,\mathrm{MeV}}
\newcommand{\keV}{\,\mathrm{keV}}

\def\half{\frac{1}{2}\,}
\def\Re{{\rm Re\,}}
\def\Im{{\rm Im\,}}
\def\Tr{{\rm Tr\,}}
\def\tr{{\rm tr\,}}
 \def\ra {\rightarrow}

\newcommand{\arXiv}[2]{\href{http://arxiv.org/pdf/#1}{{\tt [#2/#1]}}}
\newcommand{\arXivold}[1]{\href{http://arxiv.org/pdf/#1}{{\tt [#1]}}}

\def\b{\bar a}
\def\bphi{\overline \phi}

 \def\al{\alpha}
 \def\b{\beta}
 \def\ga{\gamma}
 \def\de{\delta}
 \def\De{\Delta}
 \def\eps{\epsilon}
 \def\ep{\varepsilon}
 \def\ze{\zeta}
 \def\th{\theta}
 \def\la{\lambda}
 \def\La{\Lambda}
 \def\si{M}
 \def\ph{\varphi}
 \def\Ph{\Phi}
 \def\om{\omega}
 \def\Om{\Omega}
 \def\df{\delta\phi}
\def\lra#1{\overset{\text{\scriptsize$\leftrightarrow$}}{#1}}
\def\bma#1{\mbox{\boldmath{$#1$}}}


%

The direct observation of gravitational waves (GWs) by the LIGO collaboration~\cite{TheLIGOScientific:2016pea} has ushered in a new era of GW astronomy. In addition to observing gravitational radiation from astrophysical sources, GW interferometers provide direct access to the physics of the early Universe. Little is known about the cosmic epoch after reheating and prior to Big Bang nucleosynthesis (BBN) as there are few observables directly or indirectly sensitive to this era. However, while opaque to light before photon decoupling, the early Universe was transparent to GWs. This allows for the exciting possibility of probing the pre-BBN Universe for the first time using GW experiments.     

First-order cosmological phase transitions (PTs) provide a particularly compelling source of GWs in the early Universe. At a first-order cosmological PT, bubbles of a new phase begin to nucleate and expand as the Universe cools.  The region inside the bubbles contains the new phase, typically characterized by a vacuum expectation value (VEV) of a scalar field that differs from its value outside the bubbles. The collision of the bubbles and the resulting motion of the ambient cosmic fluid sources a stochastic GW background that can be observable at GW interferometers. In the Standard Model (SM) of particle physics, a PT could in principle have occurred when either the electroweak (EW) gauge symmetry or the approximate chiral symmetry of QCD were spontaneously broken in the early Universe. Given the mass of the Higgs boson, it is known that the EW symmetry is broken at a cross-over in the SM~\cite{Kajantie:1995kf, Csikor:1998eu}. Likewise, absent large quark chemical potentials in the early Universe, the QCD PT of the SM is known not to be first-order~\cite{Stephanov:2007fk}. However, many compelling and well-motivated extensions of the SM predict strong first-order cosmic PTs at the EW scale and beyond. 

First order PTs associated with the EW to multi--TeV scales typically predict a stochastic background peaking at frequencies accessible by the Laser Interferometer Space Antenna (LISA)~\cite{Audley:2017drz}. These scales are particularly interesting as they are often thought to harbor new physics related to e.g.~the hierarchy problem, dark matter (DM),  or baryogenesis. Searching for a stochastic GW background from physics beyond the Standard Model (BSM)  will be an important science goal of the LISA mission. In doing so, LISA will provide a powerful probe of new physics that complements collider and other experimental tests of these scenarios. LISA has been selected by the European Space Agency as one of its flagship missions, and is currently in the planning stages, with an expected launch in the early 2030s. It is thus important and timely to study the science case for LISA as thoroughly and accurately as possible, including its prospects for detecting a stochastic GW background from cosmic PTs. 

\begin{figure}[!t]
\begin{center}
\includegraphics[width=1.1\textwidth]{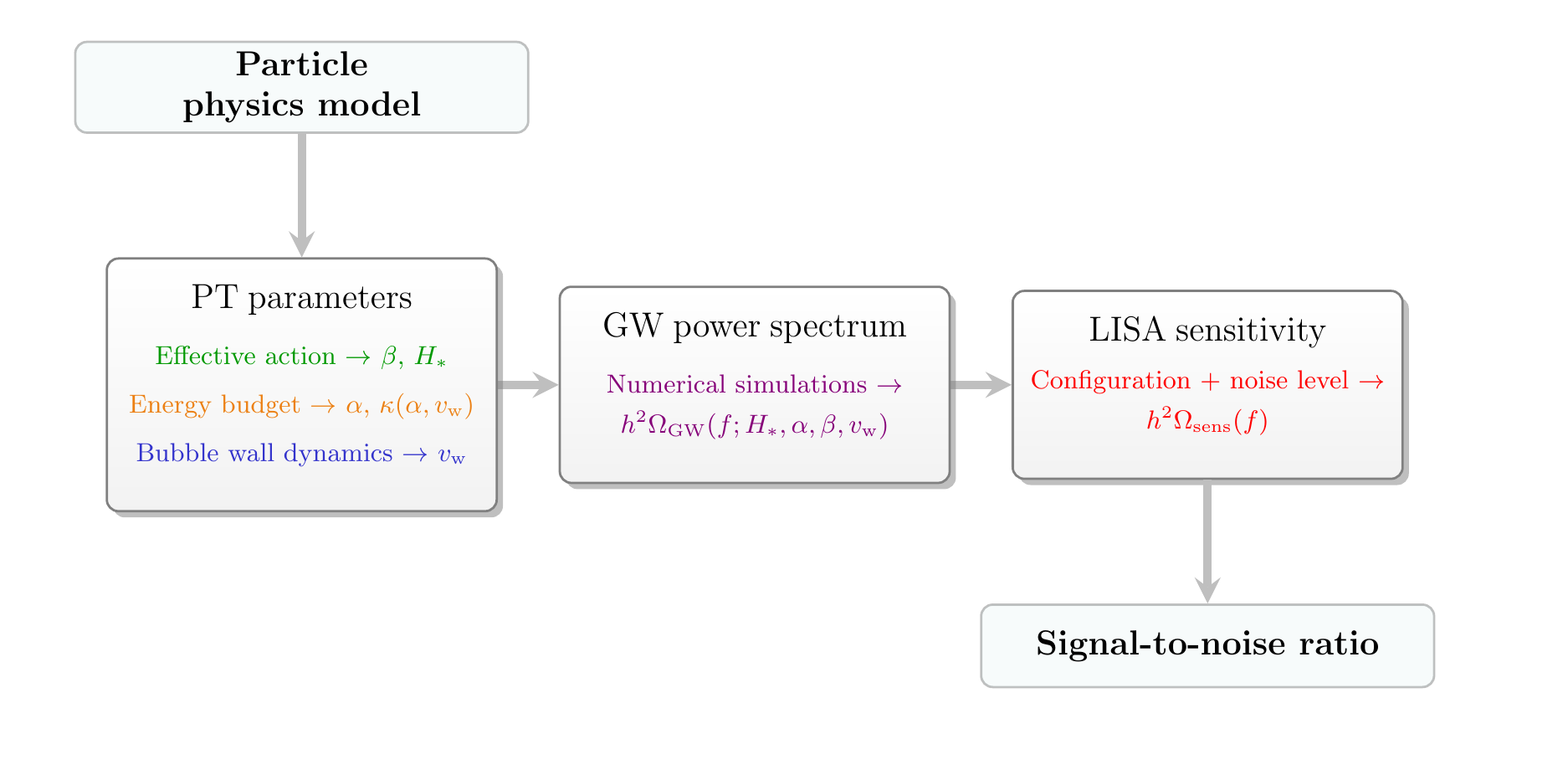}

\caption{\small Blueprint for analyzing cosmological PTs in the context of LISA. See text for details.
}
\label{fig:pipeline}
\end{center}
\end{figure} 

Several important steps are required to accurately predict the GW signal arising from a cosmological PT in a given particle physics model at LISA (and other GW experiments). Schematically, a typical analysis proceeds as illustrated in Fig.~\ref{fig:pipeline}. First, one specifies the field content and Lagrangian of a given particle physics model. From this information, the phase structure of the theory can be analyzed at finite temperature. If the model admits a first-order PT, the finite temperature effective action can be used to compute the Hubble parameter $H_*$ (or temperature $T_*$) at the time when the PT completes and the PT duration $\beta$ (defined in more detail below). Next, one can analyze the energy budget associated with the PT to determine its strength, characterized by the parameter $\alpha$ (see Sec.~\ref{sec:prelim}), and the amount of energy converted into fluid kinetic energy, often characterized by an efficiency parameter $\kappa$, again defined below\footnote{As explained in Sec.~\ref{sec:prelim}, $\kappa$ can be expressed as a function of $\alpha$ and $v_{\rm w}$, so it is not an independent parameter.}. Simultaneously, one should self-consistently solve the bubble wall equation(s) of motion (EOM) for the bubble wall speed $v_{\rm w}$. These PT parameters are then used as inputs for the determination of the stochastic GW power spectrum $h^2\Omega_{\rm GW}(f)$ in numerical simulations of the colliding bubbles and cosmic fluid. In practice, analytic expressions for the GW power spectrum derived from the simulations are  typically used in this step rather than directly simulating for each choice set of parameters. Finally, given a particular experimental configuration and knowledge of the noise level, one can obtain the predicted LISA sensitivity $h^2 \Omega_{\rm sens}(f)$ to cosmological sources. Comparing the predicted power spectrum to the LISA sensitivity for a given mission duration yields a signal-to-noise ratio (SNR) which indicates to what extent the scenario under consideration can be reconstructed~\cite{Karnesis:2019mph, Caprini:2019pxz}.

Each step of the analysis described above carries with it a set of technical challenges and open questions. A primary aim of this work is to elucidate these issues and propose a conservative approach to obtaining state-of-the-art sensitivity estimates for LISA in detecting a stochastic GW from cosmic PTs. In what follows, we will attempt to clear up common misconceptions that arise in the literature, discuss the state-of-the-art in the various steps of Fig.~\ref{fig:pipeline}, and, in cases where there remains some ambiguity or lack of concrete results, suggest how to proceed. We will then apply these results to specific models to present up-to-date prospects for detection of a corresponding stochastic GW background at LISA. We primarily focus on GWs produced by sound waves, thought to be the dominant source in most cases of interest as discussed further below (for an overview of other possible sources we refer the reader to our previous study~\cite{Caprini:2015zlo}). We will show that, in its current incarnation, LISA will provide a powerful probe of cosmic PTs in many well-motivated extensions of the SM. Given the recent progress reflected in this work, our study should be understood as an update to our previous publication~\cite{Caprini:2015zlo}. The results presented here can also be used to study GWs from cosmic PTs at other GW experiments. 

Given that there remain several open questions related to determining the stochastic GW power spectrum from first-order cosmological PTs, and that the LISA mission details have not yet been finalized, predictions for the detectability of a given model at LISA may change somewhat over the coming years. To assist the community in performing up-to-date analyses, we introduce a web-based tool, \texttt{PTPlot}, that allows users to determine whether GWs from a PT in a given model can be detected by LISA. This tool will be updated periodically to reflect predictions from the most state-of-the art simulations of GW generation at PTs, and feature up-to-date sensitivity curves for the LISA experiment. 

This remainder of work is structured as follows. In Sec.~\ref{sec:prelim} we discuss preliminaries such as the characteristics of the PT (typical bubble size, energy budget, wall velocity). In Sec.~\ref{sec:signal} we explain our assumptions regarding the GW spectrum that 
accounts only for sound shells as seen in simulations. Next, in Sec.~\ref{sec:tool} we present the tool \texttt{PTPlot} and describe its usage. The computation of the PT parameters in specific models, along with the related uncertainties, are discussed in Sec.~\ref{sec:nonpert}, with some further details provided in Appendix~\ref{sec:appendixlattice}. Finally, we present the prospects for observing the stochastic background in concrete BSM scenarios in Sec.~\ref{sec:models}. We conclude in Sec.~\ref{sec:summ}.


\section{Preliminaries: phase transition dynamics and parameters $(H_*, \beta, \alpha, \vw)$\label{sec:prelim}}

In this section, we discuss preliminaries for the determination of the GW power spectrum. In particular, we review  the properties of cosmological PTs
relevant for GW production. 
The focus is on open questions and recent developments in the literature rather than on  
an introductory explanation of the material. For a more pedagogical treatment see~\cite{Caprini:2015zlo}.
In general, the signal is a sum of contributions originating from distinct sources: the scalar field (bubble wall) kinetic energy, 
the sound waves from the bulk fluid, and turbulent motion. As will be argued below, in this paper we will eventually estimate the GW signal coming from sound waves only, to be conservative.

\subsection{Bubble nucleation and PT duration}

GW production at a first order PT in the
early Universe principally depends on four parameters, which determine
the length scale, amplitude, and 
lifetime of the shear stress perturbations.

At a first-order PT, the Universe is trapped in a
metastable state.  It can escape either by thermal fluctuation or by
quantum tunneling; in either case the escape route is via the
appearance of spherical bubbles of the stable phase with a microscopic
critical radius $\Rc$.  The bubbles are nucleated at a rate per unit
volume $\NucRatVol$, which depends exponentially on the critical
bubble action\footnote{This is the action for the smallest possible 
bubble able to expand. The word ``critical" here is unrelated to the critical temperature $\Tc$ defined below, which is the temperature 
below which bubble nucleation can occur.} $\Sc$:
\begin{equation}
  \NucRatVol = A(t) e^{-\Sc(t)}.
\end{equation}
The collisions of the bubbles, and the subsequent fluid flows, produce
the shear stresses which source GWs.

In the thermal case, the high-temperature phase becomes metastable at
a critical temperature $\Tc$.  Below $\Tc$ the bubble action decreases
from infinity and the nucleation probability increases rapidly.  The
rate of increase, defined as
\begin{equation}
\beta(T) = \frac{d}{dt} \ln \NucRatVol(T)  \, ,
\end{equation}
is an important quantity, as we will explain below.  It is a very good
approximation to take $\beta(T) = - d \Sc/dt $ as long as $S_c \gg 1$ and $\beta \gg H$.

When the bubbles appear they grow due to a pressure difference $\Delta
p$ between the interior and exterior, which is non-zero below the
critical temperature. The onset of the PT is characterized by the nucleation of one 
bubble per horizon volume on average, which corresponds roughly to
\be
\label{eq:Sc_naive}
S_c \simeq 140 
\ee
for EW-scale transitions. As the bubbles grow and more appear, the
fraction of the Universe in the metastable phase decreases
extremely rapidly~\cite{Enqvist:1991xw}, leading to bubble percolation for the PT to successfully 
complete.  
While the nucleation rate increases, the volume
into which bubbles are nucleating decreases, and so the volume
averaged nucleation rate reaches a maximum. 
The estimate~\eqref{eq:Sc_naive} can be improved by taking the precise value of the
functional determinant $A(t)$ and the expansion velocity of the bubbles into account. However, for EW baryogenesis
and the production of GWs, the time of percolation is more 
relevant than the onset of the PT. 
The criteria for the onset of the transition~\cite{Moore:1995si, Huber:2007vva} and the end of the PT (which should
more or less coincide with the time of percolation) in a radiation-dominated epoch
generally read
\bea
\label{eq:Sc_improved_TN}
S_c(\textrm{onset}) &\simeq& 141 + \log(A/T^4) - 4 \log \left(\frac{T}{100 \, {\rm GeV}}\right) - \log\left(\frac{\beta/H}{100}\right) \, , \\
S_c(\textrm{percolation}) &\simeq& 131 + \log(A/T^4) - 4 \log \left(\frac{T}{100 \, {\rm GeV}}\right) - 4\log\left(\frac{\beta/H}{100}\right) + 3 \log(v_{\rm w}) \, .
\label{eq:Sc_improved}
\eea
Here $H$ denotes the Hubble parameter and the prefactor $A(t)$ for the electroweak phase transition (EWPT) turns 
out to be $\log(A/T^4) \simeq -14$~\cite{Carrington:1993ng}. 
For moderately strong PTs, we do not need to distinguish between the Hubble parameter at the nucleation temperature $T_{\rm n}$ ($H_{\rm n}$, defined by solving~(\ref{eq:Sc_improved_TN})) and the 
Hubble parameter during percolation ($H_*$, defined by solving~(\ref{eq:Sc_improved})) as their values are very similar.

In some cases, the percolation temperature $T_*$ determined from (\ref{eq:Sc_improved}) can lie significantly below the PT critical temperature $\Tc$ 
(when the original phase becomes metastable). In this strong supercooling regime, the vacuum energy difference between the phases can come to dominate the 
energy density of the Universe. This modifies the percolation criterion above and in some cases makes it difficult for the PT to complete 
(models with non-polynomial potentials are an important exception, as discussed in Sec.~\ref{sec:warped} and \ref{sec:composite}). In a given model predicting strongly supercooled transitions, 
it should therefore be verified that percolation does indeed occur, by e.g.~comparing the predicted percolation temperature from (\ref{eq:Sc_improved}) 
and that for which the Universe becomes vacuum energy dominated. 

The first two important parameters for GW production at a thermal PT are
then the percolation temperature $T_*$ (or better, the Hubble rate at
percolation, $H_*$), and the nucleation rate parameter
at this temperature, which we call $\beta$ without indicating a
temperature argument. This nucleation rate parameter ultimately
determines the mean bubble separation. Several possible bubble expansion modes exist (see e.g.~\cite{Espinosa:2010hh}) depending on the value of the bubble wall 
velocity $\vw$. 
For bubbles expanding as detonations, the typical separation between bubbles is set by the wall velocity $\vw$.
For bubbles expanding as deflagrations ($\vw < \cs$, where $\cs$ denotes the speed of sound in the plasma), the reheating of the plasma by the
reaction front can suppress further bubble formation for large enough $\alpha$. The mean bubble separation should be corrected to account for this effect. 
In what follows we therefore estimate the mean bubble separation as
\begin{equation}
  \label{eq:rstarbetavw}
  \Rstar = \frac{(8\pi)^\frac{1}{3}}{\beta} \, \text{Max}( \vw, \cs) \, .
\end{equation}
where for deflagrations we have simply replaced $\vw$ by $\cs$ as the relevant speed. 
Careful definition of $\Rstar$ is important, since features (such as
sound shells) of size $\Rstar$ are expected to carry the majority of
the energy of the transition. Further refinement of~\eqref{eq:rstarbetavw} would allow for more
accurate predictions in the case of deflagrations and warrants future investigation.

Note that in the case of very slow PTs occurring for tuned polynomial potentials~\cite{Turner:1992tz,Huber:2006wf,Megevand:2016lpr,Ellis:2018mja}, when 
$\beta/H_*$ becomes of order unity, \eqref{eq:rstarbetavw} breaks down and the mean bubble
separation at percolation $\Rstar$ must be calculated directly from
first principles~\cite{Huber:2007vva, Ellis:2018mja}, rather than via
\eqref{eq:rstarbetavw}.

\subsection{The energy budget}

The next relevant parameter is the amount of vacuum energy released
by the PT, which is then transferred into kinetic 
energy~\cite{Kamionkowski:1993fg,Espinosa:2010hh}.   We always assume that the total energy 
density of the Universe $e$ consists of two main components: radiation energy and vacuum energy from the scalar field potential.
The kinetic energy fraction can be estimated from the study of the
self-similar flows around expanding bubbles.  The kinetic energy in
the fluid as a fraction of the total energy originally contained in
the bubble is
\begin{equation}
K = \frac{3}{\EneDen\, \vw^3} \int \, w(\xi) \, v^2 \gamma^2 \xi^2 d\xi \, ,
\end{equation}
where $v=v(\xi)$ is the fluid velocity and $w = e+p$ is the enthalpy of the fluid and $\xi = r/t$ the coordinate of the
self-similar solution.  The fraction $K$ for one bubble is a good estimate
of the average for the whole fluid once the bubbles have collided, at
least for small fractions~\cite{Hindmarsh:2015qta}. 
Since the fluid velocity $v(\xi)$ and enthalpy $w(\xi)$ depend on both the wall speed $\vw$ and the strength of the PT 
(whose precise definition we discuss below), the one-bubble kinetic energy
fraction depends on these parameters too.

To simplify matters, many analyses use the bag model. In this model,
pressure and energy density in the plasma are given as
\begin{eqnarray}
\label{eq:bag_EoS}
p_+ = \frac13 \,a_+ T^4 - \epsilon \,  ,&& \quad e_+ = a_+ T^4 + \epsilon \, , \nn \\
p_- = \frac13 \,a_- T^4 \, ,&& \quad e_- =  a_- T^4  \, ,
\end{eqnarray}
where $a_\pm$ encode the number of relativistic degrees of freedom in
the plasma in the symmetric ($+$) and broken ($-$) phases, and
$\epsilon$ is the so-called bag constant  parametrizing the jump in energy density and pressure across the phase boundary. 
In the bag model, the
relative importance of the bag constant to the thermal energy can be
used to parameterize the strength of the PT, with a strength parameter $\alpha$ defined as
\begin{equation}
\alpha = \frac{\epsilon}{a_+ T^4} \, .
\end{equation}
This is the definition of $\alpha$ which is being used in most of the literature and which is very easy to compute in a given particle physics model 
where $\epsilon$ corresponds to the potential energy difference between the metastable and true minima.
The kinetic energy fraction is then often given in terms of an efficiency
factor~\cite{Kamionkowski:1993fg,Espinosa:2010hh} by normalizing it to
the bag constant,
\begin{equation}
\kappa = \frac{3}{\epsilon \, \vw^3} \int \, w(\xi) \, v^2 \gamma^2 \xi^2 d\xi \, .
\end{equation}
The kinetic energy fraction is then
\begin{equation}
\label{e:KinEneFra}
K = \frac{\kappa \alpha}{1 + \alpha} .
\end{equation}
Notice that for a deflagration the temperature in front of the wall and the nucleation temperature differ. 
Fits to the ratio of these temperatures and $\kappa$ as a function of $\alpha$ and $\vw$ are given in~\cite{Espinosa:2010hh}.

This leaves the question how the analysis in a concrete model can be linked to the 
calculation of the energy budget above (that uses the bag equation of state). 
A more general definition, not specific to the bag model, notes that
the bag constant is related to the trace of the energy momentum tensor $\theta = (\EneDen - 3p)/4$, and the
thermal energy density (i.e.~the energy density excluding the bag
constant) is given in terms of the enthalpy, $3 w/4$~\cite{Hindmarsh:2017gnf}.  This motivates the more
general definition 
\begin{equation}
\label{e:GenAlpDef}
\al_\theta \equiv \frac43 \frac{\Delta \TraAno(\TN)}{w_+(\TN)} \, ,
\end{equation}
where $\Delta$ denotes the difference between
broken and symmetric phase.
The definition of $\alpha$ is partly a matter of convention, and
having chosen the convention, the efficiency parameter $\kappa$ can
always be defined so that the kinetic energy fraction is given by
(\ref{e:KinEneFra}). 

Another reasonable choice to match a concrete model to the 
bag equation of state is the difference in energy between the two phases, $\Delta e(T_{\rm n})$, that is the natural input parameter when 
studying the energy budget of the PT~\cite{Espinosa:2010hh}
\begin{equation}
\label{e:GenAlpDef2}
\al_e \equiv \frac43 \frac{\Delta e(\TN)}{w_+(\TN)} \, .
\end{equation}
Before the advent of full hydrodynamic simulations, this definition
was most commonly used to quantify the amount of kinetic energy in
bulk motion, since it allowed one to easily connect the energy budget
calculation to results from the envelope
approximation\footnote{Notice that the quantity $\Delta e$ is often
  called latent heat even though technically the latent heat is a
  difference in enthalpy, $\LatHea = \Delta w(\Tc)$.}. For strong
PTs (i.e.~$\alpha \gtrsim 1$), $\Delta e$ and $\Delta \theta$ are very similar
since the zero temperature potential will dominate the energy density
and free energy, thus $\alpha \sim \alpha_{\theta}\sim \alpha_e$. For weak PTs, the pressure difference
will be rather small and the energy difference $\Delta e$ will be
roughly \textsl{four times} the trace $\Delta \theta$.  Therefore, the
definition (\ref{e:GenAlpDef2}) tends to \textsl{overestimate} the
kinetic energy in this regime.

On one hand, the bag model (\ref{eq:bag_EoS}) is rather crude, and in
principle an analysis along the lines of~\cite{Kamionkowski:1993fg,
  Espinosa:2010hh} is required in every specific model, using the equation of state evaluated
in a better approximation, such as the high temperature expansion.
Such analyses are currently lacking\footnote{A general feature
  expected in more realistic models is that the speed of sound is
  smaller in the broken phase; see~\cite{Leitao:2014pda} for
  calculations of $\kappa$ in this case.}.  On the other hand, 
it was already noticed in~\cite{Espinosa:2010hh} and~\cite{Hindmarsh:2017gnf} that the main quantity entering the hydrodynamic analysis is actually $\Delta \theta$ even when the analysis does not rely on the bag equation of state. Hence, besides relatively minor effects (as the fluid velocity and enthalpy profiles around the bubble depend on the sound speed)
the hydrodynamic analysis of $\kappa$ in terms of $\Delta \theta$ should be model independent. 
At the current state of
knowledge of the GW power spectrum, an evaluation of
$\alpha$ from the effective potential and
(\ref{e:GenAlpDef}), and using the fitting functions for $\kappa$ in
\cite{Espinosa:2010hh} is sufficient for estimating the GW signal of particle physics models. In what follows
we use~\eqref{e:GenAlpDef}, dropping the $\theta$ subscript in our notation.

\subsection{The wall speed}
\label{ss:WalSpe}

The last important ingredient in the prediction of the GW spectrum is the bubble wall velocity, $\vw$. This is the speed of the 
phase interface after nucleation in the rest frame of the plasma far from the wall. The velocity is particularly important since it impacts the energy budget of the PT~\cite{Espinosa:2010hh}. 
Generally, a small wall velocity will limit the prospects for observing a GW signal regardless of the details of the GW production mechanism. 

Determining the wall velocity is one of the more involved aspects of PT calculations. While many quantities entering the energy budget can be 
determined by equilibrium or hydrodynamic considerations, the wall velocity requires an out-of-equilibrium calculation. This is done using a combination of 
Boltzmann and scalar field equations~\cite{Moore:1995ua, Moore:1995si, Konstandin:2014zta, Kozaczuk:2015owa}.  
The actual calculation depends on the scalar sector of the theory under consideration as well as the particles that obtain masses through their coupling to one 
of the scalars. Due to the dependence on the scalar sector and specific particle content, scenarios must be treated on a model-by-model basis.

Recently, a more model independent approach was presented in \cite{Dorsch:2018pat} which applies to transitions where the scalar field undergoing the transition
is identified with the SM Higgs field, which drives the bubble 
expansion and experiences the majority of the friction. The approach is based on using the criterion from bubble nucleation to reduce the number of relevant 
parameters of the bubble wall velocity to two, namely the pressure difference in units of the Higgs VEV, $\Delta V/\phi_0^4$, and the strength of the PT, 
$\xi = \phi_0/T$. Most wall velocities in our models section do not yet apply these recent results.

Another limit where the bubble wall velocities can be more easily inferred are very strong PTs. Since our last report on 
PTs~\cite{Caprini:2015zlo}, a significant new  result has
appeared related to the calculation of the wall speed in the relativistic regime~\cite{Bodeker:2017cim},
which prompts a reanalysis of the GW power spectrum
for very strong PTs.

The original analysis~\cite{Bodeker:2009qy} performed a leading order
evaluation of the friction on a PT bubble wall moving at
ultrarelativistic speeds $\gamma=\sqrt{1-v_{\rm w}^2} \gg 1$, and showed that the pressure
due to the flux of particles crossing the bubble wall saturates as
$\gamma \to \infty$. In the reference frame of the bubble wall, this
can be understood from the interplay of the density of incoming
particles $\propto \gamma \,T^3$, and the mean momentum transfer per
particle $\Delta m^2/(2E) \propto \gamma^{-1}$.  For $N$ degrees of
freedom, the maximal pressure difference is then given
by~\cite{Bodeker:2009qy}
\begin{equation}
\Delta \Pre_\text{fr} = \Delta m^2 \, N \,\int \frac{d^3p}{(2\pi)^3} \frac{f(p)}{2E_p} \sim \Delta m^2 \, N \, T^2 \, ,
\end{equation}
where $\Delta m^2$ denotes the change of mass across the
bubble wall, $f(p)$ is the particle distribution function at
equilibrium, and $E_p$ is the energy of a particle with momentum $p$.
This coincides with the leading mean-field contribution to the effective potential, and so runaway behavior is
possible if the free energy in the low-temperature phase exceeds that in the metastable phase in the mean-field approximation~\cite{Bodeker:2009qy}. 
If this criterion is met, the wall will keep on accelerating to highly
relativistic velocities.

In the original work~\cite{Bodeker:2009qy} it was noted that higher
order effects could change this picture.  Such effects have been
recently discussed in~\cite{Bodeker:2017cim}, the most relevant
being particle production when masses or interaction strengths change across
the phase boundary. These processes are analogous to transition
radiation in electrodynamics.

For the EWPT (and in general, for the spontaneous
breaking of a gauge symmetry) the dominant friction contribution at
highly relativistic bubble wall velocities comes from transition
radiation of gauge bosons acquiring mass at the interface, and scales as
\begin{equation}
\label{eq:pressure_nc}
\Delta \Pre_\text{fr} \sim \gamma\, \alpha_w\, N^\prime \, \Delta m\, T^3 \, ,
\end{equation}
with $\alpha_w=g^2/(4\pi)$ the corresponding gauge coupling and $N^\prime$ the number of species participating in the transition radiation. 
As a result, many models that were believed to exhibit runaway behavior will actually develop a
finite wall velocity, with $\gamma_\text{s} \sim ( N^\prime
\alpha_w)^{-1}$.  This has important implications for the GW spectrum
from the PT, since it implies that the dominant fraction
of the latent heat will be transformed into bulk flow of the plasma
and not into kinetic energy of the Higgs field. This follows from the fact that,
once the bubble wall reaches a terminal velocity, the energy stored in kinetic/gradient
energy at the wall only grows like $R^2$ (where $R$ is the radius of the bubble), while the 
energy in the plasma grows like $R^3$.

Thus the GW power spectrum is primarily sourced by sound waves from
bubbles producing a self-similar radial flow with $\vw \simeq 1$ in a much larger class of
PTs than previously appreciated.
There are nevertheless two scenarios where the above picture becomes
more complicated, and which may lead to much larger values of $\gamma$
and even to runaway behavior:

\vskip 0.3cm
  
\underline{No transition radiation:} For scenarios where no gauge fields are present, as in the
spontaneous breaking of an (approximate) global symmetry (e.g.~in a hidden sector), the dominant transition radiation/splitting processes 
may only involve scalars and fermions, resulting in a friction contribution growing at most 
like $ \Delta \Pre_\text{fr} \propto \mathrm{log} \gamma$. As a result, the ($\gamma$-independent)  leading order friction term 
may still provide the dominant contribution in the ultrarelativistic limit, and runaway behavior could be possible in this case. 

\vskip 0.3cm

\underline{Extreme supercooling:}
When the PT exhibits extreme supercooling, as in the case of models with (approximate) conformal symmetry (see Secs.~\ref{sec:warped} and \ref{sec:composite}), the 
temperature of the plasma at the time of bubble nucleation can be many orders of magnitude smaller than the energy scale that defines the scalar potential, $\TN \ll \Delta m$. 
It would then seem that plasma friction would play a negligible role on the expansion; however, for this to be the case, the amount of supercooling is typically
required to be extremely large. To see this,
we assume in the following that there is a relatively small number $N^\prime$ of particles in the plasma that acquire a mass $\Delta m$ in the PT
and produce transition radiation (notice that this is questionable in one of the prime examples -- the holographic phase 
transition~\cite{Creminelli:2001th,Randall:2006py,Nardini:2007me,Hassanain:2007js,Konstandin:2010cd,Konstandin:2011dr,Bunk:2017fic,
Dillon:2017ctw,vonHarling:2017yew ,Bruggisser:2018mus,Bruggisser:2018mrt,Megias:2018sxv,Baratella:2018pxi} -- since there a large tower of particles 
will obtain a mass in the transition). 
The released energy is in this case of order $\Delta m^4$ (or somewhat larger) and using~(\ref{eq:pressure_nc}) one finds that 
the Lorentz factor increases until it reaches the steady-state value
\begin{equation}
      \gamma_\text{s} \simeq \frac{\Delta m^3}{\alpha_{w} \TN^3 \, N^\prime} \gg 1 \, ,
\end{equation}
whereupon the self-similar hydrodynamic flow is established around the bubble.
At the early stage of the bubble expansion, the friction from the plasma can be neglected. 
The Higgs field behaves as in vacuum and the Higgs profile is a function of $\sqrt{t^2 - r^2+\Rc^2}$ ($\Rc$ is the bubble size at nucleation 
and also of order $\Delta m^{-1}$).  This leads to the following relation between bubble size and its wall velocity
\begin{equation}
      R = \gamma \Rc \sim \gamma / \Delta m \, .
\end{equation}
The bubble size when friction becomes important $R_\text{s}$ has to be compared with the mean bubble size at collision, $\Rstar$, which can 
be no larger than the Hubble parameter during the PT, $H \sim \Delta m^2/\mpl$ (with $\mpl$ the Planck mass). 
Hence, the bubble wall continues to accelerate until collision if
\begin{equation}
      \frac{R_\text{s}}{\Rstar} \sim \frac{\Delta m^4}{\alpha_{w}
        \TN^3 \mpl \, N^\prime} \frac{1}{\HN \Rstar} \gtrsim 1 \, .
\end{equation}
For example, if the near-conformal behavior of the models described in Secs.~\ref{sec:warped} -- \ref{sec:composite} holds down to QCD temperatures~\cite{vonHarling:2017yew ,Baratella:2018pxi}, $T\simeq1$~GeV, runaway behavior up to 
bubble percolation at the end of the PT requires $\Delta m \gtrsim 10 (\HN\Rstar N^\prime)^\frac{1}{4}$ TeV, typically corresponding to several orders 
of magnitude of supercooling. Below this bound, the bubble wall 
approaches a steady state and a finite but highly relativistic bubble wall velocity. The latent heat is then transformed mostly into bulk motion. We will assume this is
the case in our analysis, in contrast with our previous study~\cite{Caprini:2015zlo} (see also \cite{Ellis:2019oqb} for an extended analysis of the issue). 
Notice that the PT strength parameter $\alpha$ scales as $\alpha \sim (\Delta m / \TN)^4 \to \infty$, while the fraction of energy in 
bulk motion $K \sim \gamma^2_{\rm f} (\TN/\Delta m)^4 $ approaches unity~\cite{Espinosa:2010hh}.  Hence, the Lorentz factor of the fluid motion is of 
order $\gamma_{\rm f} \sim (\Delta m/\TN)^2$.  This challenging regime is so far unexplored by hydrodynamic simulations and deserves further attention. 
In addition, the bubble wall velocity in scenarios of confining PTs where a very large number of degrees of freedom acquire a mass across the bubble 
wall (see sections \ref{sec:warped} and \ref{sec:composite}) has never been explored and is an important open question. 
Results should therefore be taken with a grain of salt in these cases.

%
%

\renewcommand{\tCorr}{\tau_\text{c}}
%
%
%

\section{Prediction of the Gravitational Wave Signal \label{sec:signal}} 
\label{sec:prediction}

\subsection{Gravitational waves from bulk fluid motion}
\label{e:GraWavEst}

In this section we review estimates of the GW energy density $\OmGW$ generated by a PT occurring during the radiation era. To obtain the corresponding GW energy density parameter today $\OmGWnow$, which is the quantity relevant for LISA, one multiplies $\OmGW$, which is normalised to the Universe's energy density at production time, by the factor
\begin{equation}
  F_{\text{gw},0} = \Om_{\ga,0} \left(\frac{g_{s\,0}}{g_{s\,*}}\right)^{\frac{4}{3}} \frac{g_*}{g_0}.
\end{equation}
Here $\Om_{\ga,0}$ is the density
parameter of photons today, $g_{s\,*}$~(resp.~$g_{s\,0}$) is the effective number of entropic degrees of freedom at production time (resp.~today), and
$g_*$~(resp.~$g_0$) is the effective number of
relativistic degrees of freedom after the PT during which the GWs are produced (resp.~today).
Using the Planck best-fit value $H_0 = 67.8 \pm 0.9
\,\textrm{km}\,\textrm{s}^{-1}\,\textrm{Mpc}^{-1}$ \cite{Ade:2015xua},
and the FIRAS temperature for the CMB $T_{\gamma,0} = 2.725 \pm
0.002\, \textrm{K}$ \cite{Mather:1998gm}, setting $g_{s*}=g_*$ (true in the SM for $T > 0.1 $ MeV), and $g_0=2$, $g_{s\,0}=3.91$, we have
\begin{equation}
F_{\text{gw},0} = (3.57 \pm 0.05)\times 10^{-5}  \left(\frac{100}{g_*}\right)^{\frac{1}{3}}.
\label{eq:def_Fgw}
\end{equation}

It was originally thought~\cite{Kosowsky:1992rz,Kamionkowski:1993fg}
that GWs were mainly produced by the kinetic energy of bubble walls during bubble collisions, as this was believed to be the relevant source of anisotropic stresses in the case of strongly first-order PTs.
This thinking motivated a widely-adopted model known as the envelope
approximation~\cite{Kosowsky:1992vn}, in which the shear stresses are
considered to be localised in thin shells near the phase boundary, and
instantaneously dissipate as the walls collide.  It is now understood
that gravitational radiation from the collision phase is typically swamped by
the radiation generated by fluid motion in later phases of the
transition, unless there is so little friction from the plasma that
the bubble wall runs away (see Sec.~\ref{ss:WalSpe}).  In a
runaway, most of the available energy at the transition goes into the
kinetic energy of the bubble walls, and gravitational radiation from
the collision phase dominates, although latest numerical simulations show that the shear stresses in the collision 
region have a more complex behaviour than envisaged in the envelope approximation \cite{Cutting:2018tjt}. 

In this paper, we concentrate on the more common case of gravitational radiation generated by bulk fluid motion. In this case, the GW energy density spectrum at the time of production takes the form (see e.g.~\cite{Hindmarsh:2015qta,Hindmarsh:2017gnf,Caprini:2009yp})
\begin{equation}
  \OmGW \sim \,K^2 \,(k \fluidL)^3 \int \frac{dt_1}{t_1}\int \frac{dt_2}{t_2} \cos(k(t_1-t_2))\,\tilde\Pi(k,t_1,t_2)\,.
  \label{eq:OmGWPi}
\end{equation}
where $\OmGW$ actually denotes $d\log \OmGW (k)/d\log k$.
In the above equation, $K$ is the kinetic energy fraction of the fluid, defined to be 
\begin{equation}
\label{eq:Kdef}
K = \frac{\vev{w\gamma^2 v^2}}{\bar\EneDen} = \AdiInd \fluidV_{\rm f}^2,
\end{equation}
where $w=e+p$ is the enthalpy, $\bar\EneDen$ is the mean energy density of the fluid, $v$ is the 
spatial part of the fluid 4-velocity,
$\AdiInd = \bar{w}/\bar{\EneDen}$ is the mean adiabatic index,
and $\fluidV_{\rm f}^2 = \vev{w\gamma^2 v^2}/\bar{w}$ is the enthalpy-weighted root mean square (RMS) of $v$, 
reducing to the RMS 3-velocity for non-relativistic flows. For an ultra-relativistic fluid, $\AdiInd
\simeq 4/3$. 
The kinetic energy fraction can be estimated using a hydrodynamic
analysis of the expansion of spherical
bubbles~\cite{Kamionkowski:1993fg,Espinosa:2010hh}.  
Furthermore, in~\eqref{eq:OmGWPi}, $k$ is the comoving wave-number and $\fluidL$ denotes the flow length scale, the scale at which the kinetic energy of the fluid motion is maximal: this is set by
the mean bubble separation $\fluidL \sim \Rstar \sim \vw/\beta$.
The double integral comes from the definition of the GW energy density parameter, while $\cos(k(t_1-t_2))/(t_1t_2)$ is the Green's function of the GW equation of motion, assuming that the source is operating during the radiation era with a scale factor evolving as $a \propto t$ with respect to the conformal time $t$ (see e.g.~the derivation in \cite{Caprini:2018mtu}). Finally, $\tilde\Pi(k,t_1,t_2)$ is the unequal time correlator of the normalised fluid shear stress. It is convenient to normalise the spatial part of the fluid energy momentum tensor $T_{ij}({\bf x},t)=w \gamma^2 v_i v_j$ as $\tilde T_{ij}= T_{ij}/(\bar\EneDen K \fluidL^{3/2})$, leading to the
normalised shear stress $\tilde\Pi_{ij}({\bf k},t)=\Lambda_{ijk\ell}\,\tilde T_{k\ell}({\bf k},t)$, where $\Lambda_{ijk\ell}$ projects the transverse traceless part. The unequal time correlator appearing in \eqref{eq:OmGWPi} is then defined as $\langle \tilde\Pi_{ij}({\bf k},t_1)\tilde\Pi_{ij}({\bf q},t_2) \rangle=(2\pi)^3\delta({\bf k}-{\bf q})\tilde\Pi(k,t_1,t_2)$.

Some fairly general considerations dictate the possible scaling of $\OmGW$ from~\eqref{eq:OmGWPi}. One expects the source to be active for a given time interval $\tOn$, which sets the integration range in \eqref{eq:OmGWPi}. Furthermore, the source is expected to decorrelate over an autocorrelation time $\tCorr<\tOn$, which represents the time interval that it takes for the anisotropic stress to decorrelate. If the duration of the source is less than one Hubble time $\tOn \Hc<1$, the factor $(t_1t_2)^{-1}$ in~\eqref{eq:OmGWPi} becomes $(\Hc a_*)^2$, where $a_*$ denotes the scale factor at the bubble collision time. Usually one then finds, for $k<\tCorr^{-1}$ 
\begin{equation}
\label{e:OmGWEst}
\OmGW \sim K^2 \,(k \fluidL)^3 \, (\Hc \tOn) (\Hc\tCorr) \,\tilde P_{\rm gw}(k)~~~~~~~~~(\tOn \Hc<1\,,~~k<\tCorr^{-1})
\end{equation}
where $\tilde P_{\rm gw}(k)$ parametrizes the dependence 
on wave-number inherited directly from the source. For $k>\tCorr^{-1}$, the factor $\Hc\tCorr$ becomes $\Hc/k$. 

If the duration of the source is more than one Hubble time, $\tOn \Hc>1$, one expects decorrelation after a time $\tCorr<\Hc^{-1}$, yielding instead
\begin{equation}
\label{e:OmGWEstLong}
\OmGW \sim K^2 \,(k \fluidL)^3 \, (\Hc\tCorr) \,\tilde P_{\rm gw}(k)~~~~~~~~~(\tOn \Hc>1\,,~~k<\tCorr^{-1})
\end{equation}
and again $\Hc\tCorr \rightarrow \Hc/k$ for $k>\tCorr^{-1}$. It appears that the effective
lifetime for GW production is of order $\tOn\simeq \Hc^{-1}$ for a source persisting for longer than a Hubble time\footnote{We focus here on the case of a source that indeed decorrelates (i.e.~for which $\tCorr<\tOn$) as this is the relevant one for bulk fluid motion. Note, however, that non decorrelating sources might also exist. If the source remains coherent over its lifetime $\tOn$, the above estimates change as follows. If the lifetime is smaller than one Hubble time, the factor $(\Hc \tOn) (\Hc\tCorr)$ in \eqref{e:OmGWEst} becomes simply $(\Hc \tOn)^2$ for $k<\tOn^{-1}$. On the other hand, for a coherent source lasting more than one Hubble time, one finds $\OmGW \sim K^2 \,(k \fluidL)^3 \, \log^2{[(t_*+\tOn)/t_*]} \,\tilde P_{\rm gw}(k)$, where $t_*=1/(\Hc a_*)$ denotes the initial time of action of the source. For $k>\tOn^{-1}$, the behaviour depends on the details of the time dependence of the source\cite{Caprini:2009fx,Caprini:2009yp}.}. \cite{Caprini:2009yp,Caprini:2009fx,Hindmarsh:2015qta}. 

We will use the scalings above to determine the expected GW energy density from bulk fluid motion. To do so, one must specify the scales $\tOn$ and $\tCorr$ entering~\eqref{e:OmGWEst} -- \eqref{e:OmGWEstLong}. Since the source of the anisotropic stress is the bulk fluid motion, we set $\tCorr$ to the fluid autocorrelation time, which typically scales as the ratio between the flow length scale and the characteristic propagation velocity of the flow (c.f.~Table~\ref{tab:timescales}). The results of the simulations presented in section \ref{sec:num_sim} confirm this assumption. Two processes of the fluid motion source GWs: sound waves (the longitudinal mode) and turbulence (the vortical mode). In considering the dynamics of the fluid, there are (at least) three relevant timescales that determine the lifetime of the bulk motion. Besides the fluid autocorrelation time $\tCorr$ mentioned above, there are the shock formation time $\tau_{\rm sh}$ and the turbulence eddy turnover time $\tau_{\rm tu}$. Table~\ref{tab:timescales} collects these timescales. We will further discuss their respective role in the following subsection. 

\subsection{Relevant time scales and scaling of GW energy density}
\label{subsec:timescales}

GW power is determined by the kinetic energy fraction
in the bulk motion of the plasma $K$, the fluid autocorrelation time $\tCorr$, and the time for which the bulk motion lasts $\tOn$. As we will see in the following, the latter scale depends on whether or not shocks develop and turbulence is produced. 

At a first-order PT, the expanding bubbles generate compression and rarefaction waves
(i.e.\ sound)~\cite{Hogan:1986qda}, which continue oscillating long
after the transition is complete, generating significant gravitational
radiation \cite{Hindmarsh:2013xza,Hindmarsh:2015qta,Hindmarsh:2017gnf}. 
This acoustic phase has a lifetime
determined by the generation of
shocks~\cite{Landau1987Fluid,Pen:2015qta}, which happens on a
timescale $\tShock \sim \fluidL/\fluidV_\parallel$, where
$\fluidV_\parallel$ is the RMS longitudinal velocity associated with
the sound waves.  The shocks decay by generating entropy, on the same
timescale, setting the lifetime of the acoustic phase. The autocorrelation time in
acoustic production is set by the sound-crossing time of the flow
length scale $\tCorr \sim \fluidL/\cs\sim \Rstar/\cs$ \cite{Hindmarsh:2016lnk}. Applying these timescales, one finds for long-lasting acoustic production (c.f.~\eqref{e:OmGWEstLong}),
\begin{equation} \label{eq:OmGWSW1}
\OmGW^\text{ac} \propto K^2 \,(\Hc\Rstar/c_s) \, .
\end{equation}
For stronger flows with $ \fluidV_\parallel \gg \Hc\Rstar$, the
lifetime of the sound waves is less than the Hubble time, since shocks develop within one Hubble time. The GW energy density parameter becomes (c.f.~\eqref{e:OmGWEst})
\begin{equation} \label{eq:OmGWSW2}
\OmGW^\text{ac} \propto K^{3/2}\,(\Hc\Rstar)^2/c_s \, .
\end{equation}

\begin{table}
  \begin{tabular}{lll}
    \textbf{Symbol} & \textbf{Role} & \textbf{Approximate expression}  \\
    \hline
    $\beta^{-1}$ & PT duration &  \\
    $\tCorr$ & Fluid autocorrelation time & $\fluidL/\cs$ (turbulence: $\tTurb$)\\
    $\tShock$ &  Shock formation time & $\fluidL/\fluidV_\parallel$ \\
    $\tTurb$ & Eddy turnover time on the scale of the flow & $\fluidL/\fluidV_\perp$   \\
    $\tOn$ & Time for which bulk motion of the fluid lasts  & \\
    $\Hc^{-1}$ & Hubble time &  \\
    \hline
  \end{tabular}
  \caption{\label{tab:timescales} Summary of timescales discussed in
    the text,  approximately arranged by expected duration.}
\end{table}

To arrive at a corresponding estimate for the contribution from turbulence, one requires the fraction of kinetic energy converted into vortical motion.
Numerical simulations of intermediate-strength transitions have shown that during the acoustic phase
approximately 1-10 \% of the bulk motion from the 
bubble walls is converted into vorticity: this can be estimated from the ratio $\bar{U}_\perp/\fluidV_{\rm f}$, given for example in 
Table II of \cite{Hindmarsh:2015qta} and Table III of \cite{Hindmarsh:2017gnf}. Meanwhile, transitions featuring deflagrations with larger $\alpha$ were recently found to exhibit more efficient generation of vortical motion and a suppression of the kinetic energy transferred to the plasma~\cite{Cutting:2019zws}. Based on the findings for detonations and small-$\alpha$ deflagrations, in our last report~\cite{Caprini:2015zlo} 
we set $\ep = K_\perp/K = 0.05$. This estimate is accurate at low RMS fluid
velocities, $\fluidV_{\rm f} \lesssim 0.05$, but $\ep$ is likely larger than 0.05  at higher fluid velocities. Moreover, when shocks develop within a Hubble time
their collisions and other non-linear effects may also generate turbulence
in the transverse (vortical) component of the fluid flow. Nevertheless, the kinetic energy fraction of the turbulent flow cannot exceed the total 
kinetic energy fraction in bulk flows of the plasma, setting an upper bound on the total possible contribution to the GW energy density from turbulence. The
relative importance of acoustic and turbulent GW
production depends on how efficiently turbulent kinetic energy
$K_\perp$ is generated from the decaying sound waves and is a subject of ongoing study.

In addition, several timescales are also relevant for determining the contribution from turbulence to the GW energy density. The turbulent cascade is expected to set in after one eddy turnover time at the flow length scale, $\tTurb \sim \fluidL/\fluidV_\perp$. Afterwards, in the absence of stirring, turbulence decays. The net turbulent kinetic energy is dissipated when the Reynolds number at the flow length scale becomes of order one \cite{Landau1987Fluid}. In the early Universe fluid this can take several Hubble times, so the turbulence lifetime could in principle satisfy $\tOn \gg \Hc^{-1}$ \cite{Caprini:2009yp}. However, turbulence decorrelates, which effectively reduces its lifetime to $\tOn\simeq \Hc^{-1}$, as reflected in~\eqref{e:OmGWEstLong}. At the integral scale $\fluidL$, the autocorrelation time of the turbulent flow corresponds to one eddy turnover time \cite{Davidson}; it therefore features the same parametric dependence as the shock appearance time, $\tTurb \sim \fluidL/\fluidV_\perp$. At smaller scales, the autocorrelation time is the eddy turn-over time on that scale \cite{Davidson}. Turbulence decorrelation is usually modelled with a Gaussian functional form \cite{Kraichnan}.

Different approaches to describe turbulent decay adopted in the literature lead to different 
parametric dependences for the resulting GW power, as well as different predictions for the 
GW power spectrum.~\cite{Gogoberidze:2007an} considered stationary turbulence with a fixed lifetime, and a scale-dependent velocity decorrelation time, corresponding to the eddy turn-over time on a given scale. 
Reference~\cite{Caprini:2009fx} attempted to include the free-decay, but to guarantee positivity of the GW power spectrum introduced a ``top-hat'' modelling of the shear stress two-time correlation function that sets the autocorrelation time of the shear stresses to the light-crossing time on a given scale \cite{Caprini:2009yp,Caprini:2009fx}. A direct consequence of this assumption is that the GW power spectrum changes slope at frequencies higher than $\Hc$, absent in \cite{Gogoberidze:2007an}. This model provides the basis of the turbulent GW power spectrum used in our last report \cite{Caprini:2015zlo}. More recently, flows with both compressional and rotational modes, and with both helical and non-helical magnetic fields, have been modelled in detail~\cite{Niksa:2018ofa}. The new ingredient is the Kraichan sweeping model for the velocity decorrelation, and a variety of power laws and peak shapes are found depending on the relative amount of turbulence. The first numerical simulations~\cite{Pol:2019yex} of GW production by MHD turbulence have produced power spectra which remain to be understood in the context of the models. There are interesting indications that turbulent flows are less efficient at producing GW than sound waves.

The preceding discussion highlights that the turbulence contribution to $\OmGW$ is the subject of ongoing study. Nevertheless, scaling predictions for $\OmGW$ can be derived from~\eqref{e:OmGWEstLong} under different assumptions. For turbulence lasting longer than the Hubble time at nucleation \cite{Caprini:2009yp}, setting the autocorrelation time to the eddy turnover time on the characteristic scale of the flow $\tCorr=\tTurb \sim R_*/\fluidV_\perp$ yields 
\begin{equation}
\label{eq:omturb1}
\OmGW^\text{tu} \propto K_\perp^{3/2}\, (\Hc\Rstar).
\end{equation}
This scaling is in agreement with the result of \cite{Caprini:2009yp}, adopted in \cite{Caprini:2015zlo}. If, on the other hand, one neglects the free decay and assumes that the turbulence lifetime is much less than the Hubble time at nucleation, one has the estimates $\tOn \sim \tCorr\sim \tTurb$ \cite{Caprini:2009yp}, leading to 
\begin{equation}
\OmGW^\text{tu} \propto K_\perp\,(\Hc\Rstar)^2. 
\label{eq:omturb2}
\end{equation}

We have discussed how the characteristics of the fluid dynamics influence the scaling of the GW power spectrum amplitude with the relevant fluid flow parameters. 
As presented in the next section, in the case of GW production from sound waves, the results from numerical simulations allow one to also infer the shape of the power spectrum. On the other hand, further work is required to assess the amplitude and spectral shape from turbulence. In the remainder of this study we therefore neglect GWs from turbulence altogether, in order to be conservative in estimating the corresponding GW signals at LISA.

\subsection{GW power spectra: numerical simulations}
\label{sec:num_sim}

Our expressions for $\OmGWnow$, and in particular its spectral shape,
are motivated by large-scale simulations of bubble collisions and the resulting bulk fluid motion.
In~\cite{Hindmarsh:2013xza,Giblin:2014qia,Hindmarsh:2015qta,Hindmarsh:2017gnf}
GW production from a thermal PT was
studied through numerical simulations of a scalar field coupled to a 
relativistic ideal fluid. The best results for the amplitude and
spectral shape of the GW power spectrum come from
~\cite{Hindmarsh:2015qta,Hindmarsh:2017gnf}. While the bubbles in
these simulations are nucleated simultaneously (as opposed to the distribution of bubble sizes expected from a first order PT),  
it is expected that
for a given mean bubble separation $R_*$ the results can be used to
infer the results for a given $\beta$ via~(\ref{eq:rstarbetavw}). 

The ansatz below, based on the simulations in \cite{Hindmarsh:2015qta,Hindmarsh:2017gnf} 
and the analytic understanding in \cite{Hindmarsh:2016lnk, Hindmarsh:2019phv},
is most reliable when the sound
shell surrounding the bubble wall is large. This means, in turn, that
the wall speed must be far from the Chapman-Jouguet speed, in which case the fluid just reaches sonic velocity in the frame of the shock. For
deflagrations (wall speed smaller than the speed of sound) this means
$\vw < \vCJ = \cs$, while for detonations (wall speed faster than the
speed of sound) we require $\vw > \vCJ \simeq \cs (1 +
\sqrt{2\alpha_\theta} ) $ so that the fluid flow behind the wall is well away from the speed of sound.

Using~\eqref{eq:OmGWSW1}, for wall speeds away from the Chapman-Jouguet speed, we have
\begin{equation}\label{eq:OmGWmaster1}
\boxed{
\frac{d \OmGWnow}{d \ln(f)} = 
0.687
F_{\text{gw},0} K^2 (\Hc\Rstar/c_s) \OmGWscaled \fitfun\left(\frac{f}{\fpnow}\right) \, ,
}
\end{equation}
where the numerical factor ensures that the total GW power is $K^2 (\Hc\Rstar/c_s) \OmGWscaled$
and $F_{\text{gw},0}$ is given by equation (\ref{eq:def_Fgw}). 
The kinetic
energy fraction of the fluid $K$ is as defined above in (\ref{e:KinEneFra}); $R_*$ can be related to $\beta$ and $\vw$ by~(\ref{eq:rstarbetavw}); $ \OmGWscaled \sim 10^{-2}$ is numerically
obtained from simulations; and the spectral shape function is
\begin{equation}
\label{e:FitFun}
\fitfun(s) = s^3\left(\frac{7}{4+3s^2}\right)^\frac{7}{2}
\end{equation}
with peak frequency
\begin{equation}
f_{\text{p},0}
\simeq 26 \left( \frac{1}{\Hc\Rstar} \right) \left( \frac{\zp}{10} \right) 
\left(  \frac{T_{*}}{100 \, \text{GeV}} \right) \left(  \frac{g_*}{100}  \right)^{\frac{1}{6}} \; \mu\text{Hz}.
\end{equation}
The quantity $ \zp \simeq 10$ is determined from simulations, and accounts for the
observed peak value of $k\Rstar$. We denote by $T_{*}$ the temperature after the PT. For weak PTs this will coincide with the nucleation temperature $\TN$ but it can differ significantly for large supercooling. In the end what controls the redshifting of the peak frequency is the Hubble scale.
Prior to a PT with large enough $\alpha$, the vacuum energy is the main component of the total energy, while after the PT it is the thermal plasma. Hence the redshifting is related to the (reheat) temperature after the PT, $T_{*}$.

When $\vw \approx \vCJ$, the value of $\zp$ may differ from 10: the
sound shells are so thin that they may set a substantially smaller
length scale $\Delta R_* = R_* |v_{\rm w} - c_s|/c_s$. More importantly,
though, the spectral shape ansatz in~(\ref{e:FitFun}) fails to
describe these scenarios well due to the more intricate nature of the
source, and a more complicated analysis is
required~\cite{Hindmarsh:2016lnk}.

In arriving at~\eqref{eq:OmGWmaster1}, the source was assumed to last for $\tOn \sim
1/H_*$. An additional modification is required, however, if the shock
formation timescale is less than a Hubble time, i.e. $\Hc\tShock < 1$.
In this case, the total GW power should be reduced by
a factor $\Hc\tShock = \Hc\Rstar/K^{1/2}$ to compensate for the
appearance of shocks and onset of turbulence (c.f.~\ref{eq:OmGWSW2}), leading to the modified expression
\begin{equation}\label{eq:OmGWmaster2}
\boxed{
\frac{d \OmGWnow}{d \ln(f)} = 
0.687
F_{\text{gw},0} K^{3/2} (\Hc\Rstar/\sqrt{c_s})^2 \OmGWscaled \fitfun\left(\frac{f}{\fpnow}\right).
}
\end{equation}
As emphasized in~\cite{Ellis:2018mja} and below in Sec.~\ref{sec:models},
most models of interest for LISA predict $\Hc\tShock < 1$, and so~(\ref{eq:OmGWmaster2}) is often the relevant expression.
This reduces the size of the GW signal from sound waves relative to~(\ref{eq:OmGWmaster1}), which is typically used in the literature. However, for $\Hc\tShock < 1$ the overall GW signal might receive additional non-negligible contributions from turbulence, which we do not account for in our study. 

In summary, the two key formulae used to derive the results of this paper, in particular the SNR for LISA detection, are (\ref{eq:OmGWmaster1}) and (\ref{eq:OmGWmaster2}). In the  figures displaying SNR contours, such as.~Fig.~\ref{fig:example_snr}, the grey-shaded regions will indicate where the shock formation time is longer than one Hubble time, and therefore where~(\ref{eq:OmGWmaster1}) is used. Elsewhere, we use~(\ref{eq:OmGWmaster2}).

\subsection{GW power spectra: discussion}

Where hydrodynamic numerical simulations can be combined with a good physical understanding of the fluid flows, we can 
obtain reliable power spectra for the generation of GWs, corresponding to~\eqref{eq:OmGWmaster1} -- \eqref{eq:OmGWmaster2} above. 
Still, there are regimes where hydrodynamic numerical simulations have not (yet) been carried out, 
or additional production mechanisms occur, the effects of which are currently difficult to quantify. 
In the following we comment on these cases. In the numerical analysis we rely only on the predictions from sound wave simulations (see Sec.~\ref{sec:tool}). We try to be conservative whenever the simulation results do not immediately apply.

One such scenario for which~\eqref{eq:OmGWmaster1} -- \eqref{eq:OmGWmaster2} do not necessarily apply consists of strong PTs with ever accelerating runaway bubble walls. In this case, the envelope approximation or scalar field simulations seem to be the more reliable setup~\cite{Cutting:2018tjt}. Even when the bubble walls do not runaway but approach a highly-relativistic terminal velocity, the expected form of the GW power spectrum is not clearly settled. Also in this scenario, one could make the case that derivatives of the envelope approximation such as the so-called bulk flow model are more realistic~\cite{Jinno:2017fby, Konstandin:2017sat}. 

Another issue concerns the appearance of turbulence and magnetic fields. In principle, turbulence should develop at late times and drain the 
energy in the sound waves into heat, as discussed in Sec.~\ref{subsec:timescales}. However, current simulations could not confirm the onset of turbulent motion, as for now they analyse 
the case of low RMS fluid velocities and the system is not evolved for sufficient time to observe the appearance of shocks and turbulence. 
As pointed out above, our understanding of the turbulence contribution to the GW spectrum is still evolving.
In the last analysis~\cite{Caprini:2015zlo}, we limited the contribution from turbulence to the 
percent level, based on the findings of simulations. In light of the uncertainties that exist, and in order to be conservative, we will neglect turbulence altogether in what follows. Note, however, that for first order PTs in specific models (for which the relation among $\fluidV_{\rm f}$ and $\Hc\Rstar$ can be predicted), 
the characteristic time of shock formation is often expected to be shorter than one Hubble time~\cite{Ellis:2018mja}, likely leading 
to turbulent flows and emphasizing the importance of future study along this direction.

\section{The \texttt{PTPlot} tool \label{sec:tool}}

To allow the reader to explore the GW power spectra
in light of the preceding discussion in specific scenarios, we have developed a web-based tool
`\texttt{\texttt{PTPlot}}' to accompany this paper. It may be accessed at
\href{http://www.ptplot.org}{\texttt{ptplot.org}}. This tool is written in
\texttt{Python} using the \texttt{Django} framework, performing server-side rendering of
power spectra and SNR contour plots. The aim is to make high-quality,
up-to-date and reliable predictions of the GW power
spectra easily available to the model-building community. In time, it
may be extended to include other cosmological sources of GWs, and other GW detectors.

\texttt{PTPlot} allows the user to display GW power spectra for
any thermal PT parameters ($\alpha$, $\beta/H_*$, $g_*$,
$T_*$, $v_\mathrm{w}$) of their choice, using the formulae presented
here, with a comparison to the public sensitivity curve for LISA given
in the Science Requirements
Document~\cite{LISA_docs}. 
The \texttt{PTPlot} tool uses  (\ref{eq:OmGWmaster1})  (shading the corresponding regions grey) and (\ref{eq:OmGWmaster2})
that, as discussed before, only take into account the contribution from sound waves.

It is anticipated that the sensitivity curve (and SNR contours) will
be updated if, in the future, an improved public estimate of the
eventual LISA sensitivity curve is made available.
The site carries a database of benchmark points for all the theories
considered in Sec.~\ref{sec:models} below, and the power spectra can be viewed
for these as well. 

\subsection{Signal to noise ratio plots}

\begin{figure}[!t] 
\begin{center}
 \includegraphics[width=0.45\columnwidth]{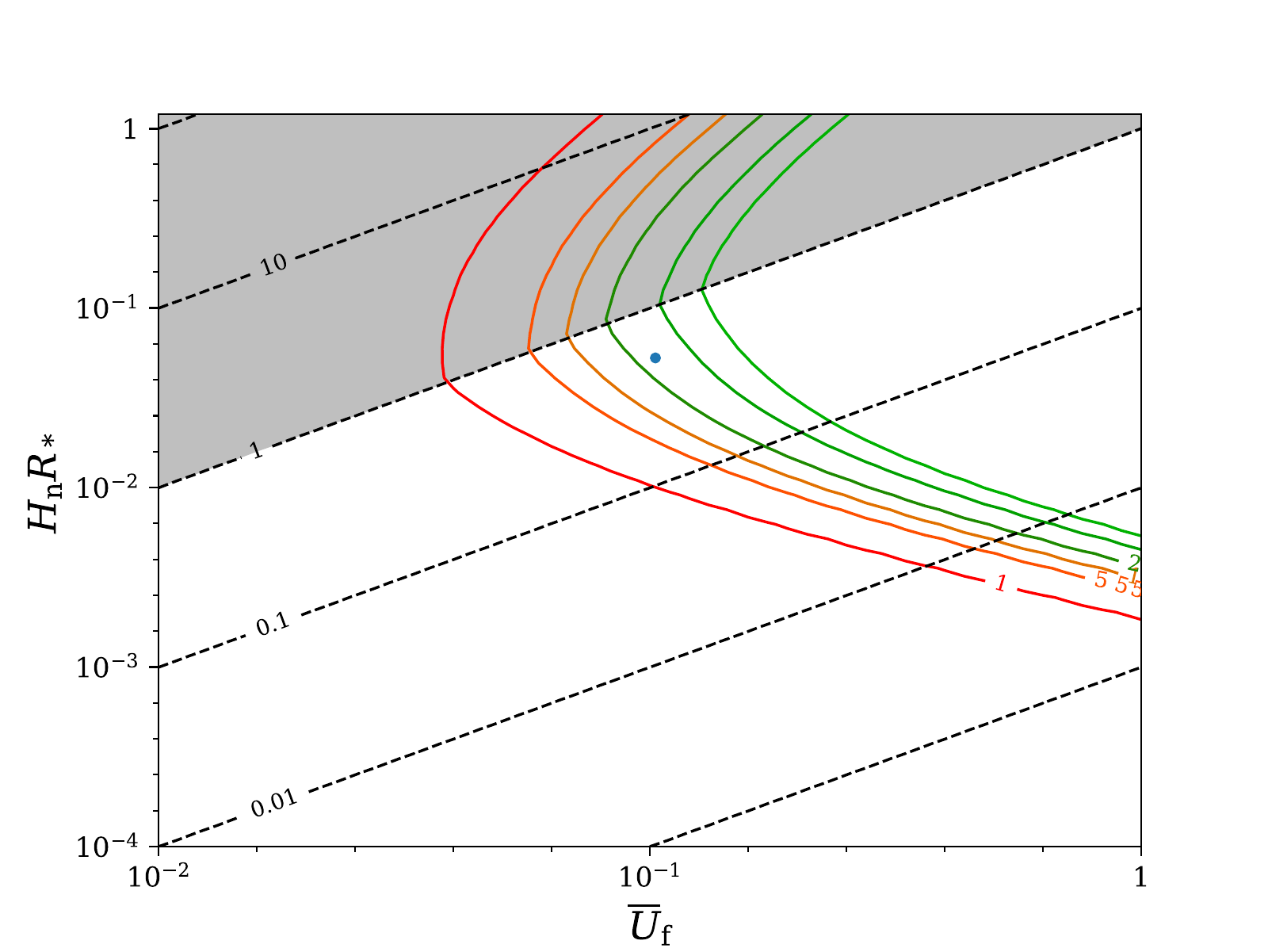}
 \includegraphics[width=0.45\columnwidth]{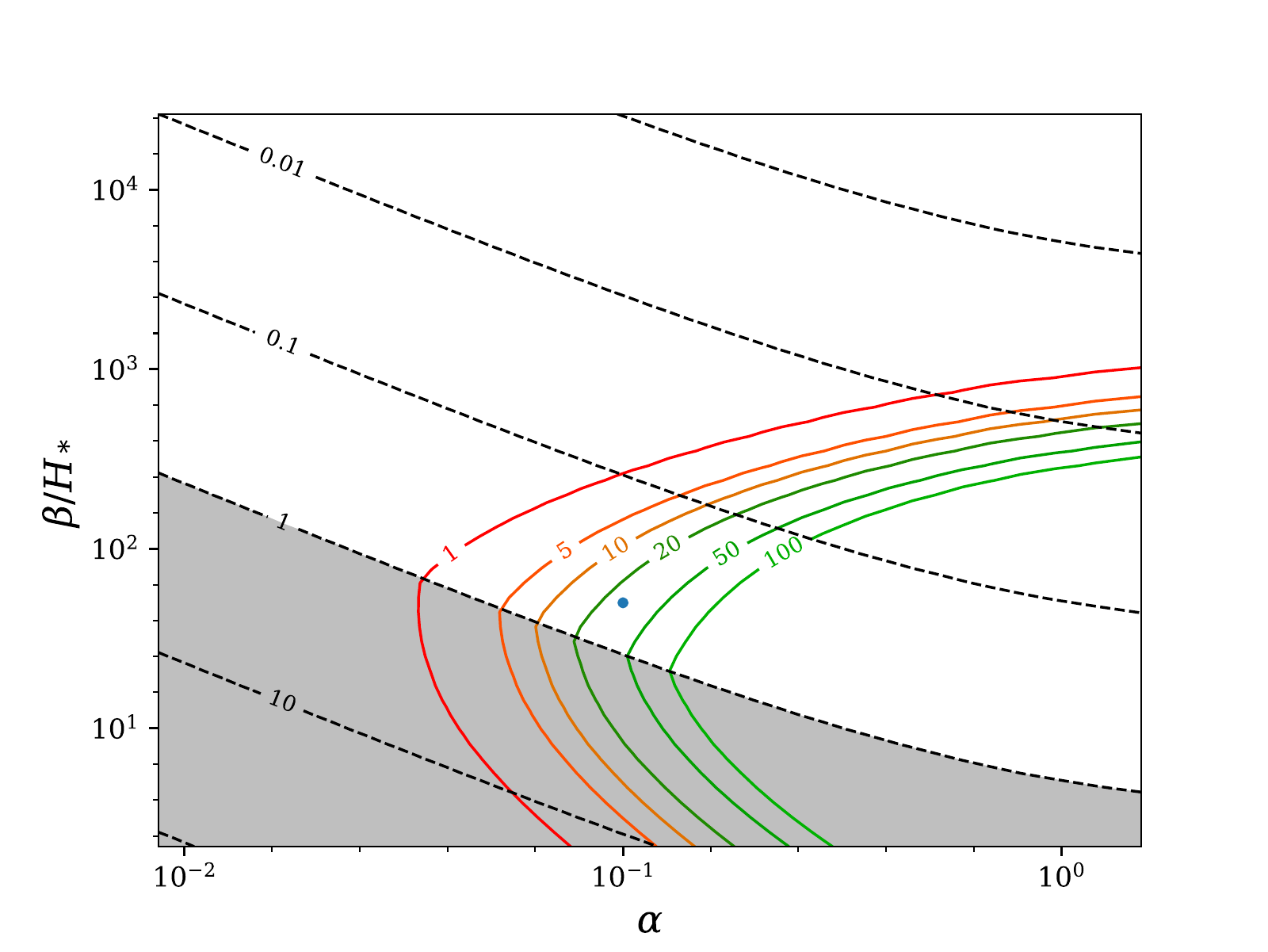}
 \caption{\small Example output of the \texttt{PTPlot} tool.
 The colored lines show the SNR that depends on $T_*$, $g_*(T_*)$ and $v_{\rm w}$. The two remaining parameters can either be $\alpha$ vs $\beta/H_*$ or $\bar U_{\rm f}$ vs $R_* H_*$.  The dotted straight lines are the contours of the fluid turnover time $H_{\rm n} R_*/\overline{U}_{\rm f}$ quantifying the effect of turbulence.
 In the gray shaded region the decay of sound waves into turbulence is less important than the Hubble damping and the SNR curve reflects this effect. The model parameters are given in the text. 
 \label{fig:example_snr}}
\end{center}
\end{figure}

In addition to the GW power spectra, SNR contour plots
can also be viewed. These show where the parameter (or parameters) lie
in either the $H_{\rm n} R_*$-$\overline{U}_{\rm f}$ or $\beta/H_*$-$\alpha$
planes~\footnote{Note that we use $H_{\rm n}$ and $H_*$ interchangeably in this section.}. 
The SNR is computed according to the standard formula,
\begin{equation}
  \text{SNR} = \sqrt{\mathcal{T} \int_{f_\text{min}}^{f_\text{max}}
    \mathrm{d}f \left[ \frac{h^2 \Omega_\text{GW} (f) }{ h^2
        \Omega_\text{Sens}(f)} \right]^2 } \, ,
  \label{eq:snr}
\end{equation}
where $\mathcal{T}$ is the duration of the mission times its duty cycle and $h^2 \Omega_\text{Sens}(f)$ is the nominal sensitivity of a
given LISA configuration to cosmological sources, obtained from the
power spectral density $S_h(f)$
\begin{equation}
  h^2 \Omega_\text{Sens} (f) = \frac{2\pi^2}{3 H_0^2} f^3 S_h(f).
\end{equation}
We take $H_0 = 100\, h \, \mathrm{km}\,\mathrm{s}^{-1}\,\mathrm{Mpc}^{-1}$ and the observed Hubble 
parameter today is given by $h=0.678\pm 0.009$ \cite{Ade:2015xua}.
For $\Omega_\text{GW} (f)$, we use   (\ref{eq:OmGWmaster1}) and (\ref{eq:OmGWmaster2}) as appropriate. For $\mathcal{T}$ we take 4 years as the mission duration and a duty cycle of 75\%, yielding $\mathcal{T}\simeq 9.46\times 10^7\,$s which is the minimal data-taking time guaranteed by the LISA mission requirements~\cite{LISA_docs}.

To give a responsive web interface, the SNR values are precomputed as
a function of $\overline{U}_{\rm f}$ and $H_{\rm n} R_*$ at fixed $T_*$ and $g_*$;
note that the SNR contours are necessarily two-dimensional slices
through a higher-dimensional parameter space and this slicing was
chosen for consistency with previous work~\cite{Caprini:2015zlo}. In
our case, $\overline{U}_{\rm f}$ and $H_{\rm n} R_*$ are calculated from
$\beta/H_*$, $v_\mathrm{w}$ and $\alpha$ using~(\ref{eq:rstarbetavw}),  (\ref{e:KinEneFra}), (\ref{eq:Kdef}) and the efficiency factor from the literature~\cite{Espinosa:2010hh}. 

Note that an SNR plot in the
$\overline{U}_{\rm f}$-$H_{\rm n} R_*$ plane was first presented in
~\cite{Hindmarsh:2017gnf}; it is a natural choice of parameters,
motivated by the results of simulations. Furthermore, contours of the
fluid turnover time $H_{\rm n} R_*/\overline{U}_{\rm f}$ are straight lines on
this plot; this combination quantifies the expected importance of
turbulence.  Regions where the acoustic period will last for a 
Hubble time are shaded on these SNR plots. Note that for producing the SNR curves the duration of the source is taken to be the Hubble time or the fluid turnover time, whichever is shorter, as the most conservative estimate possible~\cite{Ellis:2018mja, Hindmarsh:2017gnf}.

On the other hand, for an SNR plot in the $\beta/H_*$-$\alpha$ plane, which is more practical for model builders, 
the input parameters can be plotted directly, but the contours are
deformed by the inverse mapping from $\overline{U}_{\rm f}$ and $H_{\rm n} R_*$ to
$\alpha$ and $\beta/H_*$.

\begin{figure}[!t] 
\begin{center}
 \includegraphics[width=0.50\columnwidth]{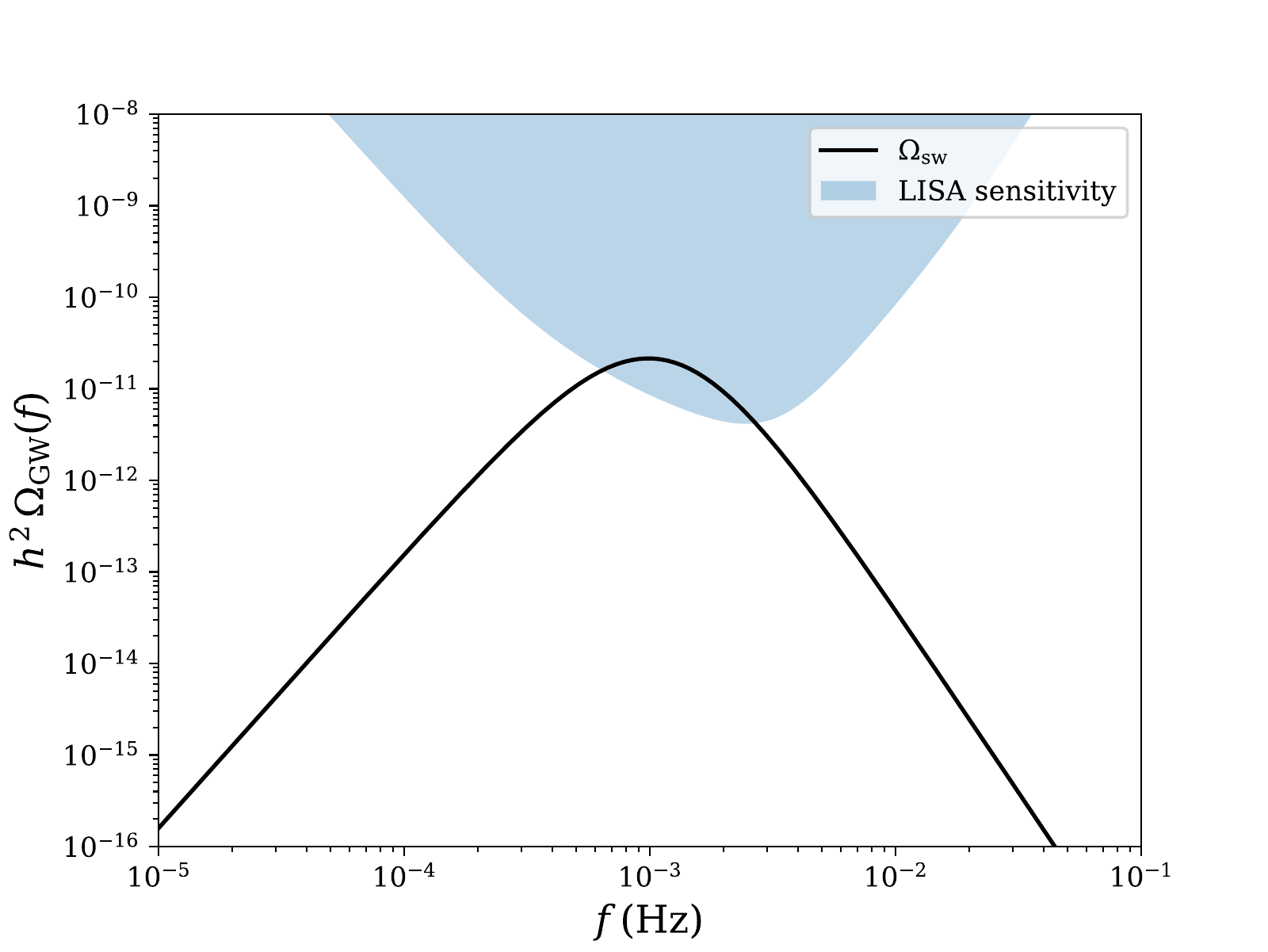}
 \caption{\small Example output of the \texttt{PTPlot} tool.
 The plot shows an example of the GW power spectrum from a first-order PT, along with the LISA sensitivity curve ($h^2\Omega_{\rm Sens}(f)$ taken from the LISA Science Requirements Document \cite{LISA_docs}). The parameters of the example model are $v_{\rm w}=0.9$, $\alpha=0.1$, 
$\beta/H_*=50$, $T_*=200$ GeV, $g_*=100$.
 \label{fig:example_ps}}
\end{center}
\end{figure}

Figs.~\ref{fig:example_snr} and~\ref{fig:example_ps} show three example plots produced by the \texttt{PTPlot} tool.
The two plots in Fig.~\ref{fig:example_snr} display the SNR in the $\bar U_{\rm f}$ vs $R_* H_*$ and $\alpha$ vs $\beta/H_*$ parameter spaces. Figure~\ref{fig:example_ps} shows the expected GW power spectrum for some example model and the LISA sensitivity curve.
All sensitivity plots presented in Sec.~\ref{sec:models} were made with \texttt{PTPlot}.

\section{Determining $\alpha$, $\beta$, and $H_*$ in specific models \label{sec:nonpert}}

When considering a specific model, the parameters $\alpha$, $\beta$, and $T_*$ (or $H_*$) entering the energy budget (and \texttt{PTPlot}) need to be computed microphysically. This is typically done using a perturbative effective potential approach (implicitly assumed in the discussion of Sec.~\ref{sec:prelim}), which can result in significant uncertainties in the predicted GW parameters. Here, we discuss methods for going beyond the standard approach and the corresponding uncertainties as they relate to LISA. 

The majority of GW predictions in specific BSM scenarios rely on the computation of the effective potential $V[\{\phi_i\}]$, through a perturbative expansion to one- or sometimes two-loop order in 4D. Here, $\{\phi_i\}$ denotes the set of scalar fields involved in the transition (the order parameters). Under the assumption that the $\{\phi_i\}$ are homogeneous, one may compute the finite temperature corrections to the classical potential. The global minimum of the effective potential then corresponds to the finite temperature expectation value of the fields. The order of the transition is determined by whether this minimum changes continuously (second-order/cross-over) or discontinuously (first-order) as a function of the temperature. The parameter $\alpha$ follows directly from the effective potential, while $\beta/H_*$ and $T_*$ can be determined by computing the action of the bounce solution, which follows from the Euclidean equations of motion for the scalar(s) again utilizing the effective potential.

An alternative method that has received renewed interest lately is to investigate the phase diagram and determine the GW parameters by computing the effective {\it action} using numerical Monte-Carlo lattice simulations. This method was instrumental in establishing that the minimal SM does not have a first-order PT at the physical value of the Higgs mass \cite{Kajantie:1995kf}. By considering the effective action rather than just the effective potential, no assumption is made about homogeneity of the fields, and mixed configurations (such as bubbles) contribute. Issues related to the well-known infrared divergences of finite temperature perturbation theory are automatically avoided in this approach, allowing for theoretically robust and accurate predictions. The computation may be done in full 4D simulations of an effective bosonic model \cite{Csikor:1998eu}, but because of the numerical effort involved, parameter scans are more feasible in simulations of effective 3D models that are matched onto the 4D theory at high temperature through a procedure known as dimensional reduction (DR). 
The status of the art regarding this method is reviewed in Appendix
 \ref{sec:appendixlattice}.

The short summary is that there remain several challenges associated with providing nonperturbative predictions for LISA in BSM scenarios. Firstly, lattice Monte Carlo simulations are still computationally expensive in 3D, and, as argued  in Appendix \ref{sec:appendixlattice}, need to be done on a model-by-model basis. Even once a model is specified, scanning the parameter space nonperturbatively is often not feasible. Instead, one typically chooses benchmark points to analyze, which further prevents extensive phenomenological investigation. Furthermore, many scenarios rely on relatively large couplings to induce a first-order PT through radiative effects. Dimensional reduction is usually done perturbatively to a fixed order in the couplings, and so large couplings limit the accuracy of the dimensionally reduced description (see e.g.~\cite{Laine:2017hdk, Kainulainen:2019kyp} for recent discussions). In this case, the perturbative matching procedure can fail, and one cannot perform 3D Monte Carlo studies of the model at finite temperature, necessitating genuine 4D simulations which are even more computationally challenging. Thus, there remains ample room for future improvements along this direction as LISA moves towards launch. In the interim, we rely mostly on results obtained from 4D perturbation theory, commenting on existing nonperturbative results for the various models discussed in Sec.~\ref{sec:models} below when applicable.

\section{Overview of specific models\label{sec:models}}

A variety of new physics models predict first order PTs in the early Universe. In~\cite{Caprini:2015zlo} several such models were considered to benchmark LISA's sensitivity to realistic BSM scenarios. This section should be considered both an update as well as an extension of the previous work. The recent developments in the computation of the stochastic GW background from PTs discussed in the preceding sections and the updated LISA sensitivity curve necessitate a re-evaluation of the detectability of specific scenarios and benchmark points. Furthermore additional data, e.g.~from the LHC, has become available and puts additional constraints on models that modify the electroweak phase transition (EWPT). With respect to these developments, the models and benchmark points of~\cite{Caprini:2015zlo} will be updated below. In addition, we will summarise the recent developments in the study of models with cosmological PTs, with a focus on new, more precise theoretical computations of the PT properties (c.f.~Sec.~\ref{sec:nonpert}) as well as on the complementarity between LISA and other experiments in probing these models. 

 The models we discuss below fall into two broad categories: those where new physics contributions modify the nature of the EWPT, and those where some other field undergoes a PT largely independent of the Higgs, with an intermediate regime of multi-field and multi-step transitions involving the Higgs and other fields. Among the simplest possibilities are extensions of the SM by a scalar singlet or EW multiplet, discussed in Secs.~\ref{sec:singlet} and~\ref{sec:doublet}, or effective operators in the case that the new physics affecting the EWPT is sufficiently heavy (Sec.~\ref{sec:eft}). The dynamics of more complex models such as supersymmetric extensions of the SM (Sec.~\ref{sec:susy}) can often be reduced to that of the SM with a small number of additional fields, and in that sense the results of Secs.~\ref{sec:singlet} -- \ref{sec:eft} can be considered representative of a large class of BSM scenarios. More qualitative changes occur if the new physics involves non-perturbative or non-polynomial potentials such as extra dimensional or composite Higgs models, or if the PT happens in a dark sector, as discussed in Secs.~\ref{sec:warped},~\ref{sec:composite}, and~\ref{sec:dark} respectively. 

\subsection{Singlet extension of the SM } 

\label{sec:singlet}

Many extensions of the SM feature new gauge singlet scalar fields that can yield strong first-order cosmological PTs in the early Universe, and hence contribute to the stochastic GW background. Such extensions can be tied to dark matter (DM)~\cite{Curtin:2014jma, Chala:2016ykx, Dev:2019njv}, the EW hierarchy problem~\cite{Craig:2014lda}, and EW baryogenesis~\cite{Morrissey:2012db, Huang:2018aja}, for example\footnote{Even though certain models with extra dimensions or a composite Higgs are technically speaking singlet extensions of the SM (with a dilaton playing the role of the singlet), we will treat them separately in Secs.~\ref{sec:warped} and \ref{sec:composite}.}. Singlet scalars can be difficult to probe experimentally despite featuring sizable couplings to the SM Higgs field, since they do not couple directly to the other SM states except through mixing with the Higgs. While these models are sometimes considered `nightmare scenarios' at colliders~\cite{Curtin:2014jma}, they provide a compelling target for LISA~\cite{Caprini:2015zlo} and highlight the complementarity between LISA and the LHC.

Consider a gauge singlet scalar field, $S$, that couples to the SM Higgs field, $H$. The most general renormalizable scalar potential is 
\begin{equation} \label{eq:singlet_pot}
\begin{aligned}
V(H,S)=&-\mu^2 \left|H\right|^2 + \lambda \left|H\right|^4 +\frac{1}{2} a_1 \left|H\right|^2 S + \frac{1}{2} a_2 \left|H\right|^2 S^2  \\
&+ b_1 S+ \frac{1}{2}b_2 S^2 +\frac{1}{3} b_3 S^3+\frac{1}{4}b_4 S^4.
\end{aligned}
\end{equation}
A redefinition of the $S$ field allows one to either remove the tadpole term from the above expression (see e.g.~\cite{Profumo:2007wc}), or shift away the VEV of $S$ at zero-temperature. We choose the latter, but emphasize that this is simply a convention and that the results can be translated between schemes with simple redefinitions of the various parameters (see e.g.~\cite{Espinosa:2011ax} for an in-depth discussion of these parametrizations). After EW symmetry breaking, in unitary gauge we can parametrize $H = [0, ~ (h + v)/\sqrt{2}]^\text{T}$, where $v=246$ GeV. The cross-terms in~(\ref{eq:singlet_pot}) then result in mixing between $h$ and $S$. The mass eigenstate fields can be written as
\begin{equation}
\begin{aligned}
&h_1 = h \cos \theta - s \sin\theta \, ,\\
&h_2 = h \sin \theta + s \cos\theta \, ,\label{eq:singlet_eigen}
\end{aligned}
\end{equation}
where the subscripts indicate the mass ordering, $m_1\leq m_2$. LHC measurements indicate that the observed 125 GeV Higgs field is quite SM-like, and currently the mixing angle is constrained to satisfy\cite{Profumo:2014opa, Robens:2015gla, Chalons:2016jeu}
\begin{equation}
 \left|\sin \theta\right| \lesssim 0.2 - 0.3 \, ,
 \end{equation}
 depending on the mass of the singlet-like state and assuming $m_1\simeq125$ GeV (i.e.~taking the singlet-like state to be the heavier scalar). We will primarily focus on this case in what follows. This approximate bound comes from a combination of direct searches at the LHC as well as EW precision measurements, and applies across the mass range $m_2 \sim 125$ GeV -- 1 TeV. For more precise mass-dependent limits, see e.g.~\cite{Profumo:2014opa, Robens:2015gla, Chalons:2016jeu}.
 
An interesting limit of the model arises by imposing a discrete $Z_2$ symmetry under which $S\to -S$.  
If the $Z_2$ symmetry remains unbroken at $T=0$, 
there is no $h-S$ mixing, and $S$ might be stable on cosmological time scales, thereby contributing to the observed DM density. The prospects for discovery at colliders in this case are much more limited, since $S$ can only be pair produced through the Higgs and can only be indirectly observed through missing energy in such events~\cite{Craig:2014lda, Curtin:2014jma}. However, from the standpoint of LISA, the prospects are more encouraging, as we will see below. 

In both the general and $Z_2$ cases, the tree-level couplings of $S$ to $H$ can yield a strong first-order EWPT. Such a transition can proceed in either one or two steps~\cite{Espinosa:2011ax, Profumo:2007wc}. In the two-step case, the EWPT is preceded by a transition in the singlet direction. The strength of the EWPT is governed by the couplings of $S$ to the Higgs field ($a_1$ and $a_2$). These parameters also govern the interactions of the singlet-like state with the SM. We discuss the prospects for LISA and other experiments to probe these models below.

After fixing the Higgs mass and VEV, as well as setting the singlet VEV to zero, the model in~(\ref{eq:singlet_pot}) contains five free parameters. In~\cite{Chen:2017qcz}, a strategy was suggested for systematically studying the parameter space featuring strong first-order transitions, which we adopt here. Specifically, with regards to the LHC, the most relevant parameters of the model are the singlet-like mass,
and the mixing angle with the Higgs, $\theta$. For a given $m_2$ and $\theta$, one can then vary $a_2$, $b_3$ and $b_4$ across the entire range of values allowed by 
\begin{itemize}
\item Absolute vacuum stability: The potential~(\ref{eq:singlet_pot}) is required to predict no zero-temperature vacua deeper than the physical minimum at $v=246$ GeV, either at tree-- or one-loop--level\footnote{Note that in~\cite{Chen:2017qcz}, tree-level vacuum stability was not required.}.
\item Perturbativity: We require $a_2$, $b_4 < 4 \pi$, so that an analysis utilizing standard perturbation theory techniques can be justified\footnote{In~\cite{Chen:2017qcz}, in addition to perturbativity, the coupling $b_4$ was bounded by perturbative unitarity. In our case, since we require absolute vacuum stability at tree-level, this requirement does not significantly impact the parameter space.}.
\end{itemize}

Imposing these requirements, we scan linearly over the corresponding parameter space for various values of $m_2$ and $\sin \theta$, searching for points with a strong first-order EWPT\footnote{Strong singlet-only transitions were not searched for, but can also produce a signal at LISA. However, the physics driving such transitions does not necessarily involve the couplings of $S$ to the SM Higgs, and so the collider prospects for discovery are not as correlated.}, defined as those for which $\vc/\Tc\geq 1$, with $\vc$ the Higgs VEV at the critical temperature $\Tc$. Our conventions for the Coleman-Weinberg and finite-temperature contributions to the effective potential are those detailed in~\cite{Chen:2017qcz}. To determine the properties relevant for the GW signal, we define the PT temperature $T_*$ as that for which the three-dimensional Euclidean action of the bounce solution, $S_3$, satisfies $S_3/T_* = 100$. This choice was inspired by the percolation requirement in~(\ref{eq:Sc_improved}). 
We compute the bounce with a modified version of the \texttt{CosmoTransitions pathDeformation} module~\cite{Wainwright:2011kj}, and use this to determine $\beta/H$ and $\alpha$ at $T_*$.

For illustration, we consider masses $m_2 = 170$, 240 GeV and $ \sin \theta = 0.01$, 0.1. These values are motivated by distinct discovery prospects at the LHC and future colliders. As discussed above, currently, values of $|\sin\theta| \lesssim 0.2$ are not constrained experimentally for these masses. At the high-luminosity LHC, direct searches for the singlet-like state will likely probe $|\sin\theta | \gtrsim 0.1$ for $m_2 > 2 m_W$~\cite{Buttazzo:2015bka}. In contrast, the parameter space with $\sin\theta = 0.01$ is unlikely to be probed by the LHC or future colliders. LISA sensitivity to these points therefore illustrates its complementarity with collider searches.

It is worth noting that resonant di-Higgs production ($p p \to h_2 \to h_1 h_1$) can provide an additional probe of the parameter space for $m_2 > 2 m_1 \approx 250$ GeV, ~\cite{No:2013wsa, Chen:2014ask, Huang:2017jws, Lewis:2017dme}, but we do not discuss this region further here, as larger singlet masses tend to be harder to detect with LISA. For lower masses, non-resonant $h_2 h_2$ and $h_1 h_1$ production can also provide a handle on the parameter space  for large values of $a_2$,~\cite{Chen:2017qcz}. It would be interesting to investigate the correlation between these observables and LISA in the future.

\begin{figure}[t]
\begin{center}
 \includegraphics[width=0.45\columnwidth]{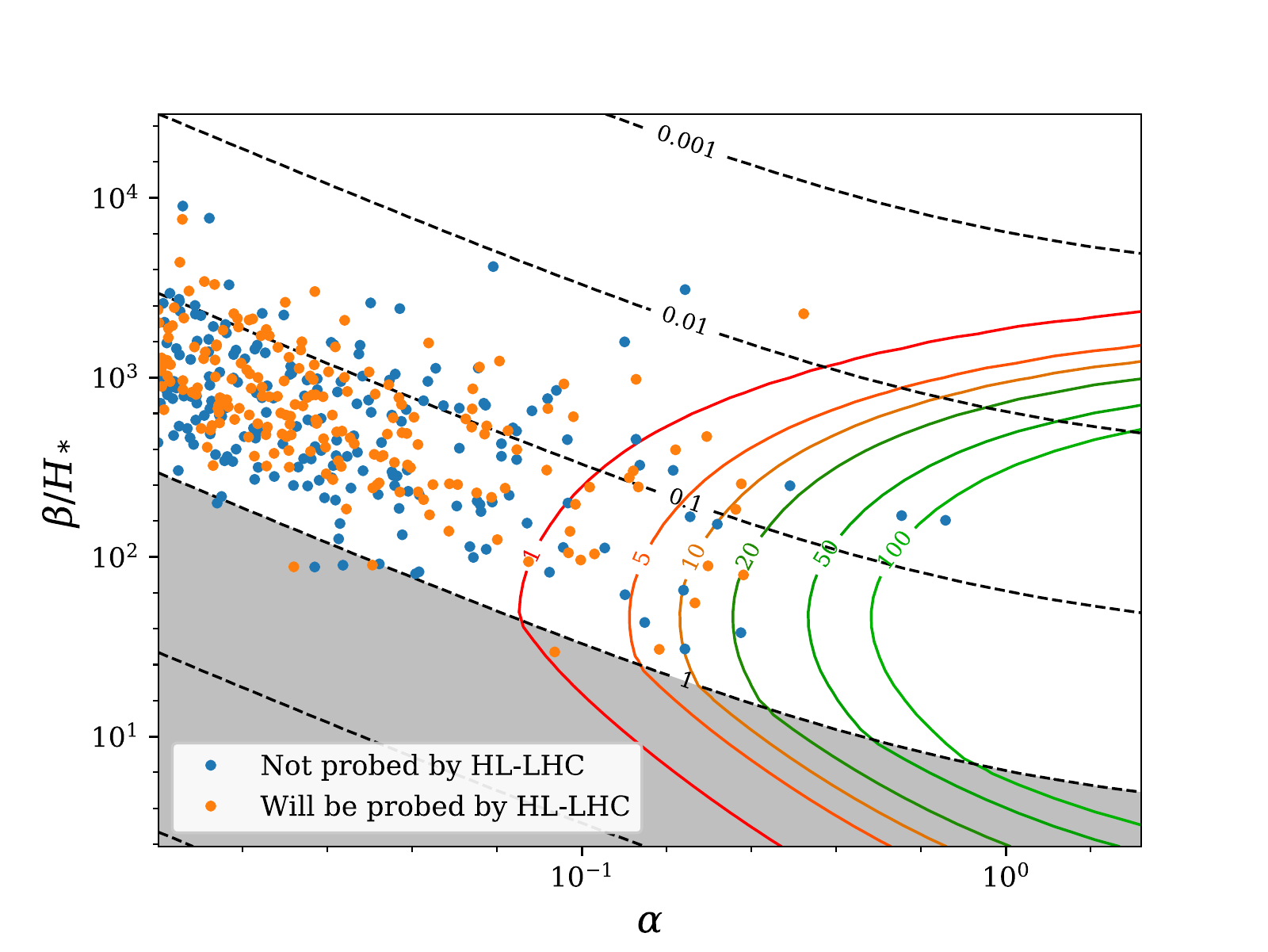}
 \includegraphics[width=0.45\textwidth]{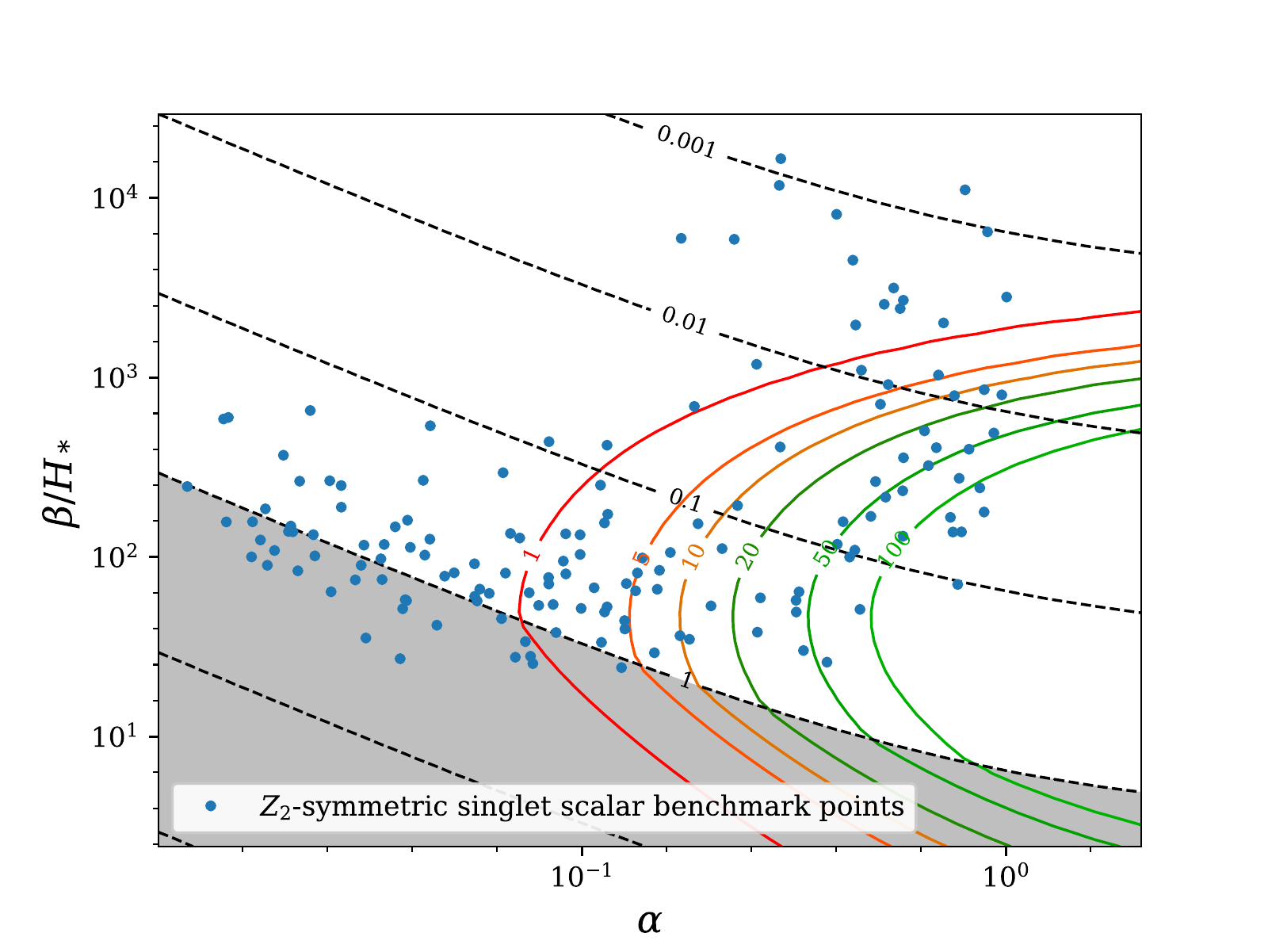}
 \caption{ \label{fig:singlet_results}\small LEFT: Predicted values of $\alpha$ and $\beta/H$ for $m_2 = 170$ GeV and 240 GeV (combined in one plot) in the general singlet model obtained by linearly varying the free parameters of the model and imposing requirements as described in the text. The mixing angles considered were $\sin \theta = 0.01$ (blue points) and  $0.1$ (orange points). The models in blue (orange) are unlikely (likely) to be probed by the high-luminosity LHC. The expected LISA sensitivities correspond to $T_*=50$ GeV. RIGHT: Sensitivity curve for the $Z_2$ symmetric singlet extension. In both the left and right panels we have taken $\vw=1$.
}
\end{center}
\end{figure}

In the left panel of Fig.~\ref{fig:singlet_results}, we show results of our analysis of the parameter space for $m_2 = 170$, 240 GeV in terms of the properties of the EWPT relevant for GW production and detection at LISA. We categorize points by whether or not they are expected to be probed by direct searches at the high-luminosity LHC with 3000 fb$^{-1}$. We find that most points feature $T_*$ between 20-100 GeV. As a result, the LISA sensitivity curves shown assume $T_*=50$ GeV. Also, for the predicted sensitivities, we have assumed a large wall velocity, $v_{\rm w} = 1$. This is expected to be an excellent assumption for points with strong PTs in singlet models~\cite{Bodeker:2009qy, Konstandin:2014zta, Kozaczuk:2015owa,Bodeker:2017cim}. 

We see that LISA will be sensitive to the strongest EWPTs in this model, while transitions of intermediate strength and below ($\alpha< \mathcal{O}(0.1)$) may not be probed. 
  Our results also show that LISA will be sensitive to regions of the parameter space not accessible to the LHC or other near-future experiments. There also exist regions to which both LISA and the LHC may be sensitive. This opens up the exciting possibility of detection both in direct collider searches, and a confirmation of a strong first-order EWPT through the GW signal at LISA.

We repeat the previous exercise in the $Z_2$ limit, corresponding to $\sin{\theta} = 0$, $b_3=0$ in our parametrization, scanning over $m_2 < 300$ GeV. We also require the singlet
to be heavier than half the HIggs mass to avoid bounds on the invisible Higgs branching fraction.
We concentrate on the parameter space producing two-step PTs, as these transitions tend to be stronger and hence can yield an observable signal at LISA (see e.g.~\cite{Huang:2016cjm}). 
We vary $a_2 < 3$ and $b_4 < 3$ searching for strong first-order PTs requiring
absolute vacuum stability. 
We show our results in the right panel of Fig.~\ref{fig:singlet_results}. Our results agree with the conclusions appearing elsewhere in the literature (see e.g.~\cite{Curtin:2014jma, Caprini:2015zlo, Huang:2016cjm, Beniwal:2017eik, Kurup:2017dzf,Curtin:2016urg}).

Comparing these results to those of the collider studies in e.g.~\cite{Curtin:2014jma, Huang:2016cjm}, it can be inferred that the LHC will probe virtually none of the parameter space shown. Future colliders will provide some sensitivity through a combination of measurements of the $Zh$ production cross-section, deviations in the Higgs self-coupling, and non-resonant $SS$ production. Reference~\cite{Curtin:2014jma} claims that a combination of these measurements might be able to cover the entire parameter space shown. Even if this is the case, LISA will likely have the first opportunity to directly probe the parameter space with strong first-order transitions in the $Z_2$ case\footnote{If the $Z_2$ symmetry is valid up to high enough scales, $S$ can be cosmologically stable, contributing to the observed DM abundance. As a result the model can also be constrained by DM direct detection experiments~\cite{Curtin:2014jma}. Note however, that these signatures are much more model-dependent, in that they are sensitive to whether or not additional DM annihilation channels are present. Furthermore they are not necessarily correlated with the existence of a strong first-order EWPT.}.

\subsubsection{Non-perturbative PT analysis}
\label{sec:nonpert_singlet}

The EWPT in the singlet model has also been studied non-perturbativey. The dimensional reduction (DR) and a preliminary investigation of the parameter space using existing SM lattice results were performed for the singlet-extended SM in \cite{Brauner:2016fla}. Subsequently,~\cite{Gould:2019qek} investigated several different regions of the model parameter space both in the $Z_2$ and non-$Z_2$ cases, comparing the perturbative and nonperturbative predictions. Requiring the self-consistency of the dimensional reduction and that the effects from higher-dimension operators and dynamics of the singlet field\footnote{In these studies, the singlet field was integrated out altogether, and so any potential impact of the singlet-Higgs vacuum structure on the transition discarded. This is a general issue for EFTs \cite{Damgaard:2015con} and not specific to DR.} are small limits the regions of parameter space that can be accurately described by the SM-like 3D effective field theory (EFT) used in these studies. Where the 3D EFT results can be trusted,~\cite{Gould:2019qek} found reasonable agreement for the scaling of the PT strength with the model parameters in both approaches. Furthermore,~\cite{Gould:2019qek} applied existing lattice results to determine $\alpha$, $\beta$ and $T_{\rm n}$ for an illustrative benchmark point in the general singlet model, finding good agreement between perturbative and non-perturbative approaches for the relative amount of supercooling, $T_\text{n}/T_c$, while the individual quantities $T_{\rm n}$, $\alpha$, and $\beta/H_{\rm n}$ featured larger (up to $\mathcal{O}(100\%)$) discrepancies. However, accounting for uncertainties due to the neglected dimension-6 operators in the 3D theory, as well as the scale-dependence in the perturbative results, brings the results into closer agreement.

Taken at face value, the findings of~\cite{Brauner:2016fla} and~\cite{Gould:2019qek} suggest that the perturbative calculations reflected in Fig.~\ref{fig:singlet_results} might predict the correct values of $\alpha$ and $\beta/H$ to within $\mathcal{O}(1)$ factors. However, none of the parameter space considered in~\cite{Gould:2019qek} features PTs strong enough to be detectable by LISA, and so future work is required to draw firm conclusions about the accuracy of perturbative predictions in these regions of the model. This will require dedicated nonperturbative studies of the singlet-extended SM, which are in progress.

\subsection{Extensions of the SM with Scalar Electroweak Multiplets} 
\label{sec:doublet}

A strong first-order EWPT may be obtained through the existence of a non-minimal scalar sector 
featuring electroweakly charged scalar fields in addition to the SM Higgs doublet. The SM gauge non-singlet nature of these fields 
helps in making them accessible at colliders, which yields a more direct connection between the 
generation of GWs from a first-order cosmological PT and phenomenological LHC studies.
Here we focus mostly on two-Higgs-doublet-model (2HDM) scenarios, in which the SM Higgs sector is enlarged 
by a second doublet. These scenarios provide a compelling benchmark of extended scalar 
sectors where the new scalars are charged under the EW symmetry of the SM. We also briefly discuss  the case of other EW representations, namely scalar triplet extensions of the SM, at the end of this subsection.

The strength of the EWPT in the 2HDM and its early Universe implications have been widely studied in the 
literature~\cite{Huet:1995mm, Cline:1996mga, Fromme:2006cm, Cline:2011mm, Dorsch:2013wja, Dorsch:2014qja, Kakizaki:2015wua,
Dorsch:2016nrg, Basler:2016obg, Dorsch:2017nza, Basler:2017uxn, Bernon:2017jgv, Huang:2017rzf}.
In the following we provide an overview of the discovery potential of such a scenario at LISA and discuss the complementarity with 
ongoing and future BSM searches at the LHC. The scalar potential for the 2HDM is given by
\begin{eqnarray}	
\label{2HDM_potential2}
V(H_1,H_2) & = & \mu^2_1 \left|H_1\right|^2 + \mu^2_2\left|H_2\right|^2 - \mu^2 \left[H_1^{\dagger}H_2+\mathrm{h.c.}\right] +\frac{\lambda_1}{2}\left|H_1\right|^4 
+\frac{\lambda_2}{2}\left|H_2\right|^4 \nonumber \\
&  & + \lambda_3 \left|H_1\right|^2\left|H_2\right|^2 +\lambda_4 \left|H_1^{\dagger}H_2\right|^2 +
 \frac{\lambda_5}{2}\left[\left(H_1^{\dagger}H_2\right)^2+\mathrm{h.c.}\right] \, , \nonumber
\end{eqnarray}
where we have assumed a (sofly broken) $Z_2$ symmetry\footnote{This is assumed for phenomenological reasons: its absence 
results in potentially dangerous flavour-changing-neutral-current contributions once the Higgs doublets are coupled to the SM fermions.}.
The breaking of the EW symmetry is shared between the two Higgs doublets $H_1$ and $H_2$, their VEVs given by 
$v_{1,2}$ (with $\sqrt{v_1^2 + v_2^2} = v = 246$ GeV, $v_2/v_1 \equiv \mathrm{tan} \beta$).
The 2HDM features three new physical states in addition to the 125 GeV Higgs $h$: 
a charged scalar $H^{\pm}$ and two neutral states $H_0$, $A_0$. 
A final relevant aspect of the 2HDM concerns the various possibilities for coupling the Higgs doublets 
$H_{1,2}$ to the SM fermions (see e.g.~\cite{Branco:2011iw}). This has implications for the phenomenology of the model, e.g.~the current and future 
LHC sensitivity prospects, but it is not relevant for the EWPT\footnote{The top quark is generically the only SM fermion whose Yukawa coupling is large enough to 
be of relevance for the EWPT, and the different possible fermion coupling choices share the same top quark Yukawa coupling assignment.}.
We will comment on the impact of 2HDM fermion coupling choices for the complementarity between the LHC and LISA when relevant.

The strength of the EWPT in 2HDM scenarios is primarily determined by a decrease in the free-energy
difference between the EW-symmetric maximum and EW-broken phase at $T = 0$~with respect to the corresponding SM value~\cite{Dorsch:2017nza, Bernon:2017jgv} 
(see also the discussion in Sec.~\ref{sec:eft}). 
This effect is mostly due to $T = 0$ one-loop contributions to the free energy. 
As a result, a strongly first-order EWPT in 2HDMs is very much correlated with the SM-like nature of the 
Higgs state $h$ when $m_{H_0} > m_h = 125$ GeV, as well as with a large mass splitting between the  
states $H_0$, $A_0$~\cite{Dorsch:2014qja, Dorsch:2017nza, Basler:2016obg, Basler:2017uxn},
which provides an important connection to LHC searches for new scalars in final states consisting of gauge and Higgs bosons.

\begin{figure}[t]
\begin{center}

\includegraphics[width=0.36\textwidth]{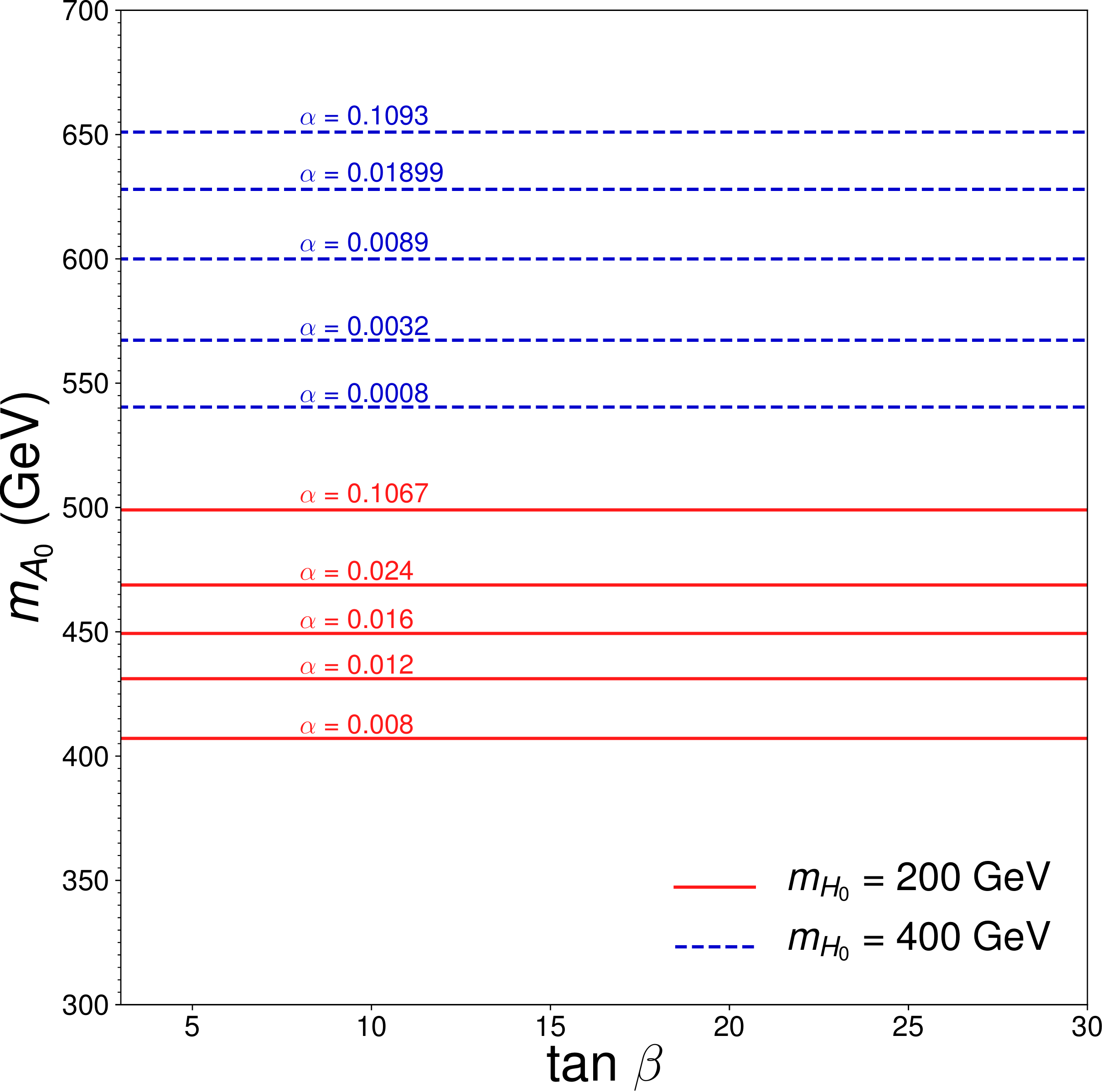}
\includegraphics[width=0.55\textwidth]{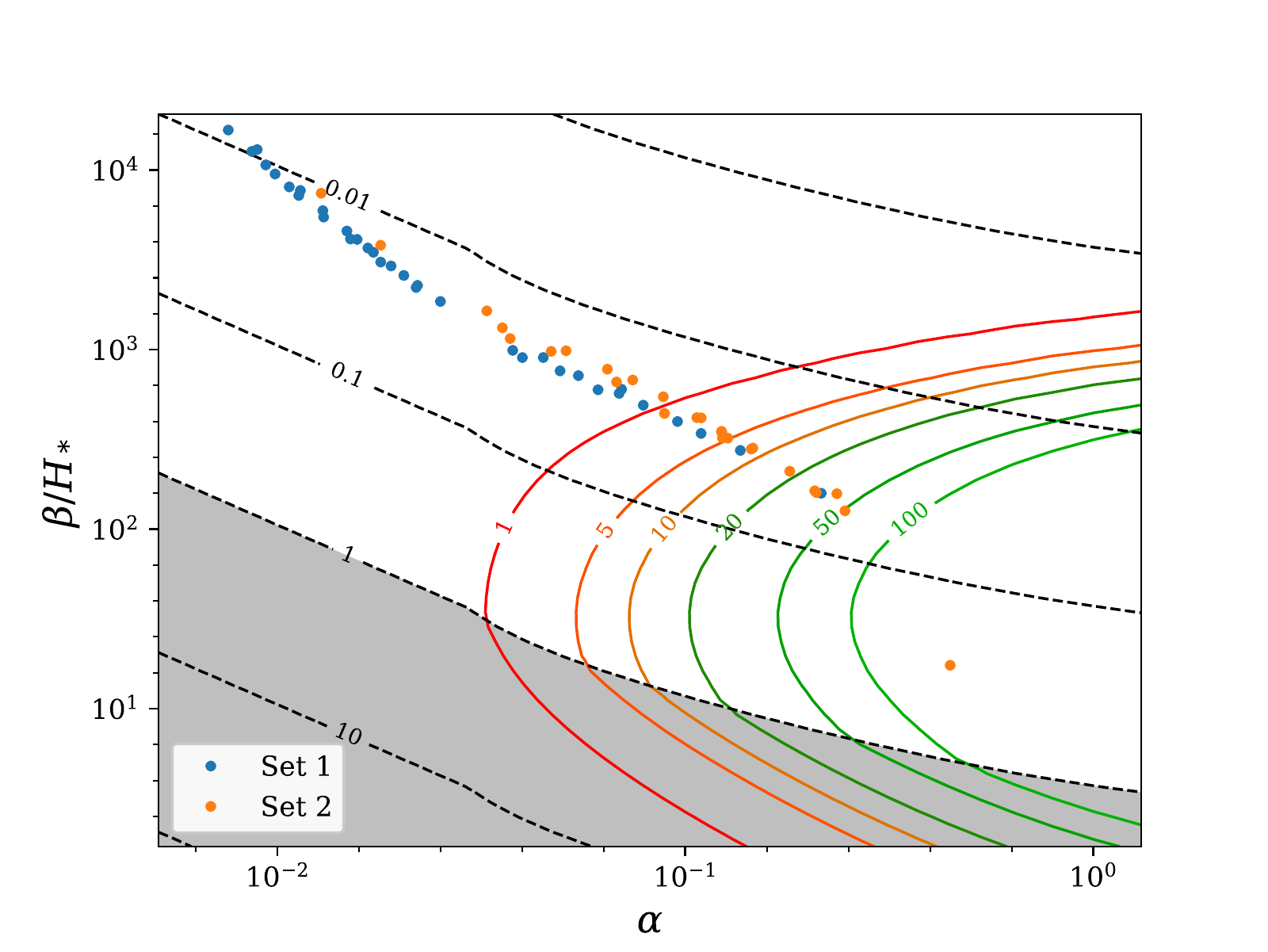}

\caption{ Results in the 2HDM. LEFT: Strength of the EWPT as given by $\alpha$ 
 in the ($\mathrm{tan} \beta$, $m_{A_0}$) plane, for $m_{H_0} = 200$ GeV (solid red) and 
 $m_{H_0} = 400$ GeV (dashed blue). RIGHT: 2HDM benchmark points (from a scan in $m_{H_0}$, $m_{A_0}$, see text for details) 
 in the ($\alpha$, $\beta/H$) plane. The points currently excluded by the LHC for the
 {\it Type-II} 2HDM (but not for the {\it Type-I} 2HDM) are shown in orange. The expected LISA sensitivities correspond to $T_*=50$ GeV and assume $\vw=0.7$. \label{Strength_EWPT_2HDM}}
\end{center}
\end{figure}

In the following, we assume a SM-like 125 GeV Higgs, as currently favoured by LHC Higgs experimental 
data~\cite{ATLAS:2019slw}. For the 2HDM this means aligning the state $h$ with the direction of EW symmetry breaking. 
We also assume $m_{H^{\pm}} \simeq m_{A_0}$ as favoured by 
EW precision observables. Finally, we consider the squared mass scale $M^2 \equiv \mu^2 (\mathrm{tan} \beta + \mathrm{tan}^{-1} 
\beta)$, corresponding to the approximate mass scale for the non-SM scalar doublet within the 2HDM, to satisfy, for 
simplicity\footnote{Despite not being strictly required, this condition is favoured by 
the boundedness of the scalar potential from below and unitarity considerations in the 2HDM.}, $M^2 \simeq m_{H_0}^2$. 
These assumptions leave $m_{H_0}$, $m_{A_0}$ and $\mathrm{tan}\,\beta$ 
as the relevant parameters in our subsequent analysis.  
In addition, we find that $\mathrm{tan}\,\beta$  does not influence the strength of the 
EWPT as measured by the parameter $\alpha$ 
($\mathrm{tan}\,\beta$ does however play an important role in the corresponding LHC phenomenology, see the discussion below), 
which is shown in Fig.~\ref{Strength_EWPT_2HDM} (left), in the ($\mathrm{tan} \beta$, $m_{A_0}$) plane for 
$m_{H_0} = 200$ GeV (solid lines) and $400$ GeV (dashed lines).

We perform a scan in $m_{H_0}$, $m_{A_0}$, with $m_{H_0} \in [180\,\mathrm{GeV},\,\,450\,\mathrm{GeV}]$ and 
$m_{A_0} \in [m_{H_0} + 100\,\mathrm{GeV},\,\,m_{H_0} + 350\,\mathrm{GeV}]$, computing the PT 	parameters $\alpha$, $\beta/H$ and $T_*$ for points predicting a strong first-order PT. 
The corresponding 2HDM results in the ($\alpha$, $\beta/H$) plane  are shown on the right hand side of Fig.~\ref{Strength_EWPT_2HDM} together with 
several LISA fixed-SNR contours, as provided by \texttt{PTPlot}. The LISA SNR curves are computed for $T_{*} = 50$ GeV (corresponding approximately to the values of $T_{*}$ 
for the 2HDM benchmarks yielding the strongest transitions,~see e.g.~\cite{Dorsch:2016nrg}) and assuming a bubble wall 
velocity of\footnote{Preliminary attempts towards computing the bubble wall velocity in 2HDM scenarios featuring a strongly first-order PT 
were made in~\cite{Dorsch:2016nrg,Dorsch:2018pat}, where it was found that in a considerable part of parameter space bubbles expand as deflagrations, 
with only the strongest transitions being detonations.} $v_{\rm w} = 0.7$.
The blue and orange points in the figure correspond to 2HDM benchmarks for which different LHC constraints apply, as discussed below.
We stress that allowing for a slight departure from our assumptions about the 2HDM parameters  
(SM-like Higgs $h$, $m_{H^{\pm}} \simeq m_{A_0}$ and $M^2 \simeq m_{H_0}^2$) within the experimentally allowed region 
would result in a small decorrelation between $\alpha$ and $\beta/H$ as compared to the very strong correlation shown 
in Fig.~\ref{Strength_EWPT_2HDM} (essentially, a minor ``thickening" 
of the line traced out by the 2HDM benchmark data in Fig.~\ref{Strength_EWPT_2HDM}). 
We also note that in the 2HDM one tends to find larger values of $\beta/H$ for a given $\alpha$ as compared to other models 
(e.g.~the singlet model, see Sec.~\ref{sec:singlet}).

LHC searches for the BSM states $A_0$, $H_0$, $H^{\pm}$ can also probe the 2HDM region of parameter space with a strongly 
first-order EWPT, resulting in a strong level of complementarity between the LHC and LISA.
In particular, given the sizable mass splitting $m_{A_0} - m_{H_0}$ characteristic of a strong EWPT in the 
2HDM\footnote{This can be readily inferred from Fig.~\ref{Strength_EWPT_2HDM}: a strongly first-order EWPT in the 2HDM observable by LISA 
would require at least $\alpha \gtrsim 0.1$, which from Fig.~\ref{Strength_EWPT_2HDM} (left) implies $m_{A_0} - m_{H_0} \gtrsim 250$ GeV.},
LHC searches for $A_0 \to Z H_0$~\cite{Dorsch:2014qja} by CMS~\cite{Khachatryan:2016are} and ATLAS~\cite{Aaboud:2018eoy} 
yield important constraints on the allowed 2HDM parameter space. 
These constraints depend on the specific type of 2HDM Yukawa coupling assignment: 
for a {\it Type-II} 2HDM, the present bounds from the LHC $\sqrt{s} = 13$ TeV ATLAS 
$A_0 \to Z H_0$ search with $36.1$ fb$^{-1}$ of integrated luminosity~\cite{Aaboud:2018eoy} exclude 
values $m_{H_0} \lesssim 340$ GeV independently of $\mathrm{tan}\,\beta$. The points that would be currently excluded by this 
search for a {\it Type-II} 2HDM are depicted in orange in Figs.~\ref{Strength_EWPT_2HDM}.
In contrast, for a {\it Type-I} 2HDM the present limits from~\cite{Aaboud:2018eoy} yield 
a lower bound on $\mathrm{tan}\,\beta$ as a function of $m_{H_0}$ and $m_{A_0}$.
Future LHC searches (see e.g.~\cite{Cepeda:2019klc}) will be able to completely explore the region
of parameter space with a strongly first-order EWPT within the 
{\it Type-II} 2HDM, while for a {\it Type-I} 2HDM the future LHC measurements will only be able to increase the 
reach in $\mathrm{tan} \beta$ without fully exploring the model parameter space. LISA, however, can 
probe such points, making clear
the complementarity between the LHC and LISA.

\vspace{3mm}

Beyond the 2HDM discussed above, other scenarios with EW scalar multiplets added to the SM could lead to a first-order EWPTs and a GW signature potentially within the sensitivity range of LISA. 
References~\cite{Patel:2012pi,Chala:2018opy} (see also~\cite{Blinov:2015sna, Inoue:2015pza}) studied the properties of the EWPT
 within a scalar EW triplet extension of the SM. They found that this scenario allows for a multi-step 
PT analogous to that of the $Z_2$ symmetric singlet extension
of the SM\footnote{However, as opposed to the two-step transition for the $Z_2$ symmetric singlet scenario, here
the first step of the transition breaks the EW symmetry. This could impact the generation of GWs (bubble wall friction 
from the SM gauge bosons would generically result in smaller values of $\vw$ than in the singlet scenario), as well as other processes such as EW baryogenesis.} 
(recall the discussion in Sec.~\ref{sec:singlet}).
Such two-step transitions have also been explored in~\cite{Blinov:2015sna} in the context of the {\sl inert doublet model} (a particular version of the 2HDM 
with an unbroken $Z_2$ symmetry). These multi-step scenarios in BSM theories with a new scalar EW inert doublet or triplet could give rise to very strong first-order EWPTs within the reach 
of LISA (see e.g.~\cite{Chala:2018opy}), while also predicting several distinct collider signatures~\cite{Patel:2012pi, Blinov:2015sna, Blinov:2015qva}. The latter include the modification of the Higgs 
coupling to two photons and signals associated with a compressed scalar spectrum (e.g.~disappearing tracks and final states involving soft objects).

Finally,~\cite{Chala:2018ari} has also considered higher EW scalar multiplets (EW quadruplets) as ultraviolet (UV) completions of a SM EFT framework 
(see Sec.~\ref{sec:eft}), finding that strongly first-order EWPTs yielding a GW signal detectable by LISA are 
possible, and that there is a high-degree of complementarity with the LHC: quadruplet masses $m_{\Theta} \lesssim 600$ GeV would be probed by the
LHC via multi-lepton signatures, while LISA could in principle probe much higher masses depending on the value of the couplings between 
the EW quadruplet and the SM Higgs doublet.

\subsubsection{Non-perturbative PT analysis}
\label{sec:nonpert_doublet}

Some of the scenarios discussed in the previous subsection have also been studied nonperturbatively at finite temperature. 
The dimensional reduction of the 2HDM to a SM-like 3D EFT (see Appendix~\ref{sec:appendixlattice}) was performed in~\cite{Laine:1996ms,Andersen:2017ika,Gorda:2018hvi}, and the PT 
analysed using existing SM lattice results in the case 
without CP-violation and in the alignment limit assuming $m_{H^{\pm}}=m_{A_0}$. The parameters $m_{H_0}$, $m_{A_0}$ and $\mu^2$ 
were then varied for a few values of tan$\beta$, showing that a region featuring a first-order PT emerges (see~\cite{Andersen:2017ika}) and broadly confirming 
that $m_{A_0} > m_{Z}+m_{H_0}$ is preferred in this region. 
However, in parts of the parameter space, issues remain regarding the validity of one or more steps in the DR procedure, as well as the mapping to the 
3D theory in this approach. Again, some information about the two-field transition dynamics may be lost, even in the alignment limit. 

More recently,~\cite{Kainulainen:2019kyp} performed new lattice simulations including the dynamical second doublet. The results for the equilibrium properties ($\alpha$, $\Tc$) were compared with predictions from perturbation theory in both 3D (using the two-loop effective potential) and 4D using different resummation schemes. For the benchmark points considered in~\cite{Kainulainen:2019kyp}, the standard 4D treatment (used in the results of Fig.~\ref{Strength_EWPT_2HDM}, for example) reproduces $\Tc$ to within about $20\%$ and $\alpha$ to within a factor of two. The 3D two-loop effective potential approach more closely reproduces the lattice results, but $\mathcal{O}(10)\%$ uncertainties remain in some cases. The results of this section obtained in the 4D perturbative approach should therefore be understood in light of these findings.

As emphasized in~\cite{Kainulainen:2019kyp}, the PTs studied nonperturbatively to date are not strong enough to be detected by LISA. To go to larger $\alpha$, larger couplings are required, which decreases the accuracy of the dimensional reduction. Future work is therefore required to make nonperturbative predictions for LISA in the 2HDM.

Similarly to the case of a second doublet added to the SM, the scenario with an additional scalar triplet~\cite{Niemi:2018asa} may also be 
dimensionally reduced to a SM-like EFT at long distances, and predicts a region of first-order PTs. 
As for the doublet and singlet models, the viability of the DR in the first-order PT region may be an 
issue (see~\cite{Niemi:2018asa} for details). The authors argue in favour of using a 
different 3D target theory, implying future numerical work to be done to pin down the corresponding phase diagram and extract nonperturbative predictions relevant for LISA.

\subsection{SUSY-extended SM} 
\label{sec:susy}

The previous two subsections studied LISA's sensitivity to extensions of the SM with minimal new field content and motivated mostly from a bottom-up perspective. However, LISA can also probe richer UV-complete scenarios. The classic example is supersymmetry (SUSY). SUSY is one of the most investigated embeddings of the SM. Several features make SUSY compelling: it predicts gauge coupling unification, provides a variety of DM candidates, and is an important ingredient in theories of quantum gravity such as string theory. Furthermore, SUSY arises naturally in generalizing the Poincare algebra. As a result, there are no quadratic divergences in the scalar sectors of SUSY models, while the remaining logarithmic divergences are not sensitive to e.g.~the Planck cutoff but instead to the (potentially much lower) SUSY breaking scale. All radiative corrections are then small when the SUSY breaking scale is close to the TeV scale, in which case the predicted EW scale is of the order of the $W$ and $Z$ boson masses without requiring tuning between the low-energy parameters. Intriguingly, low-scale SUSY breaking implies light new physics, with obvious implications for EW symmetry breaking and, possibly, the dynamics of the EWPT.

The MSSM is the simplest supersymmetric extension of the SM. In this setup the EWPT can be strongly first-order if the two stops 
(scalar top partners) feature small mixing and one of the two is lighter than the top quark~\cite{Laine:2012jy, Carena:1996wj, Delepine:1996vn, Carena:2008vj}. Interestingly enough, naturalness also favors light stops. However light stops are in strong tension with current LHC measurements: a light stop with small mixing boosts the ratio between the Higgs gluon-fusion and vector-boson-fusion signal strengths. Furthermore, light stops are highly constrained by direct searches at ATLAS and CMS~\cite{Menon:2009mz, Cohen:2012zza, Curtin:2012aa, Carena:2012np, Katz:2015uja}. Despite potential loopholes in the experimental limits (see e.g.~\cite{Liebler:2015ddv}), it seems unfeasible to overcome all current constraints simultaneously and so a strong first-order EWPT in the MSSM can be considered essentially ruled out.

\begin{table}[]
\begin{center}
\begin{tabular}{|c||c|c|c|c||}
\hline
\multirow{2}{*}{Benchmark} & \multicolumn{4}{c||}{PT parameters}  \\ \cline{2-5} 
               & $\alpha$      & $\beta/H_*$    &  $v_{\rm w}$ & $T_*$/GeV  \\ \hline \hline
    SUSY$_1$ (1A) 
      &    0.037   &  277     &  0.95     &  112     \\ \hline
SUSY$_1$ (1B) 
&  0.066     &   106  &  0.95 &  95 \\ \hline 
SUSY$_1$ (1C) 
      &  0.105     &   33.2  &  0.95 &  82 \\ \hline
     SUSY$_1$ (1D) 
          &  0.143     &   6.0  &  0.95 &  76.4 \\ \hline
     SUSY$_2$ (2A) 
        &  0.050     & 830      &  0.73     &  135    \\ \hline
     SUSY$_2$ (2B) 
      &   0.040     &  2914    &  0.72 & 146  \\ \hline
     SUSY$_3$ (3A)  
         &    0.062   &  214     &   0.5    &  74     \\ \hline
     SUSY$_3$ (3B) 
         &  0.045    &  200     &  0.5 &  79 \\ \hline
      SUSY$_4$ (4A) 
                  &        0.22    &  57      &   0.95    &   48    \\ \hline
\end{tabular}
\caption{\small Benchmarks scenarios for SUSY embeddings. Note that the listed values of $v_{\rm w}$ are quoted from the original references and are not necessarily realistic (see the main text). For the analysis of our benchmark points we simply take $v_w\simeq 0.95$ absent more detailed microphysical calculations of $\vw$ in these models.}\label{tab:susy}
\end{center}
\end{table}

Non-minimal supersymmetric scenarios, on the other hand, can still feature strong first-order EWPTs, the singlet extension of the MSSM being a prominent example. Numerous analyses have investigated the EWPT in this setup before the LHC constraints on SUSY became as stringent~\cite{Pietroni:1992in, Davies:1996qn, Apreda:2001us, Kozaczuk:2013fga}. As a proof of principle, one can however check that, although tuned, there exists a region of parameter space compatible with a very strong EWPT and that would avoid detection at the LHC. To arrive at such scenarios, one must build a superpotential allowing all new physics irrelevant for the EWPT to  be above the LHC energy reach. This implies that an exact $Z_2$ symmetry cannot be enforced on the singlet~\cite{Huber:2015znp}. The 125 GeV Higgs and the singlet therefore mix. This mixing must be tuned to a small value to mimic the SM-like Higgs signal strengths. As a result of these considerations, at low energy the theory looks like the singlet extension of the SM with an approximate $Z_2$ symmetry (see Sec.~\ref{sec:singlet}). In particular EW symmetry breaking can proceed via a two-step EWPT with ultra-relativistic bubbles~\cite{Huber:2015znp}. The benchmark points in Table~\ref{tab:susy} labeled SUSY$_1$ are taken from~\cite{Huber:2015znp} and represent this setup.

From the standpoint of LISA, it is interesting to note that singlet extensions of the MSSM  do not necessarily predict ultra-relativistic bubbles. In contrast to the aforementioned split scenario, one can instead assume the electroweakinos to be light and obtain a EWPT strong enough to be accessible by LISA through the mechanism discussed previously~\cite{Bian:2017wfv, Demidov:2017lzf}. The presence of these new degrees of freedom in the thermal bath increases the friction between the bubble wall and the plasma. Moreover, they can provide a new source of CP-violation. Thus, at least in principle, all the ingredients required by electroweak baryogenesis could be present~\cite{Menon:2004wv , Akula:2017yfr, Kozaczuk:2013fga, Kozaczuk:2014kva, Demidov:2016wcv}. In Table~\ref{tab:susy}, the benchmark models labeled SUSY$_2$ and SUSY$_3$  correspond to this light electroweakino scenario and are taken from~\cite{Bian:2017wfv} and~\cite{Demidov:2017lzf} respectively\footnote{As discussed in the original articles, the obtained values of the bubble velocities are not computed microphysically and therefore are subject to large uncertainties.~\cite{Bian:2017wfv} determines $v_{\rm w}$ as the Chapman-Jouguet speed, which underestimates $v_{\rm w}$ in the detonation regime (see Sec.~\ref{sec:num_sim}).~\cite{Demidov:2017lzf} instead does not compute $v_{\rm w}$ and provides benchmark points assuming an optimistic and pessimistic value of $v_{\rm w}$ in the deflagration regime. Notice also that~\cite{Demidov:2017lzf} evaluates $\alpha$ as the normalized latent heat. We thank Ligong  Bian, Huai-Ke Guo and Jing Shu for communications about the benchmark scenarios used in their study~\cite{Bian:2017wfv}.}. It remains an open question whether such benchmark models, characterized by large CP-violating sources and light electroweakinos, would be able to produce the observed baryon asymmetry of the Universe while remaining consistent with bounds from electric dipole moment (EDM) experiments~\cite{Andreev:2018ayy, Panico:2018hal} and collider searches~\cite{Martin:2014qra, Arina:2016rbb, Chupp:2017rkp, Athron:2018vxy, Domingo:2018ykx}. Such limits are already very stringent. If they become stronger, these benchmark points would most likely be ruled out before LISA flies.

A further step into non-minimality consists of introducing custodial triplets. As a result of the custodial symmetry, these fields  can acquire large VEVs without spoiling EW precision tests~\cite{Georgi:1985nv, Cort:2013foa}. As a consequence, the structure of the scalar potential can differ significantly from that of the MSSM, and a tree-level barrier between the different phases could exist. Very strong PTs with ultra-relativistic bubble velocities can then occur~\cite{Garcia-Pepin:2016hvs}. Even if light, the triplets can evade current LHC bounds as well as relax some of the tension between light stops and LHC data~\cite{Sirunyan:2017sbn, Vega:2017gkk}. The benchmark scenario  SUSY$_4$ in Table~\ref{tab:susy} corresponds to this setup~\cite{Garcia-Pepin:2016hvs}. 
 
Finally, we remark that relaxing minimality can also result in modifications of theory at high energies. Such extensions may not impact the phenomenology at experimentally-accessible scales but instead could explain the origin of the ``accidental" tunings perceived at low energy in the examples we have discussed\footnote{For instance the so-called $\mu$ problem, which implies that heavy Higgsinos are not natural, is not a generic feature of SUSY embedding~\cite{Dimopoulos:2014aua, Delgado:2016vib, Garcia:2015sfa}.}. In other words, perhaps the lack of tuning at low energy is not a robust guiding principle. Here we refrain from judging this argument or assessing the plausibility of the SUSY benchmark scenarios discussed above; we  simply take the SUSY benchmark points that the literature has identified as interesting from the standpoint of GW signatures, and update their detection prospects in light of the new, more precise predictions discussed in Sec.~\ref{sec:prediction}.

\begin{figure}[t]
\begin{center}
\includegraphics[width=0.60\textwidth]{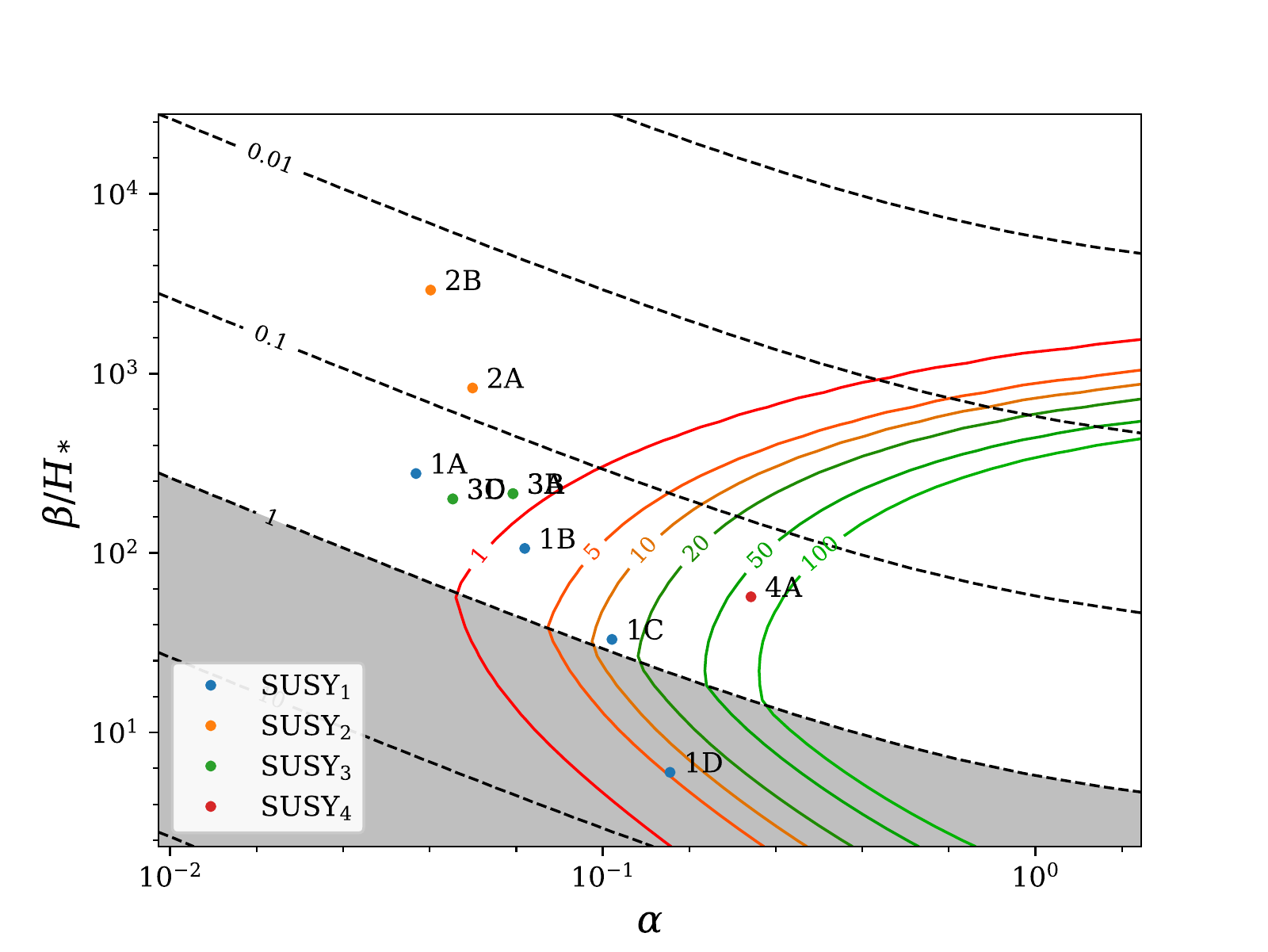}
 \caption{\small SUSY benchmark points in the ($\alpha, \beta/H$) plane for the benchmark models listed in Table \ref{tab:susy}, with $T_*$ rounded 
to 80\,GeV and $g_*$ to 108.75. For the wall velocity we assume $v_{\rm w} = 0.95$.} \label{fig:susy}
\end{center}
\end{figure}

Fig.~\ref{fig:susy} summarizes our findings for SUSY models summarized in Table~\ref{tab:susy}. These results show that LISA is capable of probing many of the proposed benchmark scenarios, provided $v_{\rm w}$ is large enough. In fact, were $v_{\rm w}$ below the speed of sound, our conclusions would be radically different. Existing calculations, however, suggest that whenever the EWPT is strong and the plasma does not differ much from that of the SM, wall velocities larger than the speed of sound arise. This is indeed what happens in 
SUSY scenarios that reduce to the singlet model at low energies where the wall velocity has been directly computed~\cite{Kozaczuk:2015owa}. We hence assume $v_{\rm w} = 1$ in Fig,~\ref{fig:susy} which, for PTs of these strengths, is likely to be more realistic than the Chapman-Jouguet speed assumed in most analyses. We emphasize, however, that microphysical calculations $\vw$ in these models should be done to obtain more accurate results.

\subsubsection{Non-perturbative PT analysis}

Lattice studies based on the 3D EFT approach have also been pursued for supersymmetric scenarios. Due to the wide range of possible spectra in SUSY models and the effort that each lattice analysis requires, non-perturbative studies of the EWPT  in the context of SUSY are not exhaustive. Those that do exist, however, suffice to highlight some overall trends when comparing between perturbative and non-perturbative predictions.

The structure of the MSSM 3D Lagrangian depends strongly on the assumed spectrum. The non-trivial dimensional reduction to 3D needs to be done on a case-by-case basis, depending on which degrees of freedom are assumed to be heavy. The MSSM 3D theory is therefore known only in a few regimes. The investigated regimes include the case in which all third generation squarks are light~\cite{Cline:1996mga}, the case where the right-handed stop is much lighter than all the other squarks~\cite{Losada:1996ju, Laine:1996ms, Bodeker:1996pc}, and the case where all the Higgs bosons are light (resembling the 2HDM) with or without a light slepton~\cite{Losada:1996ju}. 

In the MSSM, a 3D perturbative computation tends to find stronger EWPTs than those obtained in the standard 4D approach. The same feature arises from lattice simulations run on the 3D theories~\cite{Laine:1998qk, Laine:2012jy}. In particular, under similar assumptions concerning the heavy degrees of freedom, the lattice results relax the upper bound on the right-handed stop by around 30\% with respect to the 4D prediction (cf.~\cite{Laine:2012jy, Carena:2012np}). This sizable difference is however not sufficient to circumvent the LHC constraints on light stops, so that the aforementioned experimental tension persists.

Non-perturbative analyses of the EWPT in extensions of the MSSM have not been explicitly performed. The 3D EFT of the singlet extension of the MSSM can be found e.g.~in~\cite{Losada:1996ju} but we are not aware of lattice simulations based on this framework. The results from simulations run to explore other low-energy theories (see Sec.~\ref{sec:nonpert_singlet} and Sec.~\ref{sec:nonpert_doublet}) could be reinterpreted in terms of the supersymmetric singlet  extension. It is however unclear whether such an effort would not be premature without any experimental hint favoring a particular MSSM extension instead of other SUSY or non-SUSY setups. In any case, only perturbative results are currently available for predictions relevant for LISA.

\subsection{EFT approach to extensions of the SM} 

\label{sec:eft}

The models discussed in the previous subsections assumed the presence of new degrees of freedom present near the electroweak scale, however this need not be the case to obtain an observable signal at LISA. If the scale of new physics is sufficiently high, the impact of the new states on the EWPT can be parametrized and studied by means of an EFT. In this SM Effective Field Theory (SMEFT), the Higgs potential is augmented by terms of higher power in the Higgs field, suppressed by the scale of new physics, which we denote by $f$ (not to be confused with the GW frequency). This approach allows one to study the generic impact of new physics, in our case on the EWPT, without relying on specific models.

We write the Higgs doublet as $H = [0, ~ (h + v)/\sqrt{2}]^\text{T}$, with $h$ being the physical Higgs boson and $v= 246$ GeV the Higgs VEV. In the following we consider operators of mass dimension up to eight, so that the Higgs potential reads
 \begin{align}\nonumber
V_\text{tree}(h) &= -\mu_H^2 |H|^2 + \lambda_h |H|^4 + \frac{c_6}{f^2}|H|^6 + \frac{c_8}{f^4}|H|^8~\\
   &= -\frac{1}{2} \mu_H^2 h^2 + \frac{1}{4} \lambda_H h^4 + \frac{c_6}{8f^2}h^6 + \frac{c_8}{16f^4} h^8~.
\end{align}
If the effective theory is valid, operators with even higher mass dimension will not modify the results.
Here the scale of new physics, $f$, is taken to be about a TeV, and the Wilson coefficients, $c_{6,8}$, are expected to be of order one.  We take $c_6/f^2$ and $c_8/f^4$ as input parameters,. The remaining parameters, i.e.~the mass squared term and quartic coupling, are then fixed by requiring that the one-loop Higgs potential, 
 $V(h) = V_\text{tree}(h) + \Delta V_\text{CW}(h)$,
 has a minimum at $\langle h\rangle = v$, at which the physical Higgs mass is $m_h \simeq 125$ GeV. 

To study the EWPT in this model, we include the one-loop thermal corrections from the $W$ and $Z$ bosons, and the top quark (the latter is the only particle that is relevant for the zero temperature corrections). It has been shown \cite{Chala:2018ari} that, to a good approximation, key properties of the PT, such as $T_*$, $\phi_*$, $\alpha$ and $\beta$ depend mostly on an effective scale of new physics, $f/\sqrt{c}$, defined by
\begin{equation}
 \frac{c}{f^2} \equiv \frac{c_6}{f^2} + \frac{3}{2} v^2 \frac{c_8}{f^4}
\end{equation}
rather than on $f$, $c_6$ and $c_8$ separately. So effectively this model has only a single free parameter that governs the physics relevant for GW production.

The first study of the EWPT in the SMEFT goes back to the early 1990s~\cite{Zhang:1992fs}, where it was realized that the EWPT can be strongly first order if the scale of new physics is below a TeV, even for the observed Higgs mass. Later, the parameter space was mapped out in more detail~\cite{Grojean:2004xa,Delaunay:2007wb}. It was realized in particular that this scenario leads to a considerably enhanced Higgs self coupling~\cite{Grojean:2004xa} and electroweak baryogenesis was investigated~\cite{Bodeker:2004ws}.

A strong first-order EWPT in the SMEFT is made possible by the presence of the extra operators in the potential, which can result in a barrier separating the origin from the Higgs phase at zero temperature. For instance, the quartic coupling can be chosen to be negative, and the potential will be stabilized by the higher dimensional operators. There is an equivalent way of interpreting the mechanism of the PT. The higher dimensional operators break the link between the Higgs mass and the depth of the electroweak minimum~\cite{Harman:2015gif}. As the effective scale of new physics is lowered, the electroweak minimum becomes shallower.  The transition then happens at lower temperature with a larger jump in the Higgs field. This generally leads to stronger PTs, as is also observed in e.g.~the 2HDM (see Sec,~\ref{sec:doublet}) or the singlet extension (see Sec.~\ref{sec:singlet}). 
 
It is interesting to ask whether or not the effective theory correctly describes the impact of new physics on the PT in concrete models. This was done in~\cite{Damgaard:2015con} for the singlet extension, and it was found that the EFT correctly reproduces the PTs of the full model only in a small part of the parameter space. In~\cite{deVries:2017ncy} it was found that the effective theory also has limited validity when computing the baryon asymmetry. Finally, it has been shown in \cite{Chala:2018ari} that only in special perturbative UV completions, such as a custodial electroweak quarduplet, it is possible to generate the necessary operators in the Higgs potential without violating phenomenological constraints, such as EW precision tests. Despite these caveats, the effective theory is still a useful toy model to study properties of the EWPT in a simple manner.

The SMEFT is amongst the few models where the bubble wall velocity has been computed. This was first done in~\cite{Huber:2013kj}, using results from the SM. Refined calculations were presented in \cite{Konstandin:2014zta,Dorsch:2018pat}. The calculations show that, as the transition gets stronger, the velocity turns from sub- to supersonic, and eventually satisfies the ``old'' runaway criterion \cite{Bodeker:2009qy} before reaching the metastability region.

GW production at the EWPT in the effective theory has been studied initially in \cite{Huber:2007vva,Delaunay:2007wb}. It was recently revisited in \cite{Huang:2016odd} and \cite{Chala:2018ari}, and most of the results presented below are taken from the latter reference. All of these studies show that the EWPT in this setup can generate GWs that can be probed by LISA.

\begin{figure}[!t]
\begin{center}
 \includegraphics[width=0.36\columnwidth]{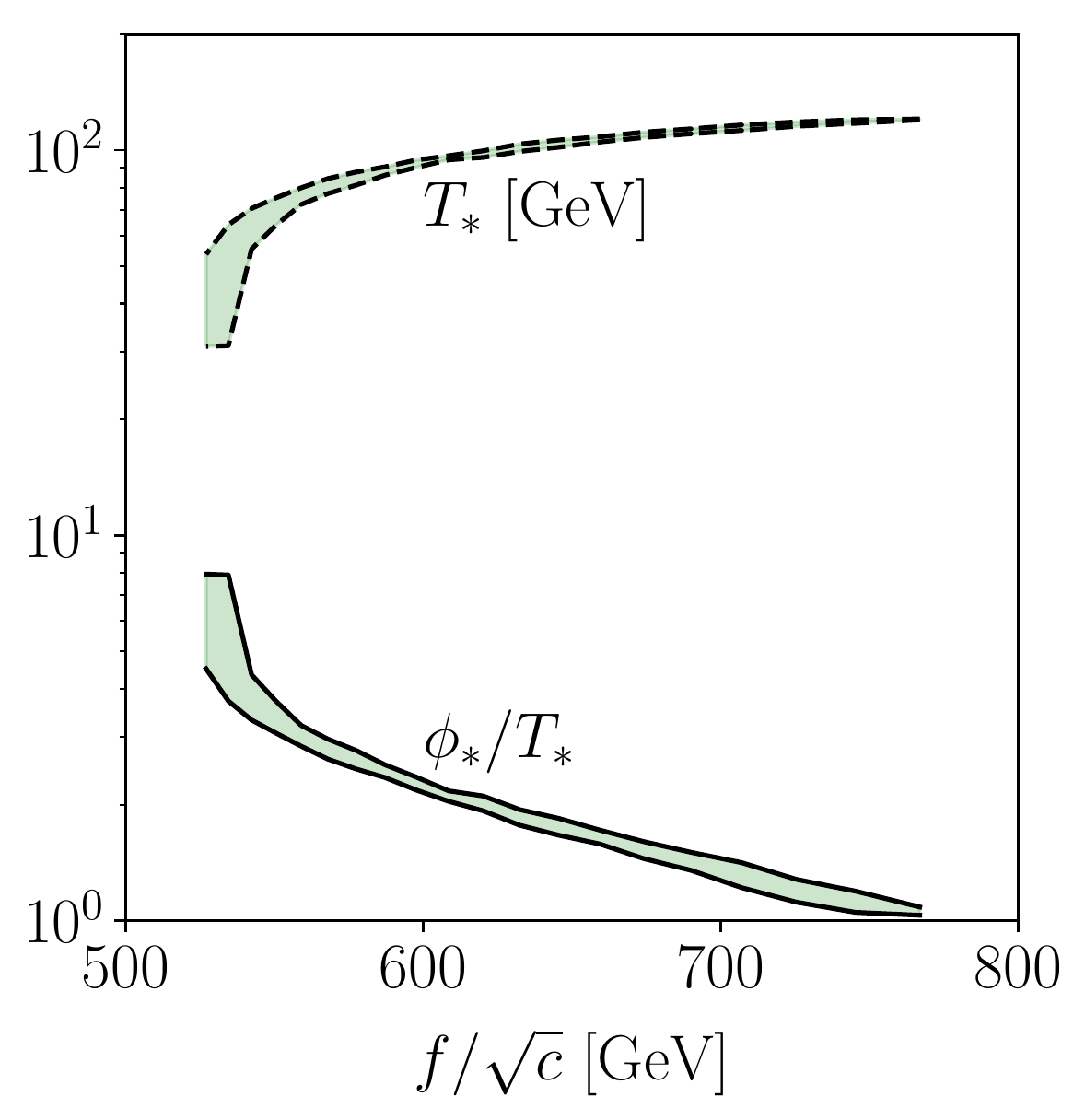}
 \includegraphics[width=0.55\textwidth]{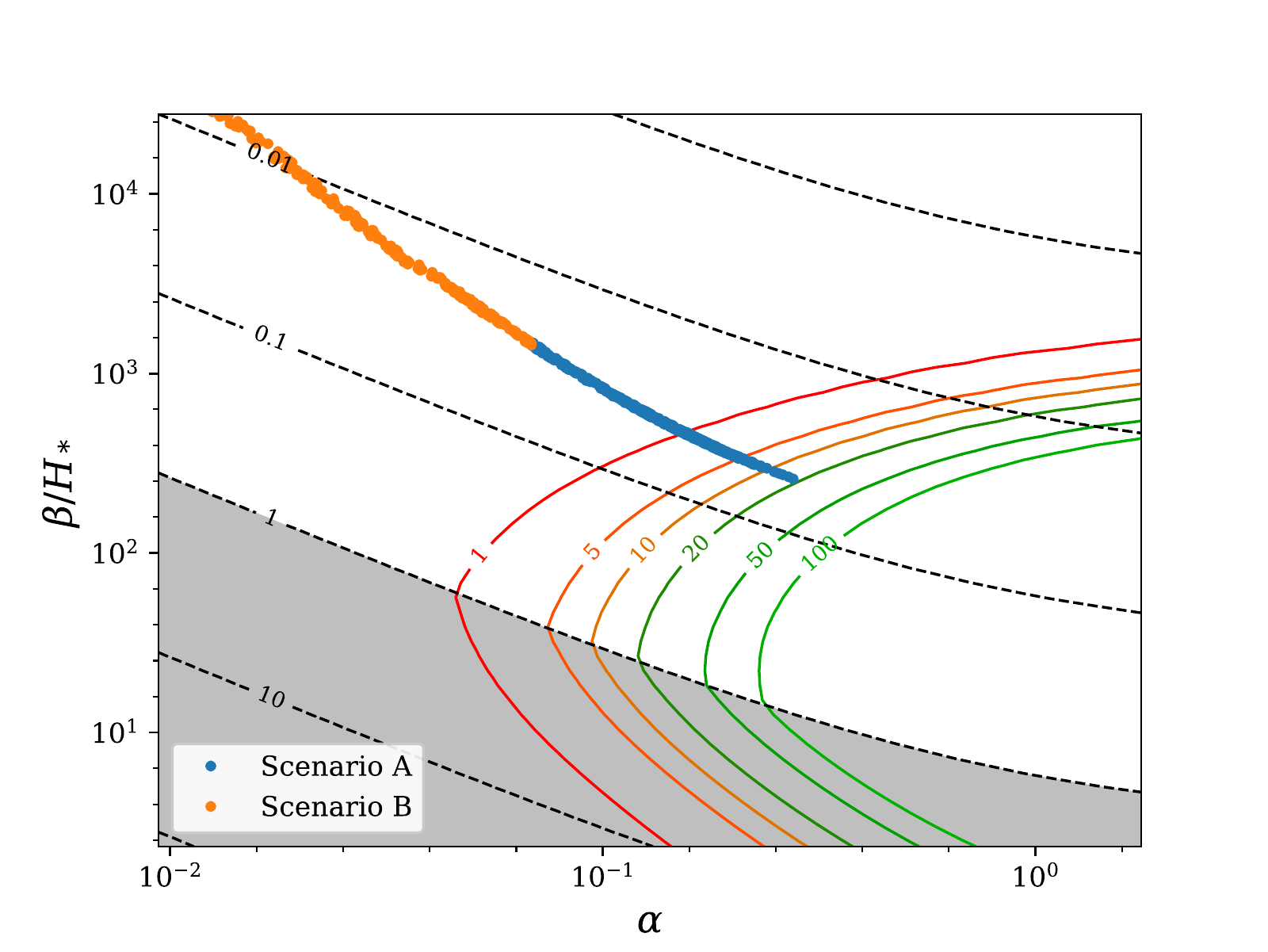}
 \caption{\small LEFT: Values of $T_*$ and $\phi_*/T_*$ as a function of the effective 
scale $f/\sqrt{c}$ (note that the parameter $f$ appearing here should not be confused with the GW frequency). RIGHT: Scatter plot on the plane of 
$\alpha$ and $\beta/H$. The blue (orange) dots, labelled ``Scenario A'' (``Scenario B''), correspond to scenarios with $T_* \sim 50$~GeV ($T_* \sim 100$~GeV). The wall velocity is taken to be $\vw=0.95$.}\label{fig:dim6results}
\end{center}
\end{figure}

On the left hand side of Fig.~\ref{fig:dim6results} we show the PT temperature $T_*$, defined by $S_3/T_*\sim 100$; as well as the value of $\phi_*/T_*$ as a function of the effective scale. In the right panel of the same figure we show a scatter plot of $\alpha$ and $\beta$ generated by \texttt{PTPlot}, along with the corresponding LISA SNR curves. Since the PT temperature changes with $f/\sqrt{c}$, we plot sensitivities for two groups of points: ``Scenario A'', for which $T_*$ is closer to $\sim 50$ GeV, and ``Scenario B'', for which $T_*=100$ GeV provides a more accurate description. 

As the effective scale of new physics, $f/\sqrt{c}$,  is lowered, the PT gets stronger, $\beta$ becomes smaller and the wall velocity increases. At $f/\sqrt{c}\sim540$ GeV, the PT no longer completes, i.e.~the symmetric phase becomes metastable. At even lower values $f/\sqrt{c}<480$ GeV, the electroweak minimum ceases to be the global one. For $f/\sqrt{c}\lesssim 650$ GeV the bubble expand as detonations. Larger values lead to deflagrations. At $f/\sqrt{c}\lesssim580$ GeV the bubble expansion satisfies the ``old'' runaway criterion of~\cite{Bodeker:2009qy}.  The relatively small value of $f/\sqrt{c}$ justifies the inclusion of operators of dimension larger than six.

The effective interactions also modify the Higgs trilinear coupling~\cite{Chala:2018ari},
\begin{equation}
\lambda_3 \sim \lambda_{3,\text{SM}}\bigg[1+\frac{v^2}{m_h^2}\bigg(2c_6 \frac{v^2}{f^2} + 4c_8 \frac{v^4}{f^4}\bigg)\bigg].
\end{equation}
Such modifications are potentially testable in double Higgs production at colliders. However, the still very small cross section makes all values compatible with a strong first-order EWPT impossible to probe at the LHC. A sizable region of this parameter space could be instead tested at a future Higgs factory, such as the FCC-ee~\cite{DiVita:2017vrr}. All values could be probed at a future 100 TeV hadron collider, such as the FCC-hh~\cite{Papaefstathiou:2015paa}. All in all, LISA will provide the first actual test of the scalar potential of the SMEFT.

\subsection{Models with warped extra dimensions}
\label{sec:warped}

The scenarios discussed thus far all rely on weakly-coupled new physics and polynomial scalar potentials which typically predict $\alpha$ significantly smaller than 1. However, there exist compelling and well-motivated frameworks that can generically give rise to large signals at LISA featuring strong dynamics and non-polynomial potentials which are the subject of the next two subsections. We start with models involving warped extra dimensions. 

Five-dimensional warped models have attracted much attention in the literature as they provide a natural and well-motivated framework for a very strong first-order PT~\cite{Creminelli:2001th,Randall:2006py,Nardini:2007me,Hassanain:2007js,Konstandin:2010cd,Konstandin:2011dr,Bunk:2017fic,Dillon:2017ctw, vonHarling:2017yew, Megias:2018sxv, Fujikura:2019oyi}. In the effective 4D language, this PT involves the radion scalar field whose potential stabilizes the interbrane distance determining the size of the slice of Anti-de Sitter space in Randall--Sundrum models.  The VEV of the radion field describes the position of the IR (i.e.~TeV) brane emerging from infinity. 
This class of models has been among the most popular solutions to the hierarchy problem. As the Higgs emerges together with the IR brane, the EW scale naturally emerges as the Planck scale suppressed by a geometric (warped) factor.  
The masses of the Kaluza--Klein resonances are around the TeV for  solving the hierarchy problem. As a result, $T_*$ is thus the TeV scale suppressed by some additional parametric factors. For $T_*$ below the EW scale, the radion PT triggers EW symmetry breaking, with intriguing implications for EW baryogenesis and LISA.  

The study of these scenarios is connected to Composite Higgs (CH) models discussed in the next section. Actually, if the conjectured AdS/CFT duality could be extended to any warped scenario, the differentiation between Randall--Sundrum and CH models would be artificial. The radion PT could be studied in the 4D language and determined as the 
deconfinement/confinement PT driven by the dilaton, the Goldstone boson of the conformal invariance of the 4D theory. 
As such, the radion is lighter than the other Kaluza--Klein resonances and it is a justified approximation  to neglect the effects of the latter in the zero-temperature 4D effective scalar potential.

The scalar potential of the radion exhibits two generic key features: it is very shallow due to near-conformal invariance, and the sizeable thermal barrier due to the large number of CFT degrees of freedom becoming massive during the  confinement PT leads to significant supercooling  and may even prevent the PT. These features are tightly connected, as observed in~\cite{Creminelli:2001th, Randall:2006py,Nardini:2007me,Hassanain:2007js,Konstandin:2010cd,Konstandin:2011dr,Bunk:2017fic,Dillon:2017ctw,Megias:2018sxv}: both can be relaxed by modulating the backreaction on the AdS metric induced by the conformality-breaking terms localized on the TeV brane. As a consequence, the PT parameters can take on a wide range of values. In particular, unless inducing extremely small or large backreactions on the metric, the hierarchy $T_{\rm n}\ll \Tc$ arises, with $\Tc$ near the TeV scale and $T_{\rm n}$ some orders of magnitude below $\Tc$.

One may wonder whether the TeV-brane terms breaking conformality are necessary to guarantee the occurrence of the PT.  
As recently noticed in~\cite{vonHarling:2017yew}, such terms are not required. 
In all Randall-Sundrum models where the SM gauge fields, in particular the $SU(3)$ gauge bosons, live in the bulk, the running of the QCD coupling is a function of the 5D radius. The QCD scale is therefore a function of the VEV of the radion. These effects, which had been ignored in the previous literature,  necessarily influence the radion potential and contribute to remove most of the thermal barrier at temperatures close to the QCD scale. The net result is that if the Universe is trapped in the conformal minimum, at some point the temperature of the (expanding) Universe reaches that of the QCD PT, the above effect becomes sizeable and the radion is released from the false minimum. It follows that the parameter region leading to a strong radion PT is far larger than that obtained by overlooking the effects of the radion on the QCD scale. In particular, the previously overlooked region of parameter space naturally exhibits a radion PT that is extremely supercooled ($T_{\rm n}\sim \Lambda_{QCD}\ll T_c$) and extraordinary strong~\cite{vonHarling:2017yew}. In this framework, the EWPT is naturally induced by QCD confinement and provides a compelling target for LISA.

\begin{figure}[t]
\begin{center}
 \includegraphics[width=0.56\textwidth]{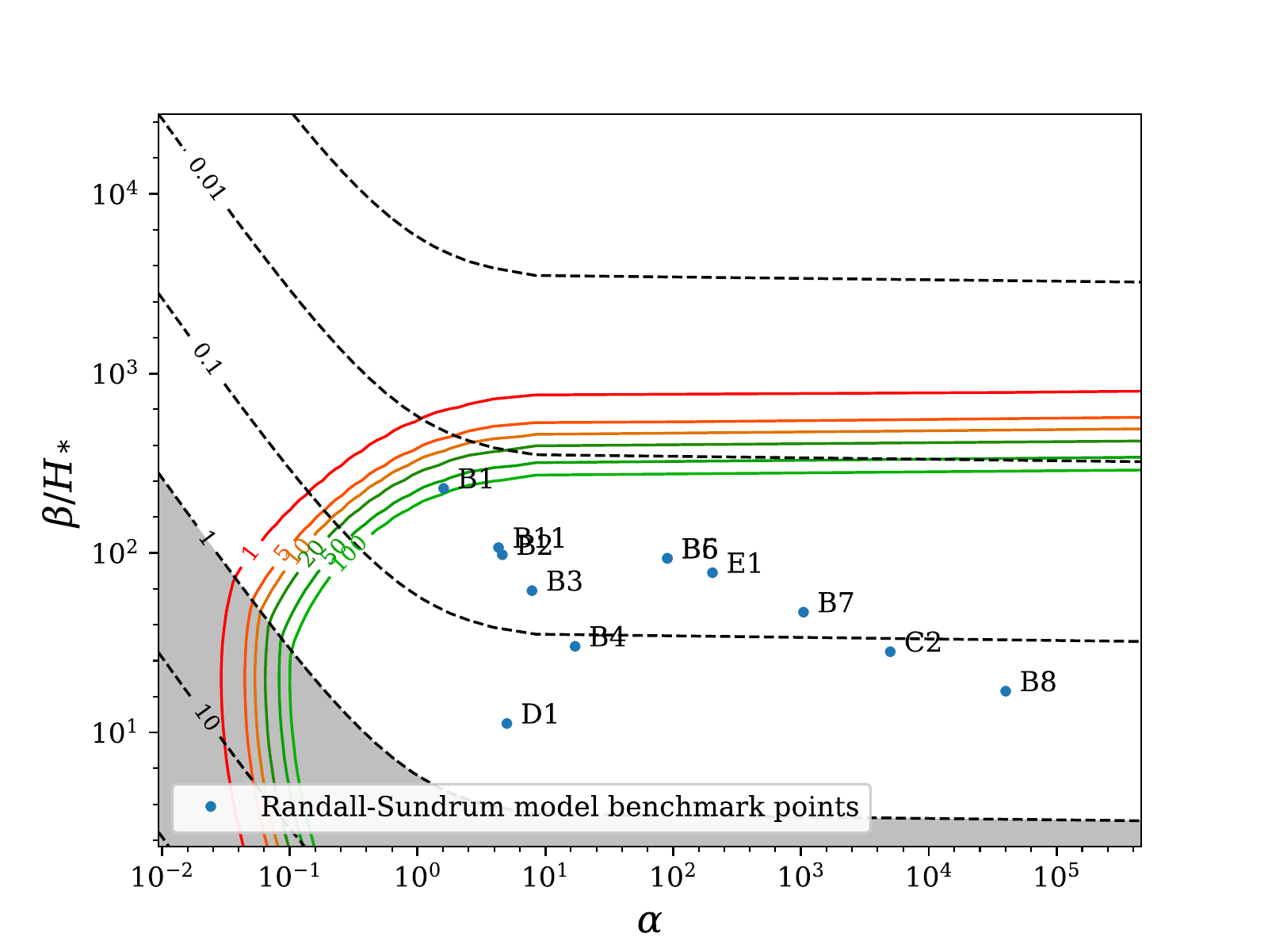}
 \caption{\small Detection prospects for some illustrative benchmark points that arise in Randall-Sundrum models. The temperature $T_*$ is rounded to 500\,GeV and $g_*$ matched the SM value. We set $\vw=0.95$. Note that  the numerical simulations of sound wave production have never been tested for the  huge values of $\alpha$ which occur in CH models. Here we present the result of a naive extrapolation of simulation results to huge $\alpha$ values,  but it should be kept in mind that there are large uncertainties for the modelling of the GW power spectrum and  that the formalism used to estimate the GW spectrum may no longer apply in that supercooled regime. }\label{fig:RSsnr}
\end{center}
\end{figure}

The preceding discussion suggests whether or not terms breaking conformality on the TeV brane are present (i.e.~whether $T_{\rm n}\sim \Lambda_{QCD}$ or $T_{\rm n}\gg \Lambda_{QCD}$) has an important impact on the corresponding phenomenological predictions. In particular, the temperature at which the PT occurs, be it near the QCD-, EW-, or TeV-scale, can have significant implications for e.g.~baryogenesis or entropy conservation in the early Universe~\footnote{For instance, if the EWPT occurs at QCD temperatures, cold EW baryogenesis can be realized with the CP violation sourced by the QCD axion~\cite{Servant:2014bla}. A QCD-scale EWPT also impacts the predictions for the QCD axion relic abundance ~\cite{Servant:2014bla,Baratella:2018pxi}.}. However, different nucleation temperatures do not lead to sizable differences in the GW predictions relevant for LISA,  since the stochastic GW background signal is sensitive to $T_*$ and not to $T_{\rm n}$. Warped models, with or without the conformality-breaking TeV-brane terms, all lead to a powerful GW signal with a similar spectrum and peak frequency. We can therefore take the benchmark points 
quoted in Table 1 of~\cite{Megias:2018sxv} and consider them as representative of a wide range of warped setups. Their detection prospects are displayed in Fig.~\ref{fig:RSsnr}, where the GW predictions derived from the simulations at $\alpha\lesssim 1$ (see Sec.~\ref{sec:prediction}) have been extrapolated well beyond their regime of validity. However, the qualitative outcome of the figure is likely independent of the SNR uncertainties, and all benchmark points are expected to lie within the LISA sensitivity region~\footnote{For LISA, the result is even more promising than the one obtained in~\cite{Megias:2018sxv, Figueroa:2018xtu} due to the updated LISA sensitivity curve and stochastic GW background predictions that we consider here.}.

\subsection{Composite Higgs models and confining phase transitions} 
\label{sec:composite}

Another class of models giving rise to signals at LISA with non-polynomial potentials are CH scenarios. CH models provide an 
alternative to SUSY for solving the hierarchy problem and are prime targets for LHC searches (see \cite{Panico:2015jxa} for a review). In this context,
the Higgs is a pseudo-Nambu Goldstone Boson (pNGB) arising from the global symmetry breaking pattern $\mathcal{G}\rightarrow\mathcal{H}$ in a new strongly coupled sector, characterised by the coupling $\widetilde{g}$ and the confinement scale $f$, whose currently preferred value is $f\simeq0.8$ TeV~\cite{Grojean:2013qca}. 
To ensure a custodial symmetry that suppresses oblique corrections to EW precision tests, the minimal possible coset is $SO(5)/SO(4)$. Larger groups are also viable, such as $SO(6) / SO(5)$ or $SO(7) / SO(6)$, in which case the Higgs is accompanied by other light pNGBs, in particular SM singlets. Light scalars in addition to the Higgs are therefore natural in these constructions.

The Higgs potential is generated from the explicit breaking of the global symmetry $G$ by an elementary sector and communicated to the composite sector from which the Higgs originates through elementary/composite interactions. The scalar potential arises at one-loop, mainly from fermionic loops.
Due to the pNGB nature of the Higgs, the potential
is a trigonometric function of the ratio $h/f$, with the generic form for the minimal CH scenario given by~\cite{Panico:2015jxa}
\begin{equation}
\label{eq:TunedHiggsPotential}
V [h] \,\sim \, f^4 \left[ \alpha  \sin^2 (h/f) + \beta \sin^4 (h/f) \right] \,.
\end{equation}
Imposing the correct EW symmetry breaking scale $h=v$ and Higgs mass leads to
\begin{equation}
\label{eq:TunedHiggsRelations}
\alpha \, = \, -2 \beta  \sin^2 (v/f) \ , \quad \, \, m_h^2 \, \approx \, 8  f^2 \sin^2 (v/f) \, \beta \, ,
\end{equation}
meaning that $|\alpha/\beta| =  2 \sin^2 (v/f)$. This quantity needs to be small since  EW precision tests and Higgs coupling measurements constrain $\sin^2 (v/f) \lesssim 0.1-0.2$ \cite{Grojean:2013qca}. Therefore, contrary to the generic expectation $\alpha \gtrsim \beta$~\cite{Panico:2012uw},  $\alpha$ has to be suppressed with respect to $\beta$, which requires an accidental cancellation. This is the irreducible tuning of CH models \cite{Csaki:2017eio}.

\begin{figure}[!t]
\begin{center}
\vspace{-0.3cm}
\includegraphics[width=225pt]{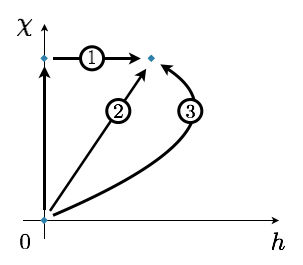}
\end{center}
\caption{\small Sketch of the different possible PT trajectories in CH models. The field $\chi$ sets the magnitude of the strong sector condensate (corresponding to $f$ today) and $h$ is the Higgs VEV. The blue points correspond to (meta)stable vacua of the theory, and the black lines show possible PT trajectories. Path (1) would correspond to the case where the reheat temperature after the new strongly-coupled sector confines is above the EWPT temperature. The EWPT then happens as in the standard case, after and independently of the strongly-coupled sector. In the other cases (2) and (3), the reheat temperature is low enough so that EW symmetry is broken at the same time as the strong-sector. While some early papers have implicitly assumed a path of the type (2) in Randall--Sundrum models, the calculation of the actual trajectory has been done only recently 
in ~\cite{Bruggisser:2018mus,Bruggisser:2018mrt} and follows (3). 
Figure taken from \cite{Bruggisser:2018mrt}.}
\label{fig:phtr}
\end{figure}

The EWPT in CH models has primarily been studied by  following an EFT approach \cite{Grojean:2004xa,Delaunay:2007wb,Grinstein:2008qi}, by expanding (\ref{eq:TunedHiggsPotential}) in $h/f$ up to dimension-six operators. In this case, it is found that a first-order EWPT sufficiently strong for  EW baryogenesis can occur provided than $f$ is below $\sim$ 1 TeV, as discussed in Sec.~\ref{sec:eft} of this report. 
The accuracy of an EFT approach to the study of the EWPT in such scenarios is therefore limited.  
More recently, the EWPT in some specific CH realisations has been investigated.  
In CH models, since the Higgs arises only when a non-zero condensate $\chi$ forms, the confinement PT and the EWPT are naturally linked. 
Nevertheless, until recently, past studies considered them separately, focusing either on the confinement PT, described by the dynamics of a dilaton/radion and relying on a 5D description~\cite{Creminelli:2001th,Randall:2006py,Nardini:2007me,Hassanain:2007js,Konstandin:2010cd,Konstandin:2011dr,Bunk:2017fic,Dillon:2017ctw,vonHarling:2017yew,Megias:2018sxv} as discussed in the previous subsection, or assumed that the EWPT takes place after the confinement of the strongly-coupled sector~\cite{Espinosa:2011eu,Chala:2016ykx,Delaunay:2007wb,Grinstein:2008qi}. The confinement PT is generically very strong, and is known to produce very large GW signals~\cite{Randall:2006py,Konstandin:2010cd,Megias:2018sxv}.
To study the EWPT in CH, there are therefore two approaches:
\begin{itemize} 
\item Assume that the confinement PT happened well before the EWPT and study the two  PTs independently.  In this case, EW baryogenesis and GW production can occur by invoking a non-minimal CH setup featuring an extra singlet~\cite{Espinosa:2011eu,Chala:2016ykx}. This case is schematically displayed as (1) in Fig.~\ref{fig:phtr}.

\item Consider the light dilaton/radion window, in which case the confinement and EW PTs can occur simultaneously. In this case, the EWPT  is automatically strongly first-order, and this scenario also features CP-violation from varying Yukawa couplings during the EWPT, allowing for viable EW baryogenesis even in minimal CH scenarios~\cite{Bruggisser:2018mus,Bruggisser:2018mrt}. This case is schematically represented by paths (2) and (3) in Fig.~\ref{fig:phtr}.

\end{itemize}
We briefly discuss these two cases below in the context of LISA.
\subsubsection{A strong first-order EWPT from an extra singlet in non-minimal cosets}

In both the $SO(6) / SO(5)$~\cite{Espinosa:2011eu, Bian:2019kmg} and  $SO(7) / 
SO(6)$~\cite{Chala:2016ykx} CH models, the EWPT proceeds mainly in two steps, 
because the new singlet $S$ is mostly pseudoscalar and consequently 
$S\rightarrow -S$ is an approximate accidental symmetry of the potential. The 
dynamics of the EWPT is therefore well captured 
by the simplified model of Sec.~\ref{sec:singlet}.
The main differences with respect to the generic extra singlet case are: 
\begin{enumerate}
 \item[(i)] In CH scenarios, not all values of $a_2$ and $b_4$ are equally 
likely. Considering explicit models,~\cite{Chala:2016ykx,Chala:2018qdf} showed 
that the most common values are in the range $\sim 0.1$--$0.5$. This result can 
be roughly understood from a term in the scalar potential $
  [ \widetilde{g}^2  (N_c y_t) \delta /(4\pi)^2] {S^2 |H|^2} $. Indeed, 
  $a_2\sim 3 \widetilde{g}^2/(4\pi)^2\delta\gtrsim 0.2$ for $\widetilde{g}\gtrsim 3$; the same reasoning applies
for $b_4$. Therefore, according to the results discussed in 
Sec.~\ref{sec:singlet}, only small $S$ masses are expected to trigger a 
strong first-order EWPT.
 \item [(ii)]Contrary to the elementary singlet model, in CH scenarios the presence of 
heavier fermionic resonances of mass $m_*\sim \widetilde{g} f$ is unavoidable. This opens 
a complementary probe of these models in addition to GWs. 
Current direct searches exclude $m_* < 1$--$1.4$ TeV, 
depending on the final state~\cite{Aaboud:2018pii}. Moreover, estimations of 
$\widetilde{g}$ and $f$ based on actual observables rather than on fine-tuning arguments 
(e.g.~in CH models with DM) hint to $m_*\gtrsim 
2$ TeV. Such values are likely out of the LHC reach, but accessible at future 
hadron colliders~\cite{Chala:2018qdf}.
\end{enumerate}
The interactions between the new singlet 
and the SM fermions induce new sources of CP violation for EW 
baryogenesis which are constrained by EDMs (generated 
at two loops) and, to a lesser extent, from LEP data. They are phenomenologically viable
 provided that the new singlet is close enough in mass to the Higgs 
boson~\cite{Espinosa:2011eu}. In summary, however,
the prospects for exploring this  class of CH models at LISA are essentially those of the singlet model, discussed in Sec.~\ref{sec:singlet}.
  
\subsubsection{A strong first-order EWPT from Higgs-dilaton interlinked dynamics}

\begin{figure}[t]
\begin{center}
\vspace{-0.3cm}
\includegraphics[width=225pt]{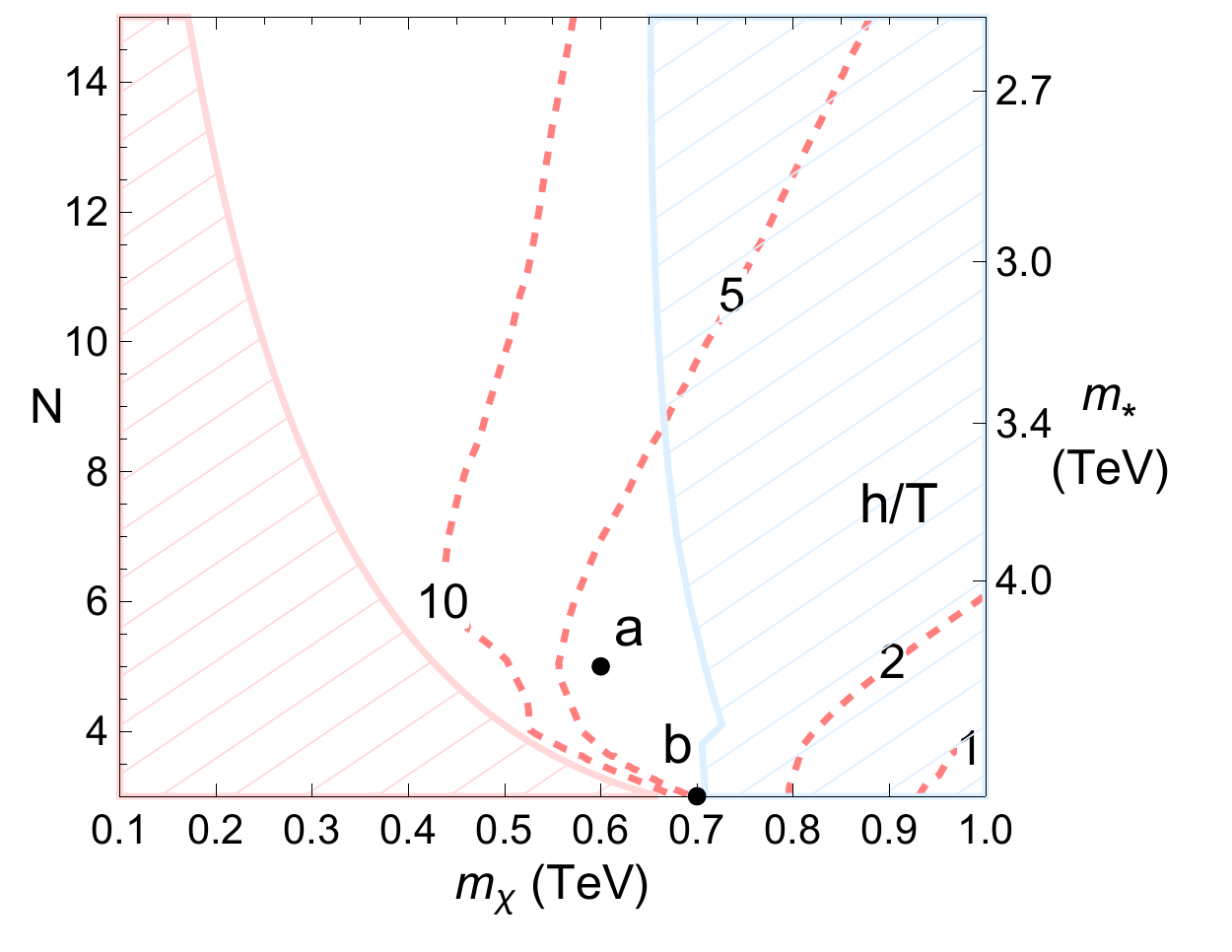}
\caption{\small PT strength ($h/T$) contours (dashed lines) in CH models for which the EWPT and the confinement PT occur simultaneously, for a meson-like dilaton. Figure taken from \cite{Bruggisser:2018mus}. In the red dashed region, no viable EW minimum can be found or the Higgs-dilaton mixing leads excessively large deviations in the Higgs couplings. In the blue region, the baryon asymmetry is washed out after reheating. The baryon asymmetry for benchmark point a (b) is $| \eta_B| \times 10^{10}\sim$ 5--5.5 (4--4.5).}
\label{fig:hoverT}
\end{center}
\end{figure}

CH scenarios do not need to feature an additional singlet to give rise to signals at LISA.
Instead, one can assume that the strongly-coupled sector is nearly conformal above the TeV scale as motivated by flavour physics~\cite{Contino:2010rs}. Confinement is then associated with the spontaneous breaking of conformal invariance. This gives rise to a pNGB, the dilaton, which we denote as $\chi$. When the explicit breaking of conformal invariance is sufficiently small, the dilaton can be significantly lighter than the confinement scale, and this justifies studying the dynamics of $\chi$ and $h$ together. This can be done by using a large $N$ expansion of the underlying strongly-coupled gauge theory
where $N$ represents the number of colors. Below the scale of the heavy resonances $m_*\sim \widetilde{g} f$, the PTs can be studied as a function of two main parameters: $N$ and $m_{\chi}$.
As shown in Fig.~\ref{fig:hoverT}, $h/T$ can be significantly larger than 1 in a large region of parameter space as the strength of the EWPT is inherited from the supercooled dynamics of the dilaton.
Additionally, it can be shown that the Yukawa couplings vary across the bubble wall, providing sufficient CP-violation for EW baryogenesis~\cite{Bruggisser:2018mus,Bruggisser:2018mrt}.
 If the dilaton is light enough, the reheat temperature can still be below the EW scale, such that the baryon asymmetry is not washed out.  

\begin{figure}[t]
\begin{center}
\vspace{0.3cm}
\includegraphics[width=149pt]{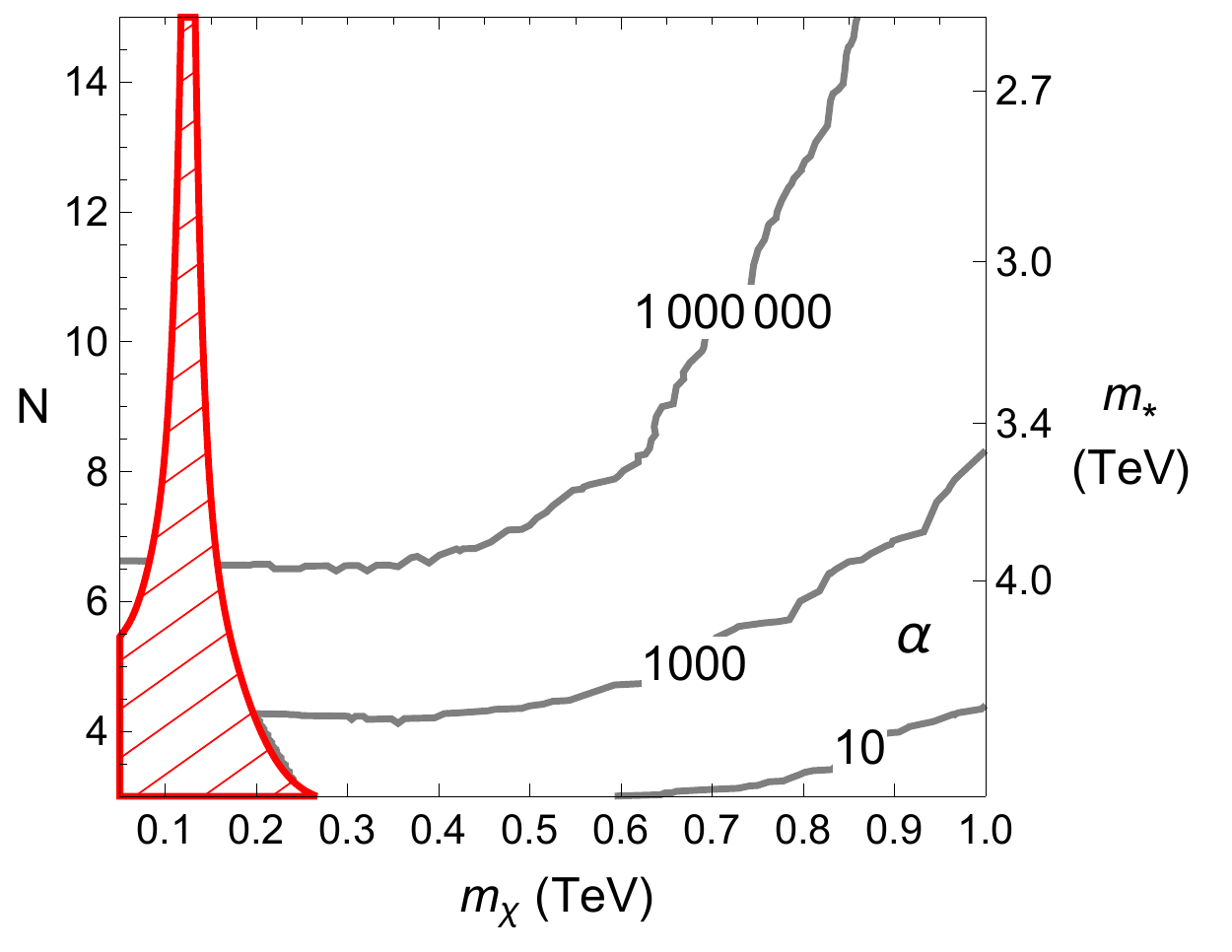}
\includegraphics[width=149pt]{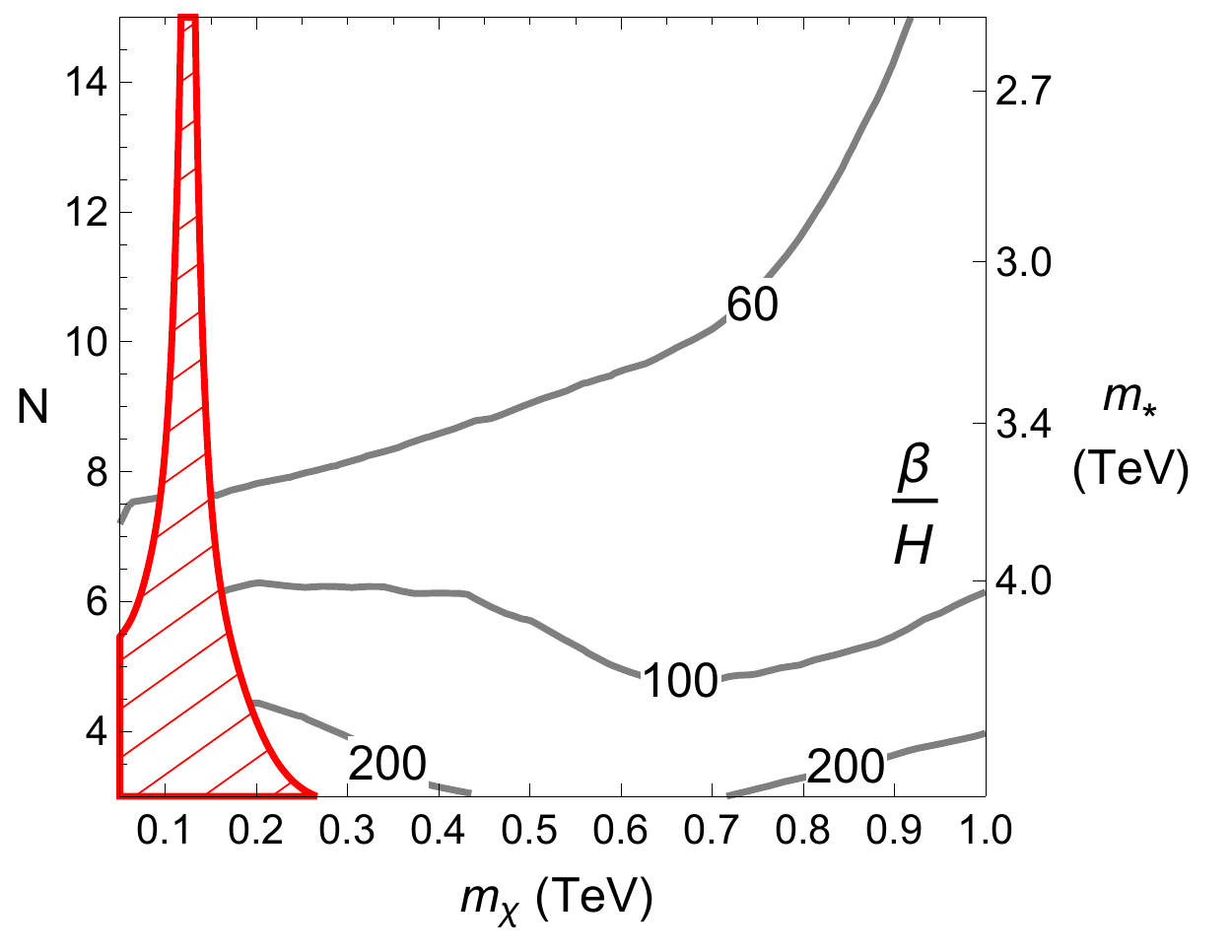}
\includegraphics[width=149pt]{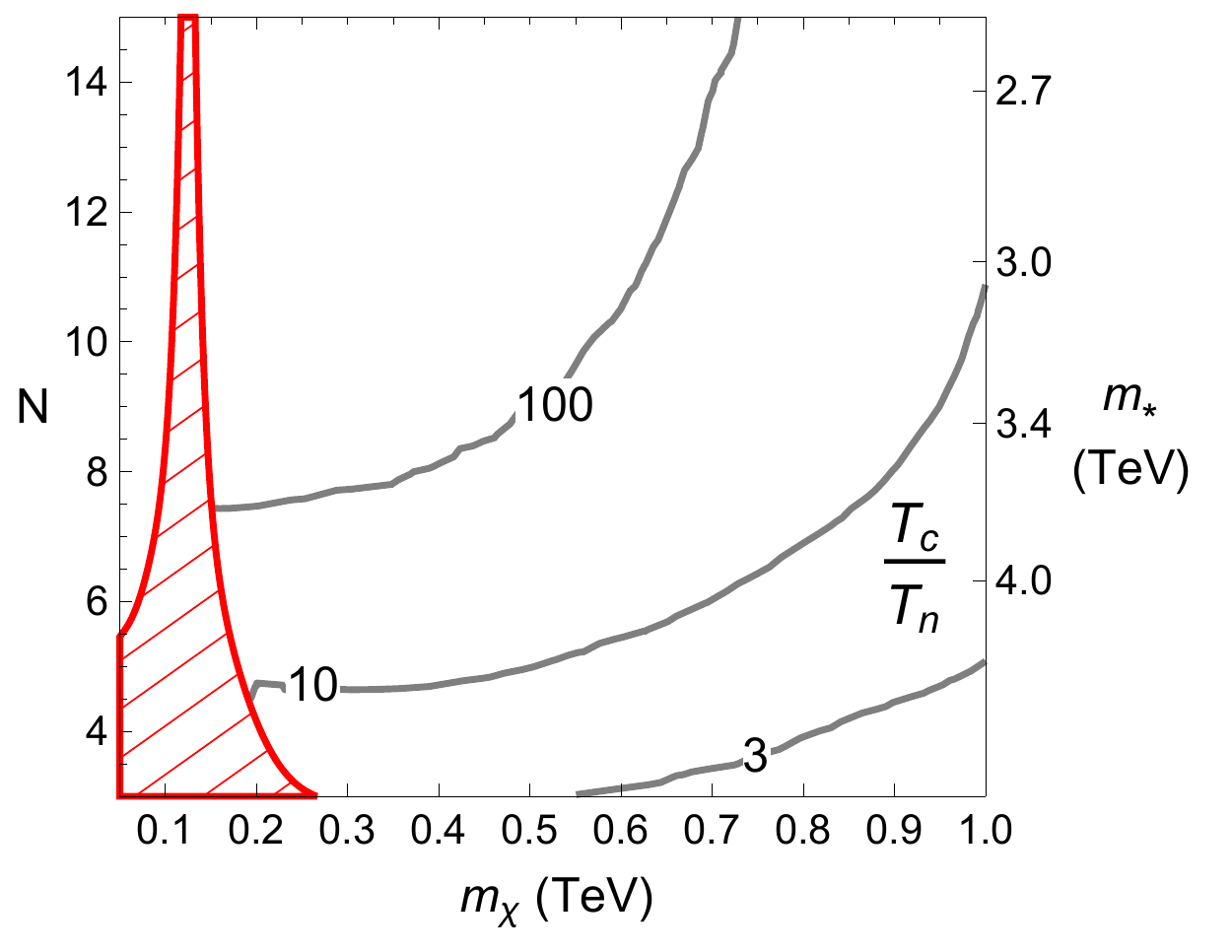}
\caption{ PT parameters relevant for LISA in CH models with Higgs-dilaton interlinked dynamics. LEFT: Contours of $\alpha$. CENTER: Contours of $\beta/H$. RIGHT: Ratio of the critical temperature to the nucleation temperature, for a glueball dilaton (figure taken from~\cite{Bruggisser:2018mrt}).
\label{fig:alphaandbeta}}
\end{center}
\end{figure}

\begin{figure}[t]
  \begin{center}
    \includegraphics[width=0.6\textwidth]{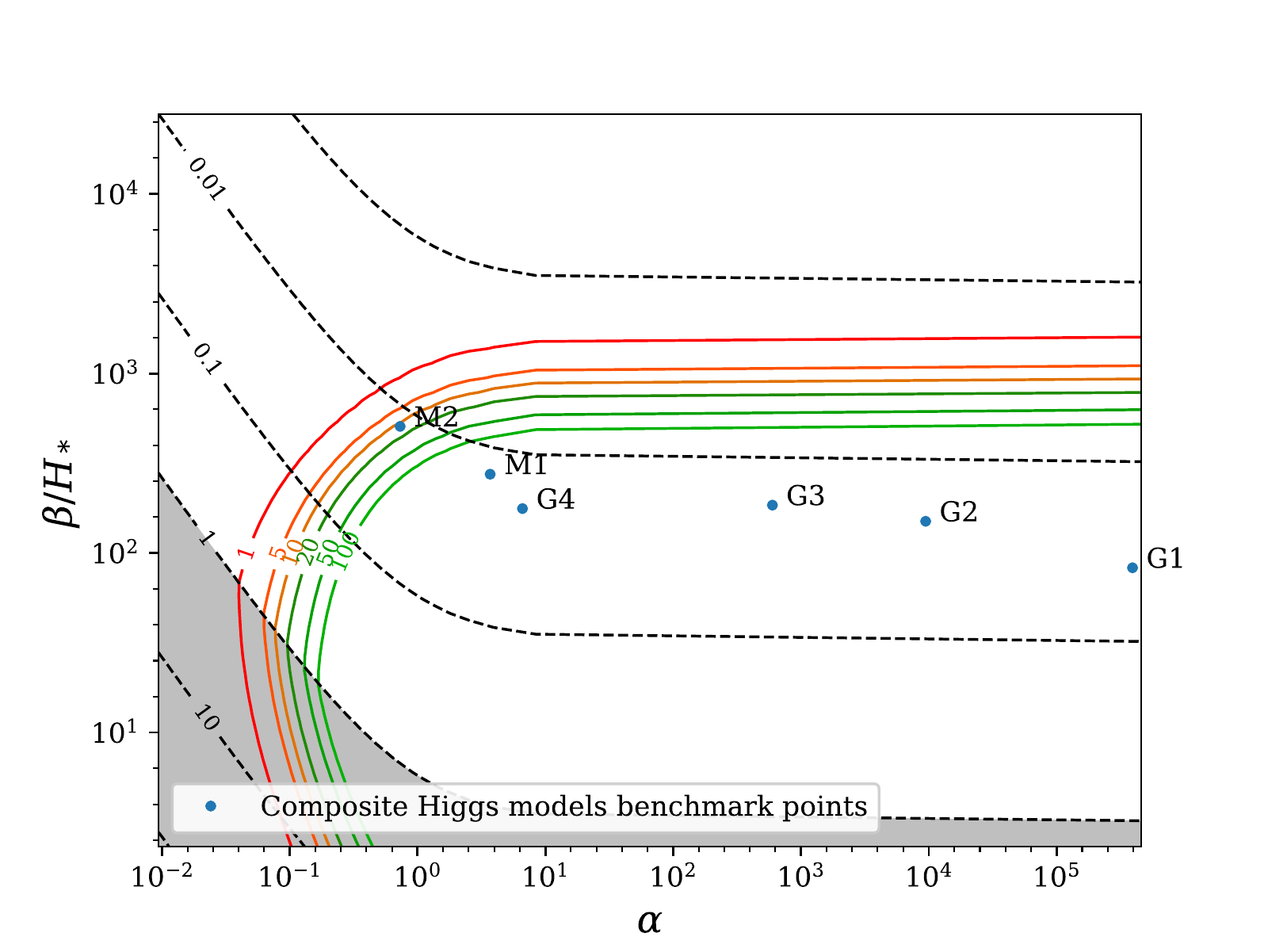}
    \caption{\small SNR for specific benchmark points of CH models. 
    The value of $g_*$ is as in the SM while $T_*\sim 150$ GeV. We set $\vw=0.95$. 
    Note that  the numerical simulations of sound wave production have never been tested for the  huge values of $\alpha$ which occur in CH models. Here we present the result of a naive extrapolation of simulation results to huge $\alpha$ values,  but it should be kept in mind that there are large uncertainties for the modelling of the GW power spectrum and  that 
    the formalism used to estimate the GW spectrum may no longer apply in that supercooled regime. 
     \label{fig:snr_composite}}
  \end{center}
\end{figure}

The key point that distinguishes these models from typical EWPT scenarios  is the fact that a large number of degrees of freedom become massive during the confinement PT. This produces a large thermal barrier in the one-loop effective potential when the tree-level potential is very shallow, making the PT naturally very supercooled~\cite{Konstandin:2011dr}. 
In most of the corresponding region of parameter space, the values of $\alpha$ are very large and $\beta/H$ can be quite small. This is illustrated in Fig.~\ref{fig:alphaandbeta} corresponding to the case of a glueball-dilaton, for which $\alpha$ can be as large as $10^6$.  Note that these results can be easily generalised to confining phase transitions not necessarily linked to the Higgs, 
leading to strong signals of GW at higher frequencies. We also stress  that the bubble wall velocity has never been computed in this setup nor in its 5D dual.
On the one hand, we expect it to be large due to the substantial supercooling. On the other hand, we also expect significant friction effects due to the large number of degrees of freedom acquiring a mass via strong coupling to the dilaton.
The typical magnitude of the bubble wall velocity therefore remains an open question in these scenarios.

In summary, CH models generically predict a strong PT at the TeV scale from confinement, leading to large GW signals at LISA (see Fig.~\ref{fig:snr_composite}).
In the light dilaton regime, the confinement PT and the EWPT happen simultaneously, a scenario which opens new opportunities for baryogenesis and is testable at the LHC by searching for the light dilaton and also in electron EDM experiments.
Non-minimal CH models predict light singlet scalars playing a role in baryogenesis with additional signatures at the LHC or EDM experiments.

This section ends our overview of PTs tied to the EW scale.
There is one qualitatively different scenario which we have not discussed, and concerns the non-trivial possibility that the EWPT occurs at a temperature significantly higher than the typical $\sim 160$ GeV via a high-temperature symmetry non-restoration effect: the Higgs, through the coupling to other scalar fields, gains a negative thermal mass squared and hence a VEV proportional to the temperature. This possibility has been discussed recently in \cite{Baldes:2018nel,Glioti:2018roy,Meade:2018saz}.  The GW peak frequency associated with the EWPT would thus be shifted to frequencies higher than usual. Specific UV completions and their implications for GW signals still remain to be investigated in this context.

\subsection{First order PTs in a dark sector} 
\label{sec:dark}

LISA can also probe sectors of new physics not tied to the EW scale. More than 95\% of the energy density in our Universe today is invisible, in the form of DM and dark energy. It is therefore quite plausible that a dark or hidden sector exists which today is fully decoupled from the SM, and which might explain either or both phenomena. Hidden sectors are also ubiquitous in UV complete theories of quantum gravity such  as string theory. Here we take a more pragmatic approach, and include in our definition of ``dark sectors'' all theories that are sufficiently weakly coupled to the SM such that they remain undetected experimentally and leave the EWPT largely unaffected. A cosmological PT in such a hidden sector can in principle occur at any temperature, independent of the weak scale, i.e.~the peak of the GW signal can be shifted almost arbitrarily (we will revisit this statement below). 

The possibility of detecting GWs from a PT in a dark sector was explored in~\cite{Schwaller:2015tja} in models of composite dark sectors, and further explored in a variety of other DM scenarios in~\cite{Jaeckel:2016jlh,Chala:2016ykx,Addazi:2016fbj,Baldes:2017rcu,Addazi:2017gpt,Tsumura:2017knk,Aoki:2017aws,Croon:2018new,Croon:2018erz,Baldes:2018emh,Madge:2018gfl,Breitbach:2018ddu,Fairbairn:2019xog}. While of course any hidden sector could be engineered to feature a cosmological PT, scenarios in which the PT serves a specific purpose can yield specific predictions for GW signals at LISA. Examples include models where the baryon asymmetry is generated by a PT in the dark sector~\cite{Baldes:2017rcu}, scenarios with spontaneous breaking of a gauge symmetry that guarantees DM stability~\cite{Baldes:2018emh,Madge:2018gfl}, scenarios where the PT is crucial for obtaining the correct relic abundance~\cite{Baker:2016xzo,Baker:2017zwx}, or models explaining flavour hierarchies~\cite{Greljo:2019xan}. 
Another interesting opportunity that arises in this context is the study of anisotropies in the stochastic GW background that could be detectable for the strongest PTs~\cite{Geller:2018mwu}. If the stochastic GW background remains the only observable signal from the dark sector, it would furthermore be important to understand which of its properties can be inferred from the amplitude and shape of the signal in the future, which was explored for example in~\cite{Croon:2018erz}.

 \begin{figure}[t]
\centering
 \includegraphics[width=0.6\textwidth]{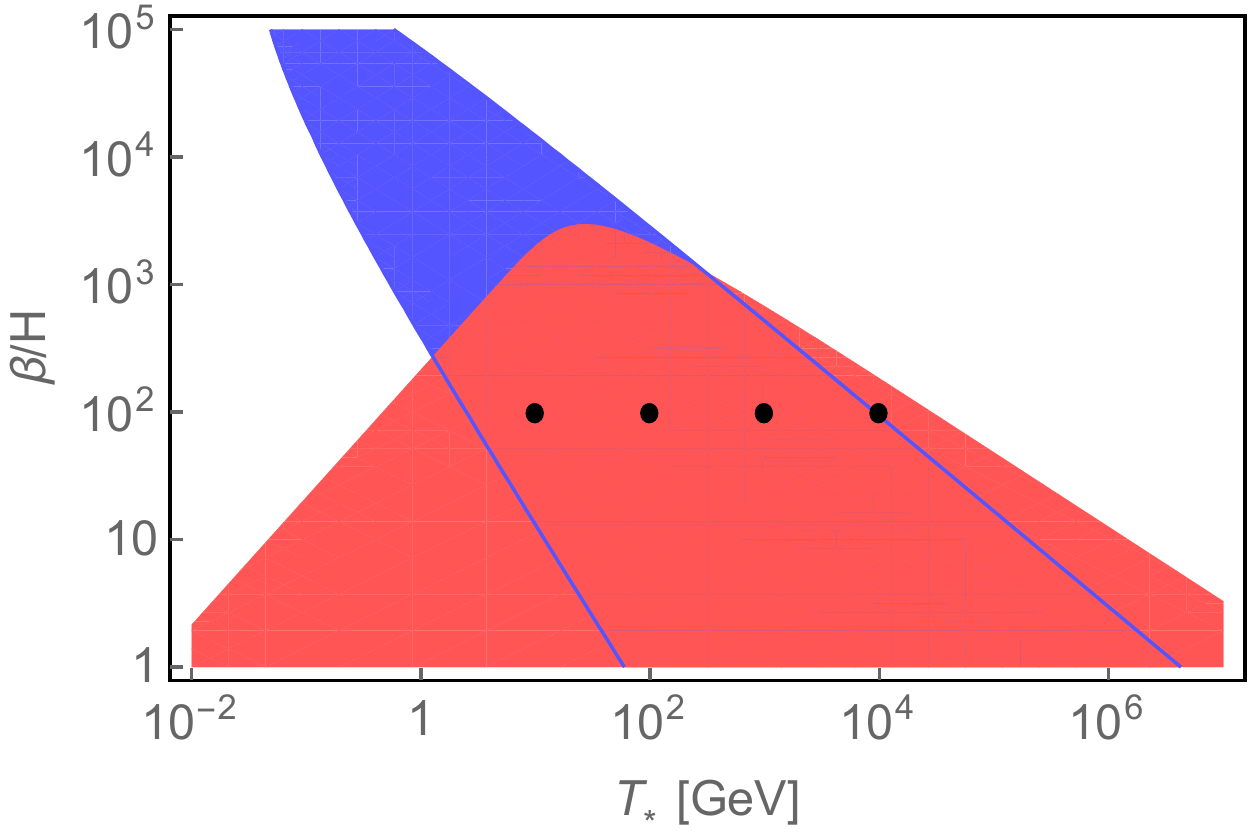}
 \caption{\small\label{fig:dark1}Sensitivity of LISA to PTs in a dark sector. Shown are regions with an SNR $\geq 10$ in the $T_*$--$\beta/H_*$ plane, assuming runaway vacuum bubbles (red, foreground) or sound wave contributions only (blue, background) with $\alpha = 0.1$, and $v_{\rm w}=1$ in both cases. Black dots show the location of the benchmark points. Note that in this figure slightly different LISA sensitivities were used.
 }
\end{figure}

While LISA is most sensitive in the frequency range corresponding to weak- to TeV-scale PTs, a sufficiently strong signal can be detected for a large range of PT temperatures. In Fig.~\ref{fig:dark1} we show the regions detectable by LISA with an SNR $\geq 10$. Very strong signals arising from PTs occurring at temperatures from below the GeV scale up to the PeV scale ($10^6$~GeV) are detectable, with the highest sensitivity is to PTs around the weak scale, as expected. 
Shown in Fig.~\ref{fig:dark1} are the predicted sensitivity considering only the envelope approximation, assuming that the PT exhibits runaway behaviour in the dark sector (red region) and the case of a very strong transition in a plasma ($\alpha = 0.1$, $v_{\rm w} =1$) with sound waves dominating the GW spectrum. Before discussing some concrete examples of models that fall in the detectable region, a few subtleties regarding PTs in the hidden sector should be noted. 

\subsubsection{Subtleties with dark sector PTs}

Fig.~\ref{fig:dark1} includes the implicit assumption that the temperature of the hidden sector at the time of the PT is equal to that of the SM photons. While this is a reasonable assumption for most BSM scenarios, the dark sector might instead be thermally decoupled from the SM at the time of the PT. The ratio of temperatures $r_T = T_{\rm dark}/T_{\rm \gamma}$, evaluated at $T_{\rm dark} = T_*$, could in principle take any value and is in general not conserved throughout the cosmic history. The effect of $r_T \neq 1$ on the amplitude and frequency of the GW signal is somewhat subtle and was discussed for the first time in~\cite{Breitbach:2018ddu} (see also \cite{Fairbairn:2019xog}). Here we summarise the main results. 

Obviously in the limit $r_T \to 0$ the energy density in the dark sector is negligible and thus no observable signal can be produced. In the limit of $r_T \leq 1$ and for hidden sectors a small number of new particles, one finds~\cite{Breitbach:2018ddu}
\begin{align}
	\alpha &= \alpha_0 r_T^4\,,\label{eqn:darksector1}\\
	\beta/H_* &= (\beta/H_*)_0 \,,\label{eqn:darksector2}
\end{align}
where $\alpha_0$ and $\beta_0$ are the values of $\alpha$ and $\beta$ for $r_T=1$. The energy budget $\alpha$ scales with the energy density in the hidden sector, as one would expect. Note that while both $\beta$ and $H_*$ depend on $r_T$, this dependence drops out of their ratio, i.e. the duration of the PT relative to the age of the Universe at the time of the transition remains unchanged. Overall this results in a strong suppression of the potential GW signal from a cold dark sector. These are only the leading effects: further effects, such as a typically small amount of additional redshifting, are discussed in~\cite{Breitbach:2018ddu} .

\begin{figure}
  \begin{center}
    \includegraphics[width=0.49\textwidth]{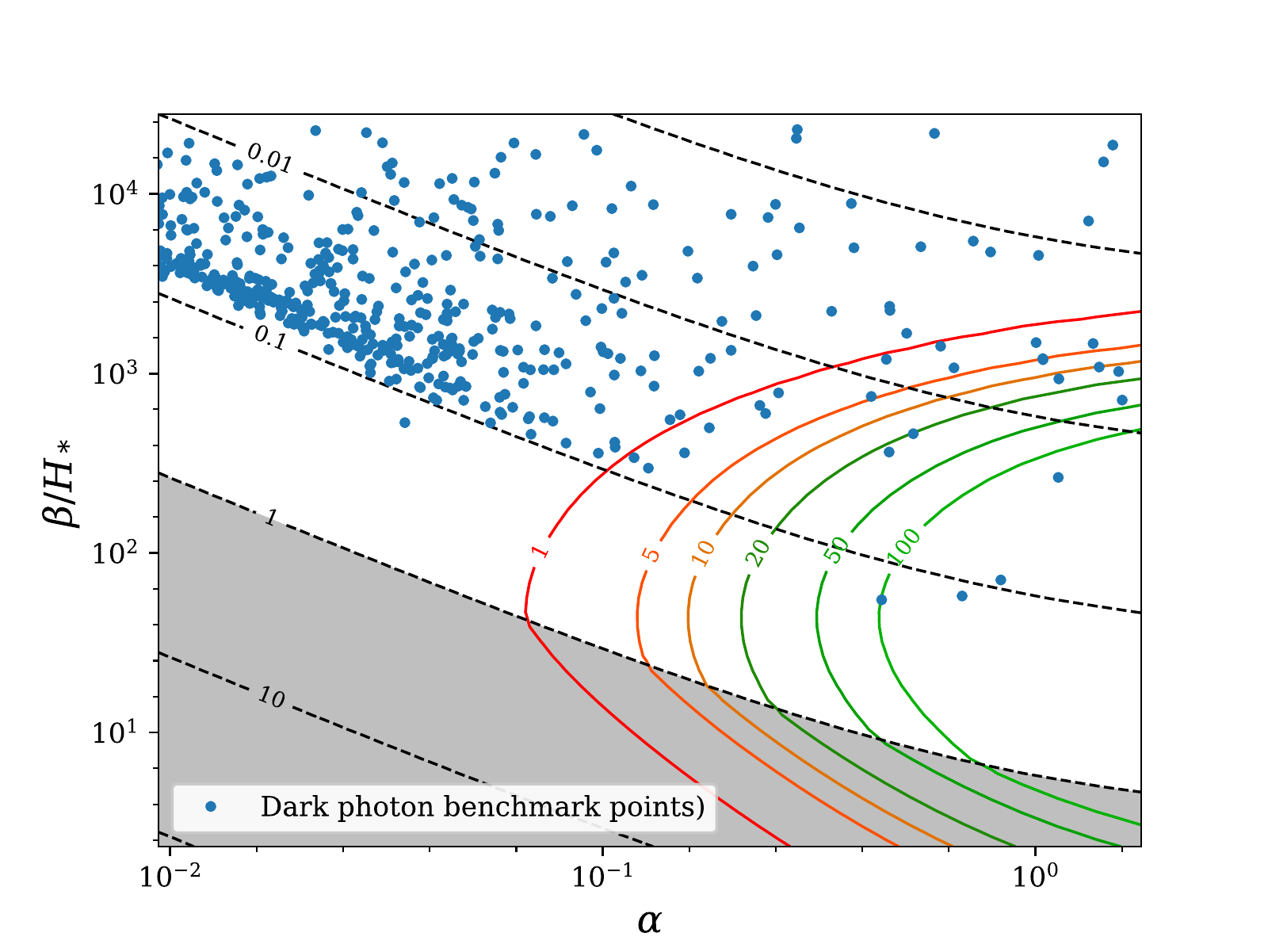}
    \includegraphics[width=0.49\textwidth]{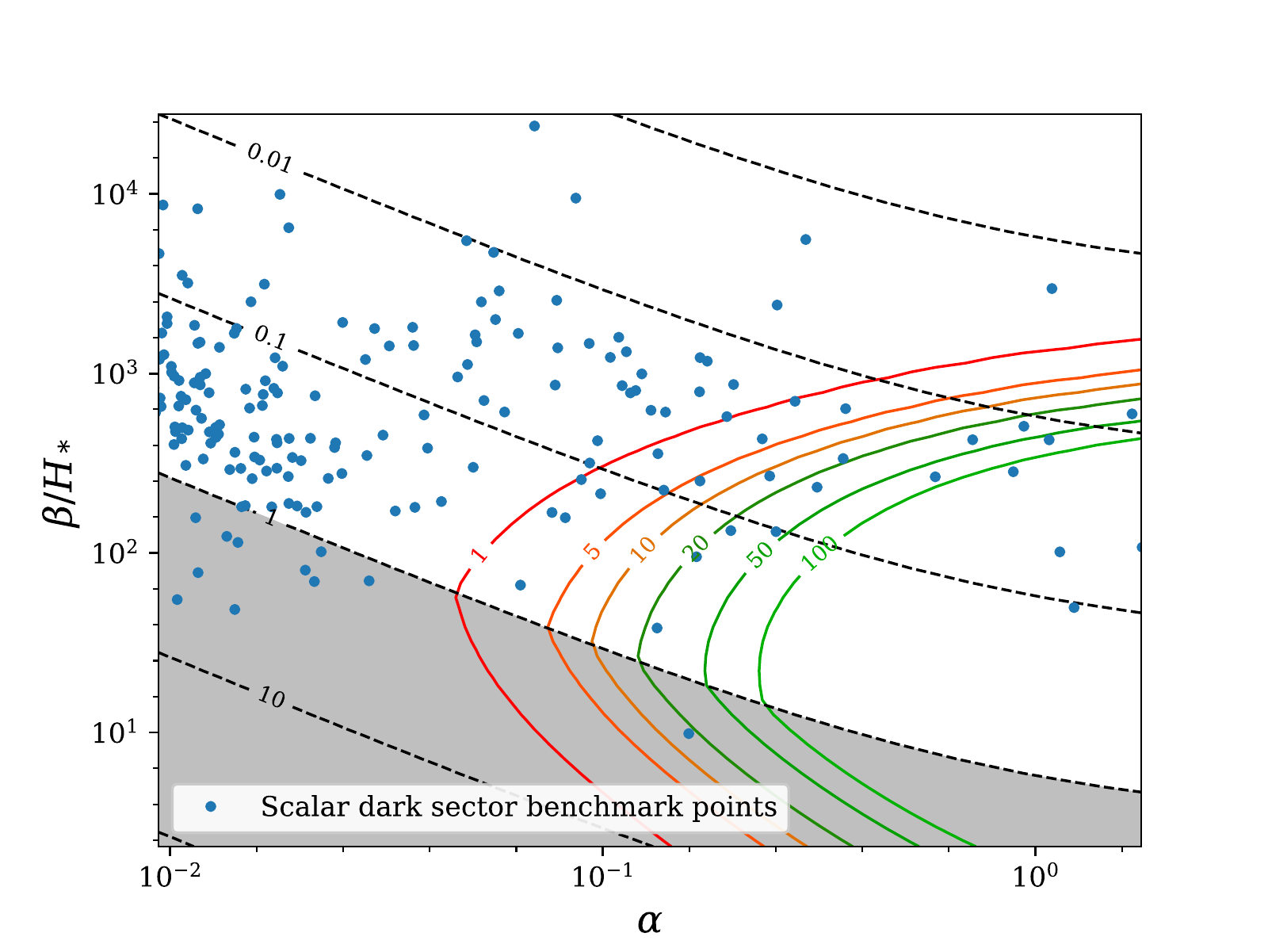}
    \caption{\small SNR plots for the data of~\cite{Breitbach:2018ddu}. 
LEFT: Benchmark points for a model with a spontaneously broken
      $\mathrm{U}(1)$ gauge symmetry in a hidden sector. RIGHT: Results for
      a model with two gauge singlet scalars in a hidden sector. The expected LISA sensitivities correspond to $T_*=50$~GeV for the dark photon model and to $T_*=100$~GeV for the scalar dark sector. In all cases we take $\vw=1$. See text for further details. \label{fig:dark2}}
  \end{center}
\end{figure}

Another source of uncertainty in predicting GW signals from dark sectors is the applicability of the sound wave and turbulence simulations of the GW signal to decoupled dark sectors. In this case, no turbulence or sound waves can be induced in the SM plasma, and it is unclear whether existing results can easily be applied to disturbances in the dark sector plasma after the PT (if a plasma exists at all). This is why in Fig.~\ref{fig:dark1} we separately show the scalar field and sound wave contributions to the GW spectrum. Depending on the concrete nature of the hidden sector, one or the other should be closer to the truth. 
In practice even a small coupling between the dark and visible sectors is sufficient to maintain thermal equilibrium between the two sectors. Therefore if $r_T=1$ one can trust the existing simulations with the same level of confidence as for most other BSM scenarios considered here. 

As a concrete application, in~\cite{Breitbach:2018ddu} two of the simplest hidden sector models featuring a first-order PT were studied, namely a dark photon that receives its mass from spontaneous symmetry breaking, and a simple model of two real singlet scalars. These models are minimal in terms of particle content, while other models often require either higher-dimensional operators or additional fields in order to exhibit a first order PT. 
In Fig.~\ref{fig:dark2} the transition parameters $\alpha$ and $\beta$ are shown for a random scan of model parameters. The expected SNR of LISA is evaluated for $T_{\rm n}=50$~GeV (dark photon) and $T_{\rm n}=100$~GeV (two singlet model). In total, 1000 points for which the transition is first order were scanned, of which at most 5\% are detectable. On the other hand a few points reach very high SNR values, which is somewhat uncommon for renormalisable and perturbative models. 
The temperature ratio in Fig.~\ref{fig:dark2} is set to $r_T=1$, and the sensitivity for other values of $r_T$ can easily be obtained using~(\ref{eqn:darksector1})-(\ref{eqn:darksector2}). 

\subsubsection{DM and collider connection}
If the dark sector features a thermal DM candidate, it is safe to assume that both sectors were in equilibrium for most of the cosmic history. Furthermore, the WIMP miracle motivates DM masses, and therefore the scale of potentially relevant PTs, around the TeV scale, i.e. within the sensitivity range of LISA and, potentially, other present-day terrestrial experiments. It is therefore interesting to explore the complementarity with direct detection and collider experiments in searching for those models. We will here focus on one example where DM arises in a model of gauged lepton number, and where the lepton number breaking PT is the source of the GWs~\cite{Schwaller:2013hqa,Madge:2018gfl}. Similar studies appeared recently in~\cite{Chala:2016ykx,Baldes:2018emh}. 

The model features an additional $U(1)_L$ gauge symmetry under which all SM leptons carry unit charge. Additional leptons which are vector-like with respect to SM interactions, and which become massive after $U(1)_L$ is spontaneously broken by a VEV of a complex scalar $\Phi$, have to be introduced to cancel anomalies. The lightest new neutrino becomes the DM candidate, stabilised by a residual global symmetry that remains after gauge symmetry breaking. LEP data constrains the $L$-symmetry breaking scale $v_\Phi$ to be above $2$~TeV, putting the PT within reach of LISA. 

\begin{figure}[t]
\centering
 \includegraphics[width=0.49\textwidth]{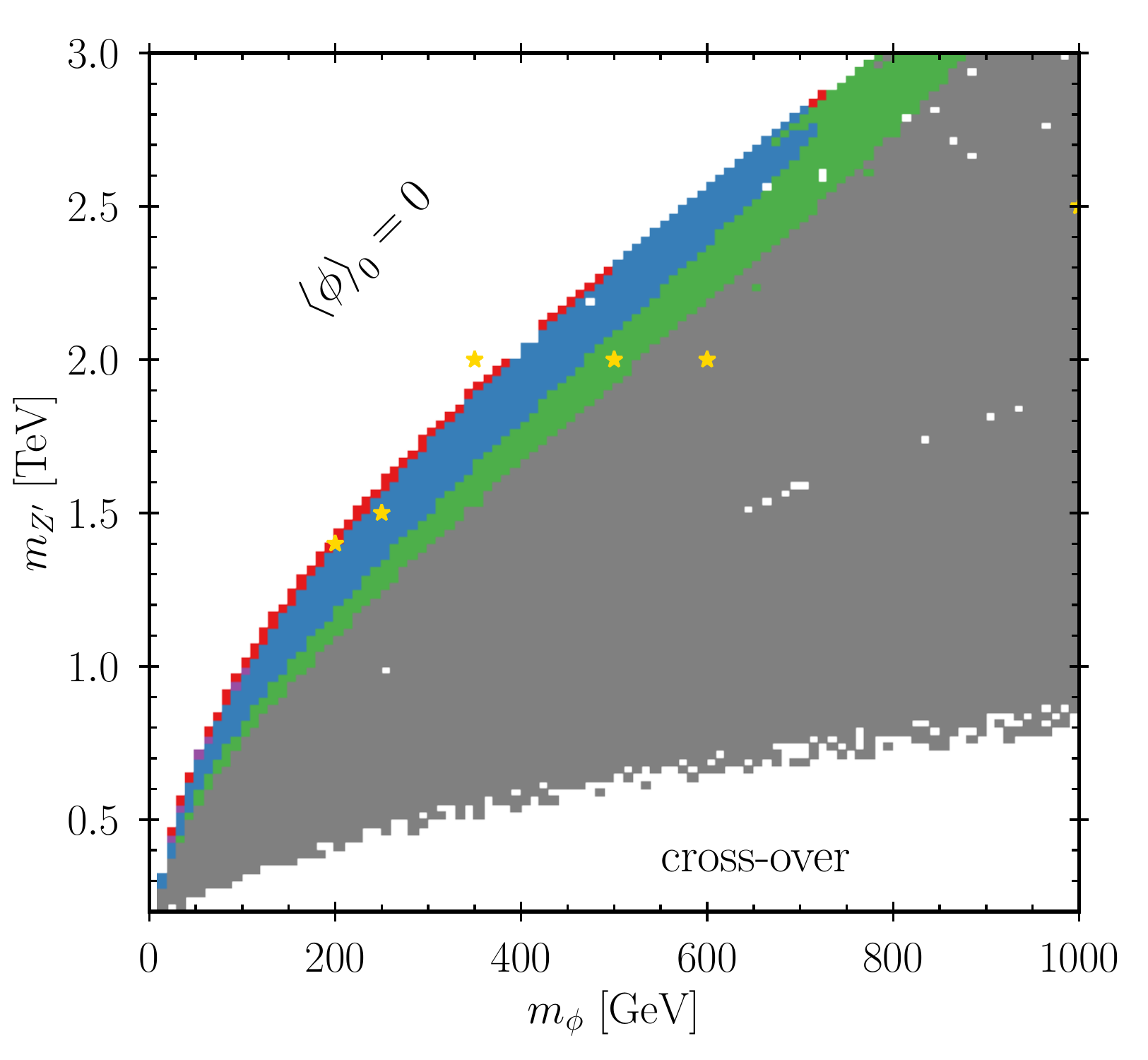}
  \includegraphics[width=0.49\textwidth]{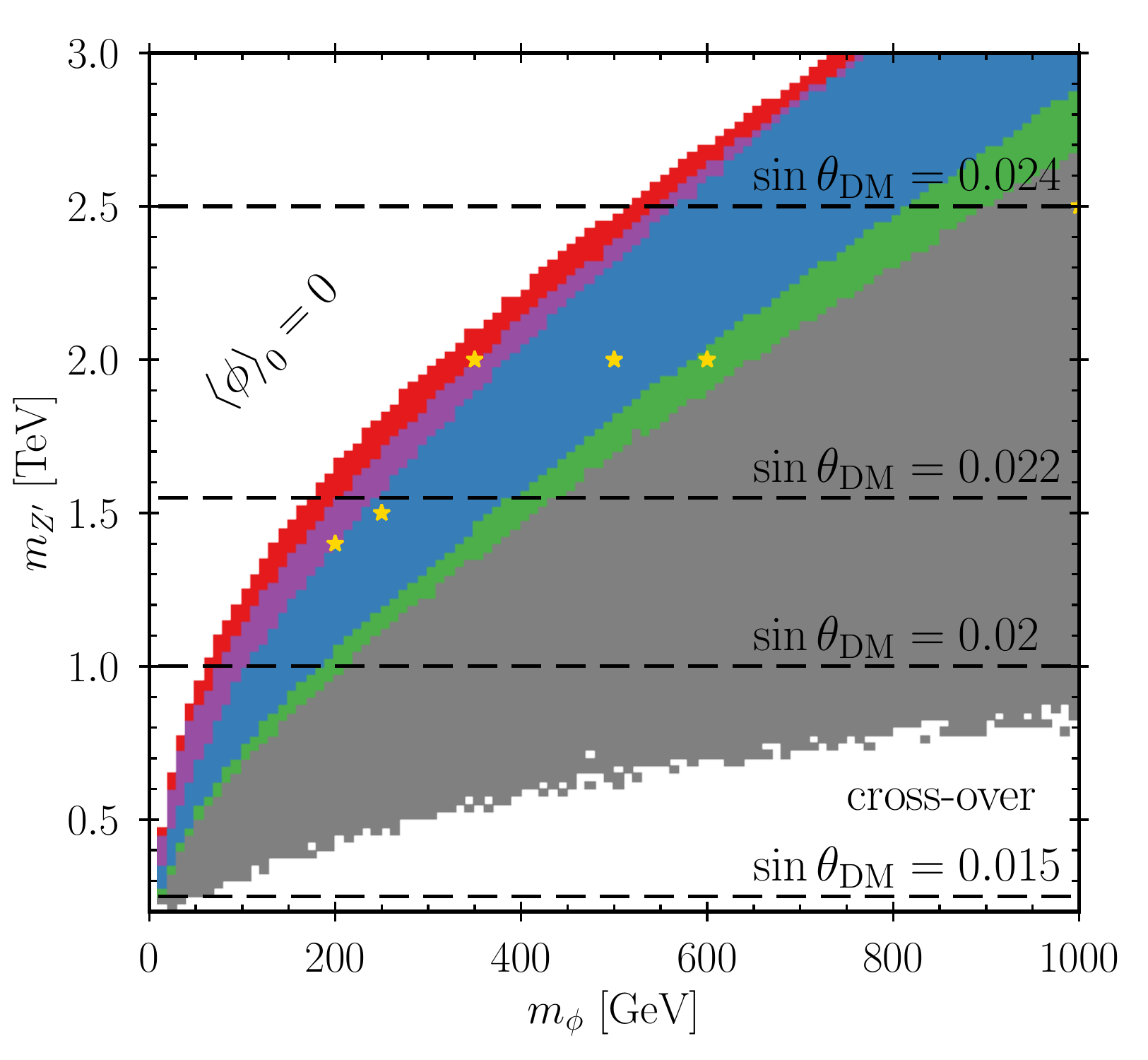}
 \caption{\small\label{fig:dark3}Reach of LISA for gauged lepton number DM model (red shaded region), from~\cite{Madge:2018gfl}. LEFT: Results for a pure abelian Higgs model with $v_\Phi=2$~TeV. RIGHT: The Yukawa couplings are chosen such that the correct DM abundance is obtained. The region below the horizontal dashed lines will be probed by Xenon1T, for the indicated value of the DM mixing angle. Notice that in this figure slightly different LISA sensitivities were used. For more details, see~\cite{Madge:2018gfl}. }
\end{figure}

In the simplest case, the effects of the Yukawa couplings of $\Phi$ to the new leptons can be neglected, and the PT depends on only three parameters, which can be taken to be $v_\Phi$ and the masses of the $U(1)_L$ gauge boson $Z_L$ and the real scalar $\phi$. A PT sufficiently strong so as to be detectable by LISA requires a sizeable hierarchy of masses $m_{Z_L}/m_\phi \approx 4-10$, with $m_{Z_L}$ typically above a TeV, as can be seen in Fig.~\ref{fig:dark3}. Such a heavy leptophilic gauge boson is difficult to detect at the LHC, but is certainly within reach of a high energy lepton collider such as CLIC or a future 100~TeV hadron collider like FCC-hh. Detectability of the scalar depends on its mixing with the SM Higgs, which here is assumed to be small in order to leave the EWPT unaffected. 

\begin{figure}
  \begin{center}
    \includegraphics[width=0.6\textwidth]{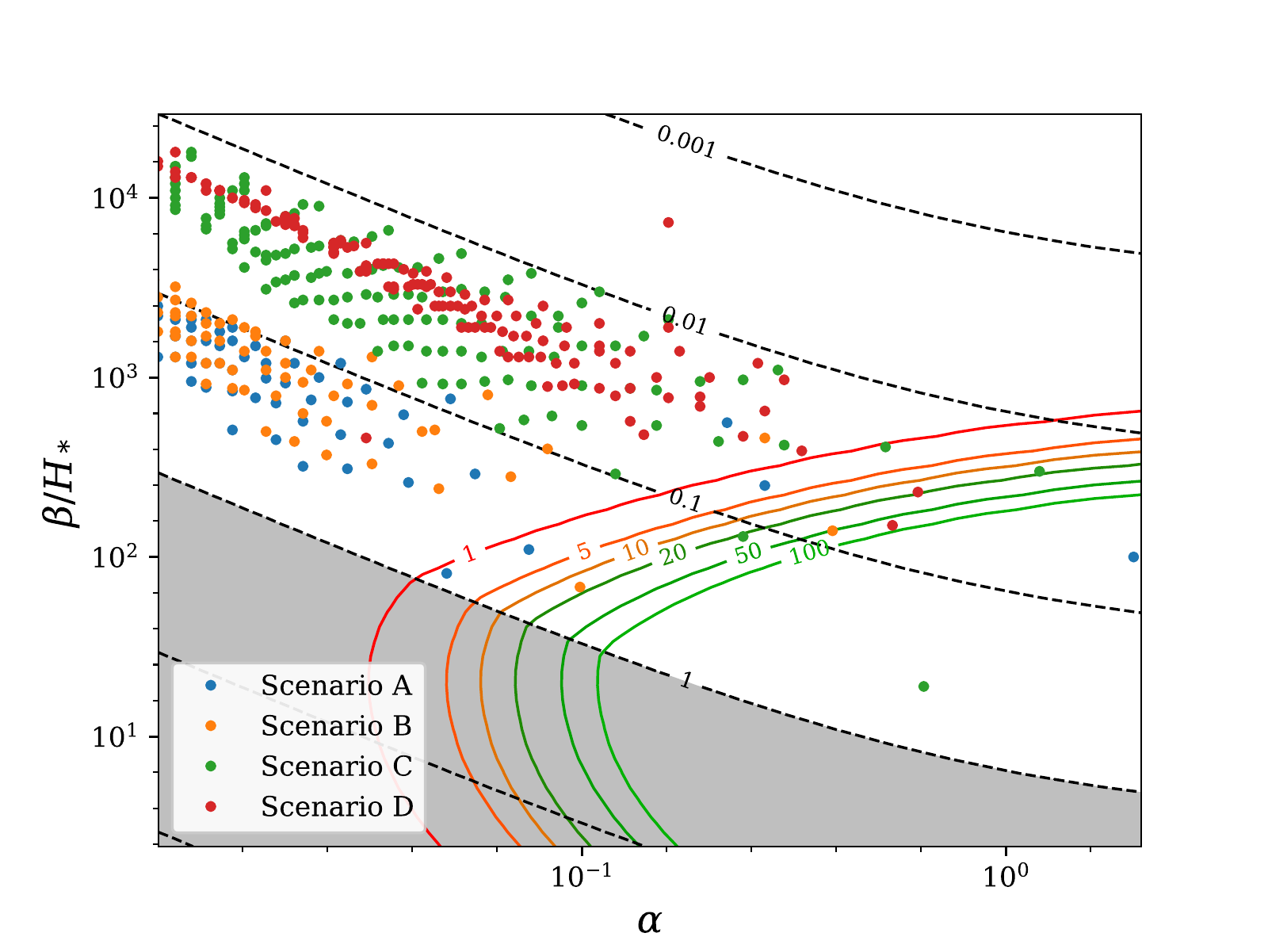}
    \caption{\small SNR plot for a gauged lepton number DM model. See text for details on the scenarios A-D. The expected LISA sensitivities correspond to $T_*=500$~GeV and $\vw=1$. \label{fig:dark4}}
  \end{center}
\end{figure}
 
Including the lepton Yukawa couplings, and in particular setting the DM mass such that it reproduces the correct relic abundance, has an interesting effect on the phenomenological predictions, namely the region in the $m_\phi$--$m_{Z_L}$ plane that can be probed by LISA is increased significantly, as shown in Fig.~\ref{fig:dark3}. While searches for leptophilic DM at colliders is notoriously difficult, the whole parameter space is in principle detectable by Xenon1T~\cite{Aprile:2018dbl} or other future direct detection experiments. As far as  complementarity with LISA is concerned, here the PT is not essential for producing or setting the DM relic abundance, and as such one cannot conclusively probe the model with GWs alone. However, one could for example envision a scenario where $Z_L$ is found at a future collider but the corresponding scalar $\phi$ remains elusive. Observation of a signal in LISA could then point to the correct mass range for the scalar, and thus aid in its discovery. The SNR values for several benchmark scenarios are shown in Fig.~\ref{fig:dark4}. The different scenarios are as follows: for case A, the effects of the fermions in the model are neglected, while for B and C the fermion masses are fixed, and in scenario D the fermion masses are chosen such that the correct DM relic abundance is obtained. LISA can in principle be sensitive to all of these cases.

\section{Conclusions and outlook}
\label{sec:summ}

In this study, we have re-assessed the prospects for observing gravitational waves (GWs) from cosmological phase transitions (PTs) at LISA in light of several recent developments. Our analysis differs in several important ways from that in our previous study~\cite{Caprini:2015zlo}:
\begin{itemize}
\item The scalar field contribution to GW production is now known to be negligible in most cases of thermal PTs. This impacts the detectability of certain BSM scenarios.
\item The limited duration of the sound wave source when the timescale for shock formation is shorter than a Hubble time is accounted for in our updated projections. This reduces the strength of the signal in many scenarios relative to the case where Hubble damping is assumed to cut off the source.
\item Our analysis conservatively neglects GW production from turbulent flow, as current simulations cannot yet observe the onset of turbulent motion and the contribution from shocks is subject to quite large uncertainties, especially for large fluid velocities. Note that these contributions, once included, will likely improve the detection prospects at LISA.
\item We have investigated a wider range of specific models, taking into account recent experimental results and constraints that impact the parameter space relevant for LISA. 
\end{itemize}
As our understanding of GW production at cosmological PTs is rapidly evolving, we have presented an online tool, \texttt{PTPlot}, to allow the BSM community to accurately determine the detectability of a given model at LISA under the above assumptions. This tool was developed by D.~Weir and will continue to be updated as new results become available.

In addition to the advancements described above, we have attempted to clarify common misconceptions in the literature (e.g.~the definition of $\alpha$, the percolation temperature, determination of $\beta/H_*$,  appearance of runaway bubble walls, etc.) and identify important open questions relevant for LISA. Our analysis is mostly based on the GW spectrum as observed in hydrodynamic simulations. These simulations are the state-of-the-art when it comes to PTs with moderate wall velocities and latent heat. One open question is the form of the GW spectrum generated at very strong PTs, as well as the impact of turbulence and magnetic fields.  Furthermore, when it comes to the dynamics of the bubble walls, the wall velocity in these simulations is determined from a phenomenological friction term. {\em Ab initio} calculations of this friction is in many models  still lacking. The stability of the bubble wall front is also a topic that is debated in the literature~\cite{Huet:1992ex, Link:1992dm, Abney:1993ku, Ignatius:1993qn, KurkiSuonio:1995pp, KurkiSuonio:1995vy, Rezzolla:1996ey, Fragile:2003bw, Megevand:2013yua, Megevand:2014yua, Megevand:2014dua}. Additionally, there remain several open questions regarding perturbative calculations of the PT parameters. Computations of $\alpha$, $\beta$, and $T_{\rm n}$ relying on the conventional effective potential approach can be subject to large uncertainties. Nonperturbative results are becoming available for more models, however PTs strong enough to be detected by LISA remain challenging to reliably simulate on the lattice.

We stress that a large part of our discussions are not tight to the LISA frequency window and can be readily applied to other frequency bands. In fact, the strength of the phase transition does not depend on the energy scale of the PT but only on the shape of the scalar potential. Instead of the Higgs field, one could consider other scalar fields responsible for symmetry breaking at higher energies.  In particular, the strong signals predicted for instance from nearly-conformal potentials can be readily applied to higher energy scales and therefore translated to higher frequencies, as to be detectable at either LIGO or at  future ground-based observatories Einstein Telescope and Cosmic Explorer. 

In light of these developments and uncertainties, we have argued that LISA remains a powerful tool for exploring BSM scenarios predicting first-order cosmological PTs. The models we have considered are motivated by various compelling arguments (e.g.~the existence of dark matter, the hierarchy problem) and represent only a fraction of scenarios that can be probed by LISA. We have found that in some cases, LISA will be able to explore parameter space also accessible by future experiments, such as the high-luminosity LHC or dark matter searches, allowing for an exciting chance of simultaneous discovery at multiple experiments. In other models, LISA is likely to provide the most sensitive probe. Our results emphasize LISA's importance for exploring the early Universe and BSM physics, as well as the Mission's complementarity with other experimental efforts.

\section*{Acknowledgements}
MC was funded by the Royal Society under the Newton International Fellowship program. 
GD would like to thank CNPq (Brazil) for financial support. 
MH was supported by the Science and Technology Facilities Council (grant number ST/P000819/1), and the Academy of Finland (grant number 286769). 
SJH was supported by the Science and Technology Facilities Council (grant number ST/P000819/1).
The work of JK was supported by  Department of Energy (DOE) grant DE-SC0019195 and NSF grant PHY-1719642. 
TK and GS are funded by the Deutsche Forschungsgemeinschaft under Germany‘s Excellence Strategy – EXC 2121 "Quantum Universe" – 390833306. 
JMN is supported by Ram\'on y Cajal Fellowship contract  
RYC-2017-22986, and also acknowledges support from the Spanish  
MINECO's ``Centro de Excelencia Severo Ochoa" Programme under grant  
SEV-2016-0597, from the European Union's Horizon 2020 research and  
innovation programme under the Marie Sklodowska-Curie grant agreements  
690575 (RISE InvisiblesPlus) and 674896 (ITN ELUSIVES) and from the  
Spanish Proyectos de I+D de Generaci\'on de Conocimiento via grant  
PGC2018-096646-A-I00.
KR is funded by the Academy of Finland grants 308791, 319066 and 320123. 
PS is supported by the Cluster of Excellence “Precision Physics, Fundamental Interactions, and Structure of Matter” (PRISMA+ EXC 2118/1) funded by the German Research Foundation(DFG). PS would like to thank Moritz Breitbach and Eric Madge for providing the data for Figures (14), (15) and (16), and Eric Madge for checking the ptplot code.
DJW (ORCID ID 0000-0001-6986-0517) was supported by an Science and Technology Facilities Council Ernest Rutherford Fellowship, grant no. ST/R003904/1, by the Research Funds of the University of Helsinki, and by the Academy of Finland, grant nos. 286769, 324882 and 328958.

\appendix

\section{Phase transition parameters from the lattice \label{sec:appendixlattice}}

In this appendix, we discuss nonperturbative determinations of the PT parameters relevant for LISA. As mentioned in the main text, the starting point for nonperturbative studies of a finite temperature PT is dimensional reduction, in which a 4D theory is matched onto a corresponding 3D theory. This matching can be done perturbatively and has been performed at one to two loop accuracy in the SM and several other models to date \cite{Laine:2017hdk,Farakos:1994kx,Kajantie:1995dw,Laine:1996ms,Bodeker:1996pc,Laine:2012jy,Brauner:2016fla,Gorda:2018hvi,Niemi:2018asa}. In this procedure, a choice of 3D theory must also be made, specifying which degrees of freedom should be integrated out, and which field operators should be included. Once a model is dimensionally reduced, it can be studied either in 3D perturbation theory or, more accurately, on the lattice. The latter constitutes a completely non-perturbative approach, and allows a detailed (in principle exact) investigation of the phase diagram of the DR theory. This procedure involves the ``constrained" effective action
\begin{eqnarray}
e^{-\Gamma[\{\mathcal{\bar{O}}_i\}]} = \int \mathcal{D}\{\Phi_i\} e^{-S_3[\{\phi_i\}]}\delta(\{\mathcal{O}_i-\mathcal{\bar{O}}_i\}),
\end{eqnarray}
where $\mathcal{O}(\{\phi_i\})$ are now certain gauge-invariant operators used as approximate order parameters of the transition ($\mathcal{O}=\phi^\dagger\phi$, \ldots). The method allows us to compute the PT order and strength, the critical and nucleation temperatures\footnote{Note that strictly speaking, existing lattice studies provide values of the nucleation temperature~\cite{Moore:2000jw}, not the transition temperature reflected in~\eqref{eq:Sc_improved}. However, for PTs that have been studied nonperturbatively to date, these temperatures are not expected to differ significantly. Therefore, in discussing nonperturbative results throughout our study, we do not distinguish between $\TN$ and $T_{*}$. } ($H_{\rm n}$), latent heat ($\alpha$) and bubble nucleation rate ($\beta$). Once the phase diagram for a given 3D effective model has been determined, different 4D models may be mapped onto it, allowing us to extract non-perturbative information about 4D PTs in a wide class of models from simulations of the same 3D theory.

The EWPT in the SM was first studied using these methods a few decades ago. For the SM, the 3D action is
\begin{eqnarray}
\label{eq:3Dtheory}
S_3=\int d^3x\left[ \frac{1}{2}\textrm{Tr} \,W_{rs}W^{rs}+(D_r\phi)^\dagger D^r\phi+m_3^2\phi^\dagger\phi+\lambda_3(\phi^\dagger\phi)^2\right]
\end{eqnarray}
(here $W_{rs}$ is the 3D SU(2) gauge field strength and the ``3'' subscripts refer to 3D quantities). Through the matching procedure, the parameters $g_3$ (in the covariant derivative), $\lambda_3$ and $m_3$ can be expressed as functions of the 4D parameters and $T$. The effective action in~(\ref{eq:3Dtheory}) neglects higher-dimensional operators, such as $(\phi^\dagger \phi)^3$, introduced by the dimensional reduction because in the SM their effects are small~\cite{Kajantie:1995dw}. Defining $x=\lambda_3/g_3^2$ and $ y=m_3^2/g_3^4$, lattice studies have shown that the 3D theory (\ref{eq:3Dtheory}) features a first order PT for $y\simeq 0$, $0<x<0.11$, while it predicts an equilibrium cross-over elsewhere ($y\simeq 0$, $x>0.11$.) \cite{Kajantie:1995kf}. In addition to the equilibrium properties of the PT (such as $\alpha$ and $\Tc$), it is also possible to compute the nucleation rate non-perturbatively in a 3D DR theory. This was done some time ago for the SM with an unphysically light Higgs mass \cite{Moore:2000jw} and for the cubic anisotropy model (similar to a 2-scalar field model) \cite{Moore:2001vf}. 

For the observed value of the Higgs mass, $x>0.11$ in the SM and the methods described above predict a smooth cross-over. However, in certain regions of parameter space, scenarios beyond the SM can be matched on to the 3D effective theory described by (\ref{eq:3Dtheory}). One can thereby extract information about the phase structure in BSM theories from existing SM lattice studies. In order for~(\ref{eq:3Dtheory}) to be an accurate description of the long distance physics, the new degrees of freedom introduced in the model should be heavy at the PT (so that they can be integrated out), new contributions to higher-dimensionalal operators induced by integrating out the new fields should be small enough to neglect, and couplings should be small enough so that the perturbative matching procedure of DR accurately described the underlying theory. This matching onto the theory described by~(\ref{eq:3Dtheory}) has been performed for the singlet, doublet and triplet models (discussed further below) and existing lattice results used to study the phase structure in certain regions of parameter space. In each case, these studies~\cite{Brauner:2016fla, Gould:2019qek, Andersen:2017ika, Gorda:2018hvi, Niemi:2018asa} have found phenomenologically viable regions of the parameter space predicting a first-order PT.

The 3D theory described by~(\ref{eq:3Dtheory}) can only go so far in describing BSM scenarios. As emphasized in e.g.~\cite{Niemi:2018asa, Gould:2019qek}, requiring higher-dimensional operator effects to be small and new light fields to not play a dynamical role at the PT limits the parameter space in extensions of the SM that can be accurately studied by~(\ref{eq:3Dtheory}). In fact,~\cite{Gould:2019qek} recently argued that \emph{any} model mapping on to the 3D description of (\ref{eq:3Dtheory}) and for which higher-dimensional operators can be safely neglected results in PTs that are too weak to yield detectable GW signals at current and planned GW experiments. Instead, nonperturbative predictions for LISA require simulations of models involving either additional light dynamical fields or sizable modifications of (\ref{eq:3Dtheory}) from higher-dimensional operators.

To date very few 3D theories besides (\ref{eq:3Dtheory}) have been studied nonperturbatively.  Examples include the 2HDM, which introduces an additional $SU(2)$ Higgs doublet field to~(\ref{eq:3Dtheory}), and the MSSM, for which the 3D theory includes two $SU(2)$ Higgs doublets and one $SU(3)$ triplet, corresponding to the right-handed stop field.
 In the 2HDM,~\cite{Kainulainen:2019kyp} confirmed that there are phenomenologically viable parameter space points predicting a relatively strong first-order PT. The parameters considered in the latest non-perturbative MSSM study \cite{Laine:2012jy} also yielded a strong first order PT, somewhat stronger than that indicated by a two-loop perturbative analysis, however, the parameters considered in \cite{Laine:2012jy} are already excluded by experimental data. 
 
Currently, no existing nonperturbative studies consider PTs strong enough to predict an observable signal at LISA. Virtually all BSM studies relevant for LISA have instead used the conventional 4D perturbative approach, which can yield significant inaccuracies in the predicted PT parameters. For example, the parameter space points in the singlet and doublet models studied nonperturbatively in~\cite{Niemi:2018asa, Gould:2019qek} typically feature factor of 2-3 differences in the predicted values of $\alpha$ (defined in terms of latent heat) between the 4D perturbative and 3D nonperturbative approaches. Some differences in the predictions stem from the treatment of IR divergences in the 4D theory and can therefore be cured by a 3D perturbative treatment. Inaccuracies can also arise from the neglect of higher-dimensional operators in 3D, or from large couplings limiting the accuracy of the perturbative matching procedure in DR~\cite{Gould:2019qek}. However, even predictions from the two-loop dimensionally reduced effective potential of a given 3D theory can differ substantially from the corresponding lattice results. For example,~\cite{Niemi:2018asa} found a $\sim 20\%$ discrepancy in the determination of $\alpha$ for one of the 2HDM benchmark points considered, while~\cite{Gould:2019qek} found a similarly-sized discrepancy in the predicted value of $\beta/H$ for a representative point in the singlet model.

On the one hand, these comparisons suggest that perturbative approaches can give a reasonable estimate of the PT parameters (at least at the order-of-magnitude level), which can in turn be used to roughly determine the prospects for observing GWs in a given model at LISA. On the other hand, more quantitatively accurate predictions for LISA will in many cases require new model-specific nonperturbative studies. Improving the accuracy of the predicted GW parameters will be important for constraining models in light of LISA data, and will become especially crucial for interpreting results if a stochastic GW background consistent with a cosmological PT is observed.


\bibliographystyle{JHEP}   
\bibliography{LISA}{}

\end{document}